# Continuous and discrete phasor analysis of binned or time-gated periodic decays

Xavier Michalet

*Department of Chemistry & Biochemistry, 607 Charles E. Young Drive E., Los Angeles, CA 90095, USA*

michalet@chem.ucla.edu

Time-resolved analysis of periodically excited luminescence decays by the phasor method in the presence of time-gating or binning is revisited. Analytical expressions for discrete configurations of square gates are derived and the locus of the phasors of such modified periodic single-exponential decays is compared to the canonical universal semicircle. The effects of IRF offset, decay truncation and gate shape are also discussed. Finally, modified expressions for the phase and modulus lifetimes are provided for some simple cases. A discussion of a modified phasor calibration approach is presented, and illustration of the new concepts with examples from the literature conclude this work.

## 1. Introduction

The analysis of the temporal dependence of luminescence (fluorescence, phosphorescence, scattering, etc.) is a topic of great interest in many disciplines, ranging from fundamental photophysical studies to biomedical imaging applications [1-3]. Traditionally, two different approaches have been used to access temporal information: frequency-modulation and pulsed excitation. Analysis of the latter has often relied on time-resolved recording of the resulting emission, and fitting a decay model to the observed temporal profile [4-6]. Recently, phasor analysis [7], also known as A-B plot [8] or polar plot [9] analysis, an alternative approach rooted in the analysis of sinusoidally modulated signals [10], has emerged and gained in popularity. It can be applied to signals resulting from periodic pulsed excitation and recorded with a variety of detector types, and offers a number of advantages to the user, including an intuitive visualization of luminescence lifetime information in the data and rapid computation.

Phasor analysis of signals recorded with time-correlated single-photon counting (TCSPC) hardware, which precisely time-stamps each photon arrival with respect to the excitation pulse, is well-established[11,12] and relies on the fact that the phasor of the *recorded decay* (after correction of the effects of pile-up and electronic response function) can be considered essentially identical to that of the *emitted signal* up to a rotation and/or dilation in the complex plane. However, theoretical results for systems using either sparse sampling or lower photon arrival time resolution (such as time-gated or integrating detectors) are much more limited [13-16]. In particular, a number of subtleties can arise when the time-gating scheme involves *overlapping gates*, or on the contrary, *non-adjacent gates*. In both quasi-continuous (TCSPC) and discrete (time-gated or integrated) acquisition modalities, cases of partial coverage of the laser period (‘truncated’ decays) or offset location of the excitation pulse within the recording time window (decay offset), further affects phasor calculation. With the advent of new detectors with diverse gating schemes, and in particular studies bridging the *in vitro* and *in vivo* realms and involving results obtained with diverse tech-nologies and in different conditions[17], it appears important to investigate the modifications to standard phasor analysis brought about by this type of data.

This article focuses on results for the phasor of *periodic single-exponential decays* (PSEDs), with brief mention of their extension to linear combinations of PSEDs or more general decays. The article is organized as follows: Section 2 introduces basic concepts and definitions regarding gated (as well as ungated) decays encountered in luminescence lifetime experiment involving periodic excitation. The section ends with a short review of different examples of modified PSEDs used throughout the article. Section 3 provides a concise reminder of phasor analysis concepts used in the remainder of the article, in particular the properties of the phasor of convolution products, with special attention to the phasor of decays with finite sampling (which we refer to as 'discrete' phasor). The section ends with analytical expressions for the phasor of time-gated PSEDs for the examples introduced in Section 2, and discusses basic properties of the loci of these phasors (referred to here as 'Single-Exponential Phasor Loci' curves or SEPL, pronounced 'sepal', as seems appropriate considering the diversity of shapes adopted by the curves studied in this article). In Section 4, the effect of a *decay offset*, which are non-trivial for discrete decays, and in Section 5, the effect of *decay truncation* on the phasor of PSEDs are studied. Section 6 provides a brief overview of the influence of *gate profile* (that is, profile different from the square gate used as an illustration throughout the article) on previous results. Section 7 discusses extensions of the standard phase and modulus lifetime definitions for some of the cases discussed in the previous sections.

These elementary results being established, Section 8 examines modifications to *phasor calibration* in the different situations described previously, with the goal to map these different situations to a few 'canonical' ones, in order to facilitate data interpretation. In particular, we investigate the effect of standard phasor calibration in situations where the SEPL cannot be mapped to a canonical situation, in both gated and ungated cases. The results of section 8 are what may be of most interest to a casual reader, the preceding ones preparing the theoretical ground for it. It is possible to read it without prior knowledge of the preamble to get a gist of the results established in this work.

Section 9, which briefly discusses a few recently published studies in light of the previous results, may help better grasp when their application is important.

A graphical overview of the content of these sections is provided in Fig. 1 for reference.

Finally, Section 10 summarizes the concepts introduced in the article and provides general recommendations.

As a final note, let us clarify that this article does not intend to provide either an introduction to phasor analysis, or a discussion of its merits and pitfalls, all of which have been discussed in many excellent publications (for instance ref. [18,19]) and are recommended reading. However, it does not assume prior knowledge about any analytical aspects of phasor analysis. In order to provide a self-contained study and the necessary definitions, some results previously derived by others will inevitably be repeated, with due reference to the literature. Most calculations are presented in abridged form in appendixes provided in the Supplementary Material. A list of notations is provided in Tables I and II, with mention of where they are defined in the text. Finally, links to free software used in this work and raw data available are provided at the end of the article.

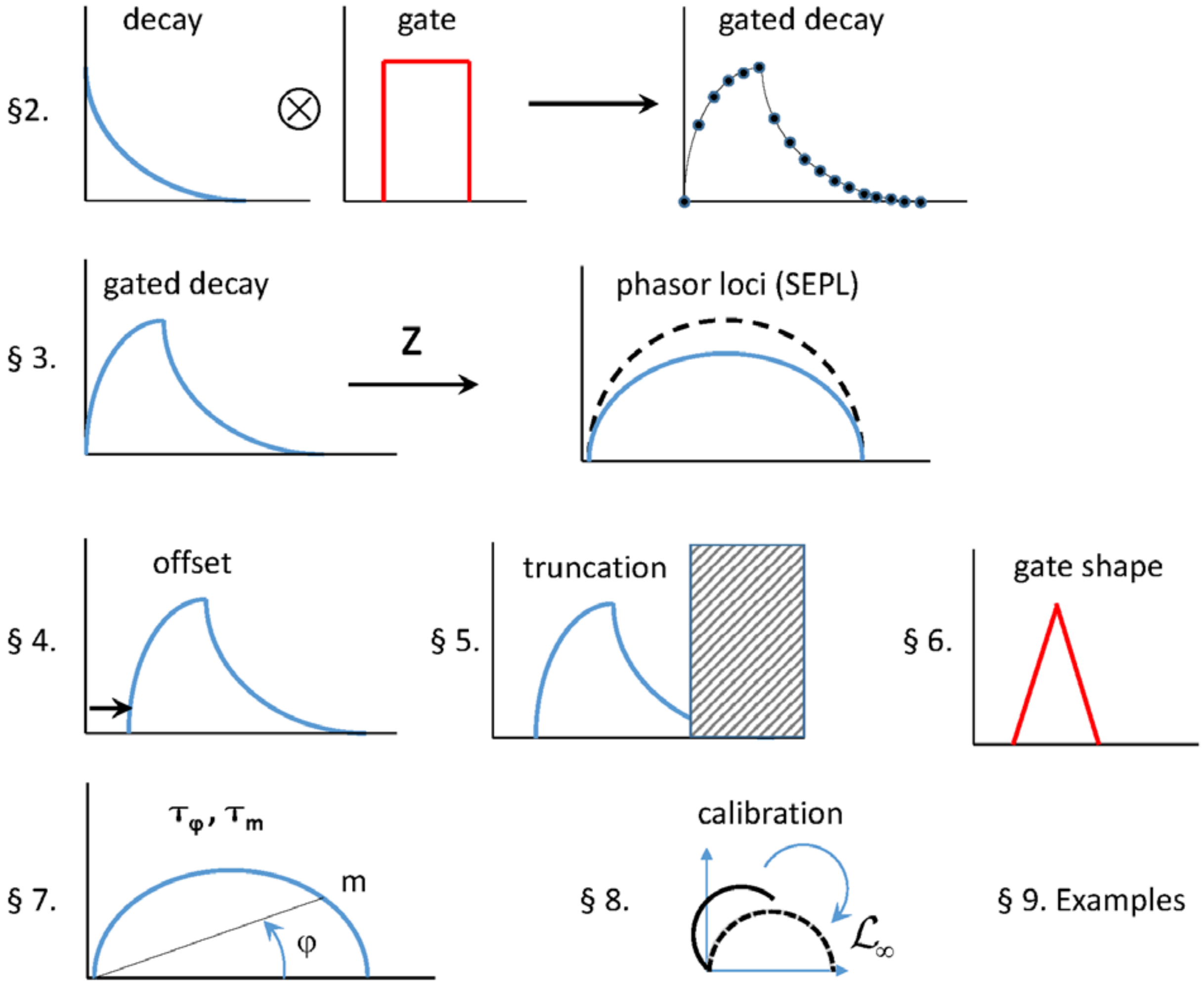

**Fig. 1**: Graphical overview of the article. The general topic of each section is schematically illustrated for easy reference.

## 2. Time-gated periodic decays

### *2.1. Periodic decays*

#### *2.1.1. Excitation pulse, pure decay and emitted signal*

Steady-state $T$-periodic excitation of a system results in a $T$-periodic emitted signal, whose temporal profile is the sum of the signals excited by individual pulses. Let $\mathcal{E}_0(t)$ be the *signal emitted by the system after a single excitation pulse* $x_0(t)$, the latter being generated nominally at time $t_0$. Since the excitation pulse will in general have a non-instantaneous profile, the definition of time $t_0$ is somewhat arbitrary. To fix ideas, we will make the reasonable assumption that $x_0(t)$ reaches a maximum at a well-defined time, which we will call $t_0$. Initially we will assume $t_0 = 0$ for simplicity, but examine the general case in Section 4.

Introducing $F_0(t)$, the *response of the sample to a single Dirac excitation pulse* $\delta(t)$, which we will refer to as the samples *pure decay,* the emitted signal $\varepsilon_0(t)$ after a *single* non-Dirac excitation pulse $x_0(t)$ is given by the convolution product:

$$\varepsilon_0(t) = \int_{-\infty}^{+\infty} x_0(u)F_0(t-u)du = x_0 * F_0(t) \tag{1}$$

By definition, $F_0(t)$ is equal to zero for $t<0$, and in general (but not necessarily) decays from a maximum value reached at $t_{max} \geq 0$, to 0 as $t \to \infty$.

*2.1.2. T-periodic summation and periodic signal*

The steady-state emitted *T-periodic* signal $\varepsilon_{0,T}(t)$ obtained by the superimposition of the responses to *infinitely many* excitation pulses separated by a period *T* is given by the *T-periodic summation*:

$$\begin{cases} \varepsilon_{0,T}(t) = \sum_{k=-\infty}^{+\infty} \varepsilon_k(t) = \sum_{k=-\infty}^{n(t)} \varepsilon_k(t) \\ n(t) = \lfloor t/T \rfloor \end{cases} \tag{2}$$

where we have introduced the definitions:

$$\begin{cases} \varepsilon_k(t) = \int_{-\infty}^{+\infty} x_k(u)F_0(t-u)du = \varepsilon_0(t-kT) \\ x_k(t) = x_0(t-kT) \end{cases} \tag{3}$$

and the notation $\lfloor x \rfloor$ denotes the 'lower' integer part of *x* (the `floor` function of programming languages):

$$\forall x \in \mathbb{R},\ x \in [n, n+1[, n \in \mathbb{Z} \Rightarrow \lfloor x \rfloor = n \tag{4}$$

while the index *T*, such as $\varepsilon_{0,T}$ in Eq. (2) indicates that it is a *T*-periodic function, as will be the convention in the remainder of this article.

$\varepsilon_k(t)$ is the systems response to the $k^{th}$ excitation pulse $x_k(t)$. The sum truncation in the right-most side of Eq. (2) is due to the fact that a system cannot respond to an excitation that has not yet taken place at time *t*.

Eq. (2) can be rewritten:

$$\varepsilon_{0,T}(t) = x_{0,T} * F_0(t) \tag{5}$$

where the *T*-periodic excitation function $x_{0,T}(t)$ is defined by the *T*-periodic summation:

$$x_{0,T}(t) = \sum_{k=-\infty}^{n(t)} x_0(t-kT) \tag{6}$$

As before, the sum is truncated at $k = n(t) = \lfloor t/T \rfloor$ as signals do not propagate back in time. For a Dirac pulse, $x_0(t) = \delta(t)$, the resulting *T*-periodic summation is a truncated *Dirac comb* (sometimes designated by the shah symbol *Ш* [20]):

$$\delta_T(t) = \sum_{k=-\infty}^{n(t)} \delta(t - kT) \tag{7}$$

An alternative way to write Eqs. (2) & (6) without the need to introduce explicitly the upper bound *n*(*t*), consists in writing $\varepsilon_0(t)$ and $x_0(t)$ as products of some function with the *Heaviside function* $H(t)$, where:

$$H(t) = \begin{cases} 0 \text{ if } t < 0 \\ 1 \text{ if } t \geq 0 \end{cases} \tag{8}$$

### *2.1.3. Cyclic convolution product*

Introducing $f_T \underset{T}{*} g_T$ the periodic version of the convolution product of two *T*-periodic functions $f_T(t)$ and $g_T(t)$ (*circular* or *cyclic convolution*, see Appendix C.1, Eq. (C2)):

$$f_T \underset{T}{*} g_T(t) = \int_0^T du\, f_T(u)\, g_T(t-u) \tag{9}$$

we can rewrite Eq. (5) as:

$$\varepsilon_{0,T}(t) = x_{0,T} \underset{T}{*} F_{0,T}(t) \tag{10}$$

where we have introduced the *T*-periodic summation $F_{0,T}(t)$ of the samples pure decay, $F_0(t)$, response of the sample to a Dirac excitation, defined by:

$$F_{0,T}(t) = \sum_{k=-\infty}^{+\infty} F_0(t - kT) \tag{11}$$

These two definitions (*T*-summation and cyclic convolution product) will be used extensively throughout this work. The connection with non-periodic functions and regular convolution product is provided by the following identities (derived in Appendix C.1):

$$f_T \underset{T}{*} g_T(t) = g_T \underset{T}{*} f_T(t) = f * g_T(t) = f_T * g(t) = g_T * f(t) = g * f_T(t) = (f * g)_T(t) \tag{12}$$

where *f* and *g* are arbitrary functions with support over $\mathbb{R}$ and $f_T$ and $g_T$ are their *T*-periodic summations.

### *2.1.4. Electronic response function*

The emitted *T*-periodic signal $\varepsilon_{0,T}(t)$ of Eq. (10) is generally detected by a series of instruments (detectors, electronics, etc.) with a characteristic response $E(t)$ to an hypothetical instantaneous incident signal $\delta(t)$, its so-called *electronic response function* (ERF)[21]. The resulting *T*-periodic recorded signal $S_T(t)$ is given by the convolution of the (non-periodic) ERF with the periodic emitted signal:

$$S_T(t) = E * \varepsilon_{0,T}(t) = \int_{-\infty}^{+\infty} du\ E(u)\varepsilon_{0,T}(t-u) = E_T \underset{T}{*} \varepsilon_{0,T}(t) \tag{13}$$

This equation introduces $E_T(t)$, the *T*-periodic summation of $E(t)$ and rewrites the recorded signal as a cyclic convolution product.

Note that at this stage, we do not specify whether the detector is time-gated or not. We will make this distinction in a later section (Section 2.2).

### 2.1.5. Instrument response function

Eq. (13) can be rewritten:

$$\begin{aligned} S_T(t) &= E_T \underset{T}{*} \left( x_{0,T} \underset{T}{*} F_{0,T} \right)(t) = \left( E_T \underset{T}{*} x_{0,T} \right) \underset{T}{*} F_{0,T}(t) \\ &= I_T \underset{T}{*} F_{0,T}(t) \end{aligned} \tag{14}$$

Eq. (14) introduces $I_T(t)$, the *T*-periodic *instrument response function* (IRF), equal to the cyclic convolution of the *T*-periodic *excitation* function $x_{0,T}(t)$ with the *T*-periodic summation of the *electronic response* function, $E_T(t)$:

$$I_T(t) = E_T \underset{T}{*} x_{0,T}(t) \tag{15}$$

In other words, the instrument response function $I_T(t)$ incorporates the details of the excitation part of the optical setup (including the laser source temporal profile) and those of the detection part (including the electronic finite response time) in a single function, as is well known[21]. While Eq. (15) is useful to understand the contribution of excitation and detection in the IRF, $I_T(t)$ is in practice the only measurable quantity in an experimental system. Eq. (14) shows that it is all that is needed to account for the recorded signal $S_T(t)$ if the source signal functional form $F_0(t)$ (or its *T*-periodic summation, $F_{0,T}(t)$) is known, a property which is at the core of the convolution properties of continuous phasors, as discussed in a later section.

### 2.1.6. Examples of periodic decays

a. Example 1: Dirac IRF and single-exponential decay

To illustrate the difference between a single-period response and the summed, *T*-periodic response, it is useful to consider the case of a Dirac IRF, $\delta(t)$, and a normalized *single-exponential decay with lifetime* $\tau$, $\Lambda_\tau(t)$, whose analytical expression is easily computed, starting from:

$$\begin{cases} I(t) = \delta(t) \\ F_0(t) = \Lambda_\tau(t) \triangleq \dfrac{1}{\tau} e^{-t/\tau} H(t) \end{cases} \tag{16}$$

where $H(t)$ is the Heaviside function and both $I(t)$ and $\Lambda_\tau(t)$ have an integral of 1 over $]-\infty, +\infty[$ (symbol $\triangleq$ will be used to indicate that the definition of the term to the left of that symbol is given by the expression on the right).

It is easy to verify that the corresponding *T*-periodic decay ($F_{0,T}(t)$ in Eq. (2) or (5)) is (Appendix D, Eq. (D1)):

$$\Lambda_{\tau,T}(t) \triangleq \frac{1}{\tau\left(1-e^{-T/\tau}\right)} e^{-\left(t-\lfloor t/T \rfloor T\right)/\tau} = \frac{1}{\tau\left(1-e^{-T/\tau}\right)} e^{-t[T]/\tau} \tag{17}$$

where we have introduced the *modulo T* operation, $t[T]$ :

$$t[T] = t - \lfloor t/T \rfloor T \quad \in [0,T[ \tag{18}$$

In other words, in this simple case, the total emitted signal is simply a scaled (and *T*-periodic) version of the original signal $\Lambda_\tau(t)$ over [0, *T*[ ($t \in [0,T[ \Rightarrow t[T] = t$). Its integral over [0, *T*] is equal to 1.

b. Example 2: Single-exponential IRF and single-exponential decay

Another simple example is provided by a single-exponential IRF with time constant $\tau_\times$ convolved with a single-exponential decay with lifetime $\tau$:

$$\begin{cases} I(t) = \Lambda_{\tau_\times}(t) = \dfrac{1}{\tau_\times} e^{-t_\times/\tau} H(t) \\ F_0(t) = \Lambda_\tau(t) = \dfrac{1}{\tau} e^{-t/\tau} H(t) \end{cases} \tag{19}$$

The result of the convolution with the *T*-periodic IRF is (Appendix D, Eq. (D3)):

$$\Psi_{\tau,\tau_\times,T}(t) \triangleq I_T * F_0(t) = I_T \underset{T}{*} F_{0,T}(t) = \Lambda_{\tau_\times,T} \underset{T}{*} \Lambda_{\tau,T}(t) = \frac{\tau\Lambda_{\tau,T}(t) - \tau_\times \Lambda_{\tau_\times,T}(t)}{\tau - \tau_\times} \tag{20}$$

for $\tau \neq \tau_\times$, where $\Lambda_{\tau,T}(t)$ is the *T*-periodic function defined in Eq. (17). A distinct formula (Eq. (D6)) needs to be used when $\tau = \tau_\times$. When $\tau_\times \to 0$, we recover Eq. (17) obtained in the case of a *T*-periodic Dirac IRF. Some properties of these functions are discussed in Appendix D. In particular, its integral over [0, *T*] is equal to 1.

### *2.1.7. General representation of periodic decays*

In the case of arbitrary IRFs, the results obtained for the simple example above can be generalized as discussed next.

a. Decomposition in bases of exponential functions

In the general case, an IRF $I(t)$ can always be expressed as the Laplace transform of some function $g(k)$:

$$I(t) = \int_0^\infty dk\, g(k) e^{-kt} \tag{21}$$

This integral transform can also be rewritten [22]:

$$I(t)=\int_0^\infty d\tau\,\eta_0(\tau)e^{-t/\tau} \tag{22}$$

where $\eta_0(\tau)$ is the weight function of $I(t)$ in the basis of single-exponential functions $\left\{e^{-t/\tau}\right\}_{\tau>0}$. As discussed in ref. [22], $\eta_0(\tau)$ needs not be positive and therefore cannot in general be considered as a probability density of lifetimes (which would also require it to be normalized).

An alternative representation is given by:

$$I(t)=\int_0^\infty d\tau\,\xi_0(\tau)\frac{e^{-t/\tau}}{\tau}=\int_0^\infty d\tau\,\xi_0(\tau)\Lambda_\tau(t) \tag{23}$$

where the basis of decomposition is comprised of the *normalized* exponential functions $\left\{\Lambda_\tau(t)\right\}_{\tau>0}$ defined in Eq. (16), and $\xi_0(\tau)$ is the weight function of $I(t)$ in this basis. This latter decomposition leads to simpler notations in some of the later results. Note that, like $\eta_0(\tau)$, $\xi_0(\tau)$ needs not be positive.

All these representations are related by:

$$\xi_0(\tau)=\tau\eta_0(\tau)=\frac{1}{\tau}g\left(\frac{1}{\tau}\right) \tag{24}$$

The $T$-periodic summation of $I(t)$ can thus be expressed as:

$$\begin{aligned}I_T(t)&=\sum_{n=-\infty}^{\lfloor t/T\rfloor}I(t-nT)=\sum_{n=-\infty}^{\lfloor t/T\rfloor}\int_0^\infty d\tau\,\eta_0(\tau)e^{-(t-nT)/\tau}=\int_0^\infty d\tau\,\eta_0(\tau)\frac{e^{-t[T]/\tau}}{1-e^{-T/\tau}}\\&=\int_0^\infty d\tau\,\eta_0(\tau)\tau\Lambda_{\tau,T}(t)=\int_0^\infty d\tau\,\xi_0(\tau)\Lambda_{\tau,T}(t)\end{aligned} \tag{25}$$

Comparing the last terms of Eq. (23) and Eq. (25), it is clear that the representation of the original function $I(t)$ in terms of $\xi_0(\tau)$ leads to a simpler form for its $T$-periodic summation $I_T(t)$, the two expressions appearing identical except for a replacement of $\Lambda_\tau(t)$ in Eq. (23) by its $T$-periodic summation $\Lambda_{\tau,T}(t)$ in Eq. (25).

b. Application to PSEDs convolved with an arbitrary IRF

Calculations similar to those detailed in Appendix D lead to the following formula for the convolution of an arbitrary $T$-periodic excitation function $I_T(t)$ and a PSED $\Lambda_{\tau_0,T}(t)$:

$$\begin{aligned}I_T\underset{T}{*}\Lambda_{\tau_0,T}(t)&=\int_0^\infty d\tau\,\xi_0(\tau)\Lambda_{\tau,T}\underset{T}{*}\Lambda_{\tau_0,T}(t)=\int_0^\infty d\tau\,\xi_0(\tau)\Psi_{\tau,\tau_0,T}(t)\\&=\int_0^\infty d\tau\,\xi_0(\tau)\frac{1}{\tau-\tau_0}\left(\tau\Lambda_{\tau,T}(t)-\tau_0\Lambda_{\tau_0,T}(t)\right)\end{aligned} \tag{26}$$

When $I(t)$ is a single exponential function with time constant $\tau_\times$, $\xi_0(\tau)=\delta(\tau-\tau_\times)$ and one recovers Eq. (20). Once again, this formula illustrates the advantage of using Eq. (23) to represent $I(t)$, since both formulas are identical save for the replacement of $\Lambda_\tau(t)$ in Eq. (23) by $\Psi_{\tau,\tau_0,T}(t)$ in Eq. (26).

Note that $\Psi_{\tau,\tau_0,T}(t)$ can be itself rewritten as:

$$\begin{cases} \Psi_{\tau,\tau_0,T}(t) = \int_0^\infty d\lambda\, p_{\tau,\tau_0}(\lambda)\Lambda_{\lambda,T}(t) \\ p_{\tau,\tau_0}(\lambda) = \dfrac{1}{\tau-\tau_0}\left(\delta(\lambda-\tau)-\delta(\lambda-\tau_0)\right) \end{cases} \tag{27}$$

which illustrates that, due to the presence of a negative sign, the weight function cannot in general be interpreted as a probability density function of lifetimes [22].

### *2.2. Time-gated or binned periodic signal*

In the previous discussion, we emphasized the fact that details of the excitation and detection processes could be encompassed in a single IRF. While this can be convenient, it is also some time useful to separate out some aspects of the data acquisition process, especially when these aspects can be experimentally controlled. This is in particular the case of the gate duration (or more generally gate shape) in a gated detection scheme, or the bin size or timing resolution in a time-tagged detection system. Separating the effect of gating or binning on the recorded signal thus allows studying their influence on data analysis. In practice, although the experimental situations are different, both gating and binning can be treated by the same simple formalism. The purpose of this section is to elaborate this point. We first briefly discuss the different possible experimental situations, before introducing the formalism allowing their general analytical description.

#### *2.2.1. Detector types*

Time-gating can be implemented in different ways depending on which detector is considered. We will distinguish between integrating and photon-counting detectors.

An example of (time-gated) integrating detector still commonly found at the time of this writing, are intensified cameras [1,23], in which the gain of the cameras intensifier is modulated (either sinusoidally, or turned on and off) periodically during the overall integration time by a camera. To a good approximation, in the case of moderate incident signal and in a range of intensifier gain values which depends on the specific detector, the signal recorded by such detectors is proportional to the applied gain (some secondary effects such as gain saturation at high intensity can come into play and would need to be corrected for). In the case of such time-gated integrating detector, the temporal variation of the gain can be identified with the gate shape discussed later in the text.

Photon-counting detectors come in many different flavors. Detectors such as single-photon avalanche diodes (SPADs), working in the so-called Geiger mode, provide a binary response (0 or 1) to an incoming instantaneous photon flux (they can detect a single photon at a time and two or more photons arriving within the duration of the avalanche are registered as a single event) [24]. By contrast, silicon photomultipliers (SiPM) [25] or hybrid photodetectors (HPD) [26] equipped with appropriate electronics, are capable of actually counting the number of impinging photons during each detection event. All these detectors are capable, with the appropriate electronics, to precisely time-tag each event with respect to a reference pulse (the so-called time-correlated single-photon counting approach, or TCSPC). Alternatively, either by design of the detector or the associated electronics, they can provides information on the finite interval of time (the “gate”), with respect to a reference pulse, during which the detection event took place. In effect, a detector working in this manner will, by accumulation over time, count the number of photons arriving during a specific window of time with respect to a reference pulse. If the gate is defined by an applied electronic signal, it will in general have a shape governed by the response of the electronics, while if it is defined digitally (as for example in the FLIMbox approach [13]) the gate shape will be essentially "square" and amount to time-binning of the equivalent time-tag information. In this respect, time binning of time-tagged data can be viewed as a form of digital time-gating and considered as a special case of time-gating, which is the perspective used in the remainder of the paper.

It is worth mentioning that, in addition to the previous detector specificities, counting electronics having their own limitation, in particular, a maximum counter value $q$ before data needs to be read out. Examples of very different values for $q$ can be found in the literature. For instance, the SPAD array used in ref. [27] is characterized by $q = 255$, while SwissSPAD [28] and SwissSPAD 2 [29] are characterized by $q = 1$. To compensate for this type of limitations, repeated measurements of finite duration (a 'frame' encompassing $L$ gates) can be performed and summed up to form an 'image' comprised of $F$ frames. The combination of all these characteristics results in general in signal saturation equivalent to the well-known pile-up effect caused by detector dead-time, which needs to and can be corrected [16,30]. In the remainder of this article, we will assume that corrected (or non-saturated) time-gated decays are used.

### *2.2.2. Gate profile*

In all cases, the detection efficiency of the detector can be modelled by a *gate function* $\Gamma_{s,W}(t)$ with finite support [$s$, $s$+$W$] such that:

$$\Gamma_{s,W}(t)\ \begin{cases} =0 & \text{if } t<s \\ \in\,]0,1] & \text{if } \in[s,s+W] \\ =0 & \text{if } t>s+W \end{cases} \tag{28}$$

where $s$ is the gates *offset* (with respect to a reference trigger, generally corresponding to the excitation pulse) and $W$ its *width*. The hypothesis of a finite support is appropriate for the specific examples discussed in this study, but can in principle be relaxed to allow for more general integration or modulation schemes or detector types, including sinusoidal ones relevant to frequency modulation techniques. In these cases, the support of the gate function covers the whole laser period (or multiple thereof, as discussed later in this section), and the notion of 'gate width' becomes useless.

The simplest example is a gate function proportional to the *boxcar* function $\Pi_{s,W}(t)$:

$$\Pi_{s,W}(t)=\begin{cases} 0 & \text{if } t<s \\ 1 & \text{if } t\in[s,s+W] \\ 0 & \text{if } t>s+W \end{cases} \tag{29}$$

In general, the detectors response to the applied voltage swing, or the voltage swing itself, is not instantaneous, and the boxcar function may need to be replaced by a function with 'rounder' edges. Herein, we will limit ourselves to the boxcar model (which we will henceforth refer to as a *square gate*), as the results derived with this model are minimally affected by small departures from it. Numerical experimentations with other gate shapes (e.g. triangle, sawtooth, logistic edge or cus-tom gate) can be performed using the accompanying *Phasor Explorer* software (Appendix F) and are discussed in Section 6.

Because gates are generally synchronized with respect to the excitation pulse, we will be interested in periodic versions of $\Gamma_{s,W}(t)$:

$$\Gamma_{s,W,T_G}(t) = \Gamma_{s,W}(t[T_G]) \begin{cases} = 0 & \text{if } t[T_G] < s \\ \in [0,1] & \text{if } t[T_G] \in [s, s+W] \\ = 0 & \text{if } t[T_G] > s+W \end{cases} \tag{30}$$

where $s \in [0, T_G[$ is again the start or *offset* of the gate with respect to the reference trigger, used to define time 0 (vide supra). $T_G = nT$ is the *gate period*, which we will allow to be equal to any multiple of the laser period ($n \geq 1$) to account for situations that may require $n > 1$. For instance, if the gate width $W > T$, it does not make sense to re-open the gate after one laser period $T$, as the previous one will still not be closed. Another possible reason could be that the detector gating electronics is not capable of responding at the laser repetition rate, forcing a decimation of the incoming laser triggers. In the particular case of a square gate (Eq. (29)), a $nT$-periodic version can also be defined as:

$$\Pi_{s,W,nT}(t) = \Pi_{s,W}(t[nT]) = \begin{cases} 0 & \text{if } t[nT] < s \\ 1 & \text{if } t[nT] \in [s, s+W] \\ 0 & \text{if } t[nT] > s+W \end{cases} \tag{31}$$

### *2.2.3. Gate offset*

Experimentally, gate data is acquired for different values of the gate offset $s$, $\{s_k\}_{1 \leq k \leq N}$. Formulas derived next will consider arbitrary values of $s$, the specific case of a finite number of offsets (and therefore, gates) being discussed separately when needed.

Note that definition Eq. (28) and following assume that the offset $s$ in $\Gamma_{s,W}(t)$ represents the beginning of the gates support (that is, the interval over which the gate value is non-zero). It is possible to extend this definition to account for cases where the gates offset is not exactly known and different from the index $s$ used to refer to it:

$$\Gamma_{s,W,s_0}(t) = \begin{cases} 0 & \text{if } t < s + s_0 \\ \in ]0,1] & \text{if } t \in [s+s_0, s+s_0+W] \\ 0 & \text{if } t > s+s_0+W \end{cases} \tag{32}$$

The unknown offset delta, $s_0$, amounts to an IRF offset $-s_0$ as will be discussed in the next section. It is possible to extend this concept of gate offset to gate profiles with support covering the whole gate period.

### *2.2.4. Gated signal and mirrored gate*

The signal accumulated during a gate starting at time $s$, $S_{T,W}(s)$, is given by:

$$S_{T,W}(s) = \int_0^{nT} \Gamma_{s,W,nT}(t) S_T(t)\, dt \tag{33}$$

This function is clearly *T*-periodic, hence the index *T* in the previous notation. This equation can be rewritten as a cyclic convolution product by introducing the *nT*-periodic *mirrored gate function* of the gate function, $\overline{\Gamma}_{W,nT}(t)$, verifying:

$$\overline{\Gamma}_{W,nT}(s-t) = \Gamma_{s,W,nT}(t) \tag{34}$$

This identity is equivalent to:

$$\overline{\Gamma}_{W,nT}(t) = \Gamma_{0,W,nT}(-t) = \Gamma_{0,W,nT}(nT-t) = \Gamma_{0,W,nT}\left(-t[nT]\right) \tag{35}$$

and is a mirror image with respect to half the gate period *nT* of the gate function starting at $t = 0$.

For a square gate (boxcar) function as defined in Eq. (31), the *mirrored square-gate function* of width *W* and period *nT* is defined by:

$$\overline{\Pi}_{W,nT}(t) = \Pi_{0,W,nT}(nT-t) \tag{36}$$

or, explicitly:

$$\overline{\Pi}_{W,nT}(t) = \Pi_{0,W,nT}(nT-t) = \begin{cases} 0 & \text{if } nT-t<0 \\ 1 & \text{if } 0 \le nT-t \le W \\ 0 & \text{if } nT-t > W \end{cases} \tag{37}$$

With definition Eq. (34) of the mirrored gate function, Eq. (33) reads:

$$S_{T,W}(t) = \int_0^{nT} ds\, S_T(s) \overline{\Gamma}_{W,nT}(t-s) = \overline{\Gamma}_{W,nT} \underset{nT}{*} S_T(t) \tag{38}$$

where the cyclic convolution product is defined for a period *nT* and we have used the fact that $S_T(t)$ is also *nT*-periodic. The notation $\underset{nT}{*}$ specifies the period of the convolution product as $T_G = nT > W$, the gate period imposed by the gate width (as noted before, in most cases, $n = 1$).

Because the experimental signal accumulated during a single gate period is generally very small (less than one photon in the case of a photon-counting detector), the signals of several (*L*) gates separated by a duration $T_G = nT$, are integrated to generate what we will designate henceforth as the *integrated gate signal* $S_{T,W}(t)_L$:

$$S_{T,W}(t)_L = \sum_{l=0}^{L-1} S_{T,W}(t + lT_G) \tag{39}$$

Using the *nT*-periodicity of the emitted signal $S_T(t)$, we have:

$$S_{T,W}(s)_L = L\, S_{T,W}(s) \tag{40}$$

Since the two signals differ only by a constant experimental multiplication factor, which does not intervene in the results derived in this discussion, we will omit the distinction between both and henceforth only refer to $S_{T,W}(t)$. However, this multiplication factor would be necessary when considering effects such as shot noise.

As mentioned in the previous section, the gate definition (28), which assumes a perfect knowledge of where the gate starts, might need to be modified into Eq. (32), which introduces an offset delta $s_0$. Plugging this definition in Eq. (33):

$$S_{T,W}(s)=\int_0^{nT}\Gamma_{s,W,s_0,nT}(t)S_T(t)\,dt=\int_0^{nT}\Gamma_{s,W,s_0,nT}(t+s_0)S_T(t+s_0)\,dt$$
$$=\int_0^{nT}\Gamma_{s,W,nT}(t)\,\delta_{-s_0}\underset{T}{*}S_T(t)\,dt=\overline{\Gamma}_{W,nT}\underset{nT}{*}\delta_{-s_0}\underset{T}{*}S_T(t) \quad (41)$$

where $\delta_{-s_0}(t)=\delta(t+s_0)$ is the Dirac function with offset $t_0=-s_0$. In other words, an imperfect knowledge of where the gate starts can be incorporated into an additional IRF offset.

### 2.2.5. *Time-gated instrument response function*

Combining Eq. (38) with Eq. (14), we obtain the following expression for the signal recorded by a setup employing a time-gated detection scheme:

$$S_{T,W}(t)=I_T\underset{T}{*}F_{0,T}(t)\underset{nT}{*}\overline{\Gamma}_{W,nT}(t)=\left(\overline{\Gamma}_{W,nT}\underset{nT}{*}I_T\right)\underset{T}{*}F_{0,T}(t)$$
$$=I_{T,W}\underset{T}{*}F_{0,T}(t) \quad (42)$$

which defines the *time-gated instrument response function* $I_{T,W}(t)$ as:

$$I_{T,W}(t)\triangleq\overline{\Gamma}_{W,nT}\underset{nT}{*}I_T(t) \quad (43)$$

demonstrating that gating is simply adding a product of convolution to the computation (note however that this product involves, not the gate function itself, but a mirrored version).

### 2.2.6. *Time-gated electronic response function*

Using $I_T(t)$ s definition (Eq. (15)), we can further decompose $I_T(t)$ and write:

$$I_{T,W}(t)=\overline{\Gamma}_{W,nT}\underset{nT}{*}E_T\underset{T}{*}x_{0,T}(t) \quad (44)$$

which introduces the *time-gated electronic response function* $E_{T,W}(t)$:

$$E_{T,W}(t)\triangleq\overline{\Gamma}_{W,nT}\underset{nT}{*}E_T(t) \quad (45)$$

This definition formally separates the gating part of the electronics response from any other electronics contributions, which is an appropriate representation if the acquisition electronic does contain two distinct signal processing stages. In cases where such a distinction does not exist or cannot be made explicitly, Eq. (44) for the IRF is replaced by a form identical to that of Eq. (15):

$$I_{T,W}(t)=E_{T,W}\underset{T}{*}x_{0,T}(t) \quad (46)$$

### 2.2.7. *Examples of square-gated PSEDs*

#### a. Example 1: Square-gated PSEDs with Dirac IRF

In the special case of a *T*-periodic Dirac IRF and a periodic single-exponential decay $\Lambda_{\tau,T}(t)$, Eq. (17), and a boxcar gate with width *W* and period *nT*, Eq. (30), we can explicitly compute the definition of the corresponding *square-gated* PSED, $\Lambda_{\tau,T,W}(t)$ by evaluating Eq. (33). The result is:

$$\Lambda_{\tau,T,W}(t) \triangleq \begin{cases} (a) \quad \dfrac{1-e^{-\omega/\tau}}{1-e^{-T/\tau}} e^{-t[T]/\tau} + k, & \text{if } t[T] \in [0, T-\omega[ \\ (b) \quad \dfrac{1-e^{-(\omega-T)/\tau}}{1-e^{-T/\tau}} e^{-t[T]/\tau} + k + 1, & \text{if } t[T] \in [T-\omega, T[ \end{cases} \tag{47}$$

$$\text{where} \quad \begin{cases} k = \lfloor W/T \rfloor \\ \omega = W[T] = W - kT \end{cases}$$

The introduction of $k$ and $\omega$ accounts for cases where $W > T$, while that of $t[T]$ shifts the time argument back to the [0, $T$] interval. Note that with this definition, the integral of $\Lambda_{\tau,T,W}(t)$ over [0, $nT$] is equal to $W$. In most experimental cases, $\omega = W$, $k = 0$ and $n = 1$.

A few examples of square-gated PSEDs are represented in Fig. 2. The *Phasor Explorer* software accompanying this article (Appendix F) allows exploring other types of gate profiles (triangle, sawtooth, etc.), including user-defined ones for which analytical formulas might not be available or convenient to use. We will briefly look at the effect of other gate shapes in Section 6.

b. Example 2: Square-gated PSEDs with single-exponential IRF

The evaluation of Eq. (33) where $S_T(t) = \Psi_{\tau,\tau_\times,T}(t)$, the ungated decay, is given by Eq. (20) is outlined in Appendix D.6 and leads to the following result for the square-gated signal (Eq. (D20) ):

$$\Psi_{\tau,\tau_\times,T,W}(t) = \frac{1}{\tau-\tau_\times}\left(\tau\Lambda_{\tau,T,W}(t) - \tau_\times\Lambda_{\tau_\times,T,W}(t)\right)$$

$$= \begin{cases} \dfrac{1}{\tau-\tau_\times}\left(\tau\dfrac{1-u}{1-y}e^{-t[T]/\tau} - \tau_\times\dfrac{1-u_\times}{1-y_\times}e^{-t[T]/\tau_\times}\right) + k, & \text{if } t[T] \in [0, T-\omega[ \\ \dfrac{1}{\tau-\tau_\times}\left(\tau\dfrac{1-uy^{-1}}{1-y}e^{-t[T]/\tau} - \tau_\times\dfrac{1-u_\times y_\times^{-1}}{1-y_\times}e^{-t[T]/\tau_\times}\right) + k + 1, & \text{if } t[T] \in [T-\omega, T[ \end{cases} \tag{48}$$

where we have used the following notations:

$$\begin{cases} k = \lfloor W/T \rfloor \\ \omega = W[T] = W - kT \\ y(\tau) = e^{-T/\tau}, \; y_\times(\tau) = e^{-T/\tau_\times} \\ u(\tau) = e^{-\omega/\tau}, \; u_\times(\tau) = e^{-\omega/\tau_\times} \end{cases} \tag{49}$$

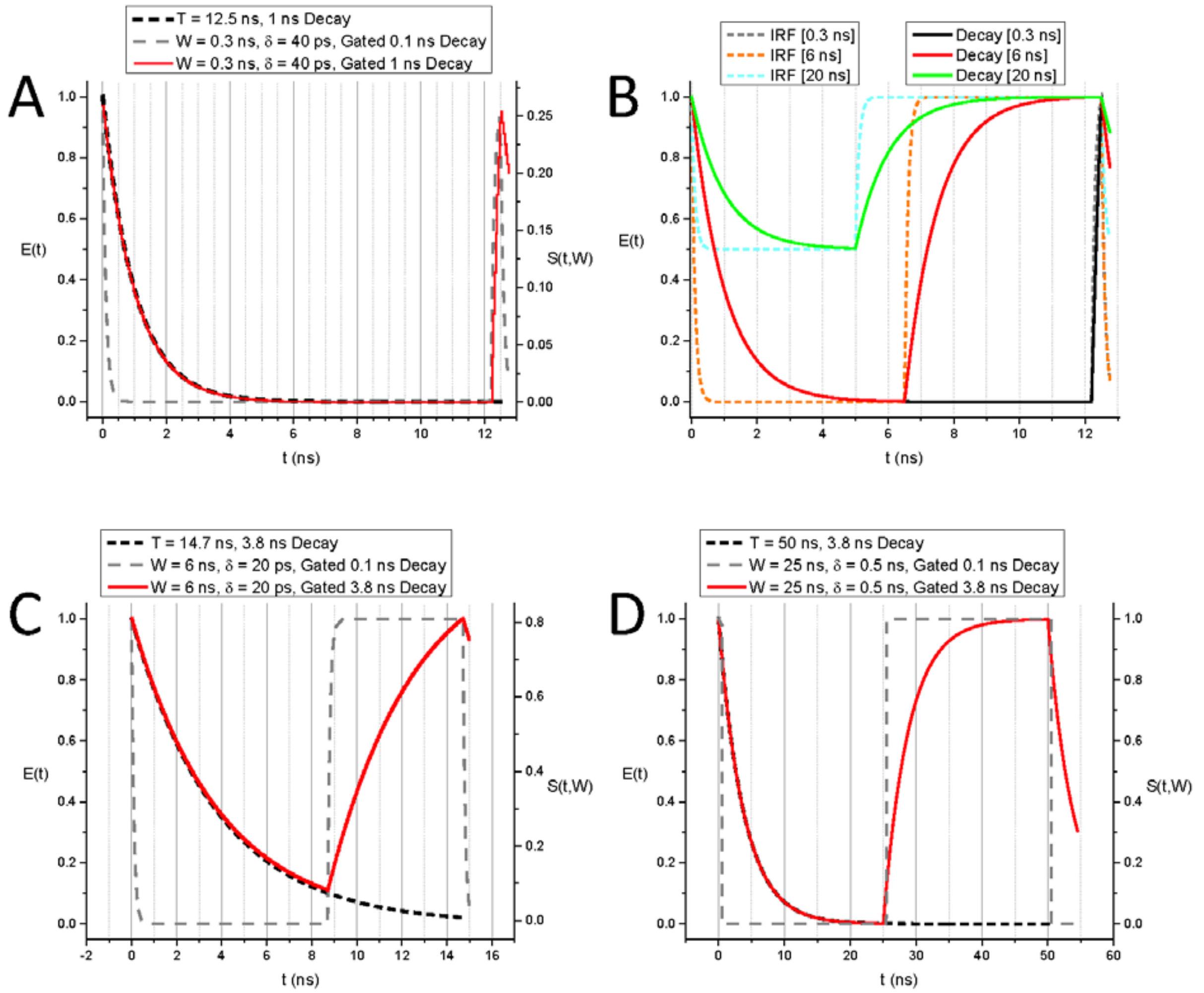

**Fig. 2**: Examples of time-gated exponential decays. A: Gated decays ($\tau_1$= 0.1 ns, representing the equivalent of an IRF, and $\tau_2$= 1 ns, laser period: 12.5 ns) with gate width $W$ = 0.3 ns and gate separation (gate step) $\delta$ = 40 ps, corresponding to settings used in time-gated ICCD measurements performed in [15]. The time-gated decay (red) is essentially identical to the ungated decay (dashed, black), except at the end of the period where the rise time of the time-gated decay is of the order of W = 0.3 ns instead of being instantaneous. B: Same decays (IRF: $\tau_1$= 0.1 ns, Decay: $\tau_2$= 1 ns), but after time-gating with parameters $W_1$ = 0.3 ns (black), $W_2$ = 6 ns (red) or $W_3$ = 20 ns ($\delta$ = 40 ps in all cases). The last two values correspond to characteristics of SwissSPAD 1 & 2 described in [28,29]. For $W_2$ = 6 ns, the main difference is at the end of the decay, where the instantaneous rise time is replaced by a mirror image of the decay, of width $W_2$. For $W_3$ = 20 ns, the resulting decay is offset vertically due to the full period integration (T = 12.5 ns) common to all gates, and the instantaneous rise time is replaced by a mirror image of the decay, of width $W_3 - T$ = 7.5 ns. C: Gated decays ($\tau_1$= 0.1 ns, representing the equivalent of an IRF, and $\tau_2$= 3.8 ns, laser period: 14.7 ns) with gate width $W$ = 6 ns and gate separation $\delta$ = 20 ps, corresponding to settings used in SwissSPAD measurements performed in [28]. The time-gated decay (red) is essentially identical to the ungated decay (dashed, black), except at the end of the period where the rise time of the time-gated decay is replaced by a mirror image of the decay, of width W = 6 ns. D: Gated decays ($\tau_1$= 0.1 ns, representing the equivalent of an IRF, and $\tau_2$= 3.8 ns, laser period: 50 ns) with gate width $W$ = 25 ns and gate separation $\delta$ = 500 ps, corresponding to settings used in SwissSPAD 2 measurements performed in [29]. The time-gated decay (red) is essentially identical to the ungated decay (dashed, black), except at the end of the period where the rise time of the time-gated decay is replaced by a mirror image of the decay, of width W = 20 ns.

### c. Square-gated PSEDs with general IRF

Using the general representation of an IRF $I(t)$ given by Eq. (23), the square-gated PSED obtained with this excitation function is given by:

$$\bar{\Pi}_{W,nT} \underset{nT}{*} I_T \underset{T}{*} \Lambda_{\tau,\tau_0,T}(t) = \int_0^\infty d\tau\, \xi_0(\tau) \bar{\Pi}_{W,nT} \underset{nT}{*} \Psi_{\tau,\tau_0,T}(t) = \int_0^\infty d\tau\, \xi_0(\tau) \Psi_{\tau,\tau_0,T,W}(t) \tag{50}$$

#### *2.2.8. General expression for time-gated PSEDs*

For any other mirrored gate shape $\bar{\Gamma}_{W,nT}(t)$ and an IRF $I(t)$ given by Eq. (23) (or equivalently, its *T*-periodic version $I_T(t)$ given by Eq. (25)), the time-gated PSED is given by the generalization of Eq. (50):

$$\bar{\Gamma}_{W,nT} \underset{nT}{*} I_T \underset{T}{*} \Lambda_{\tau,\tau_0,T}(t) = \int_0^\infty d\tau\, \xi_0(\tau) \bar{\Gamma}_{W,nT} \underset{nT}{*} \Psi_{\tau,\tau_0,T}(t) \tag{51}$$

## 3. Phasor of periodic decays

### *3.1. Definition and notations*

#### *3.1.1. Phasor and Fourier transform*

Phasor analysis is a well-documented approach to study fluorescence decays without having to resort to non-linear model fitting [7,10,19,22]. The formalism found in most discussions in the literature uses non-periodic signals $S(t)$ (equal to zero for $t < 0$) and defines the phasor $z[S](f)$ of signal *S* at harmonic frequency *f*, using infinite integrals:

$$z[S](f) = \frac{\int_{-\infty}^{+\infty} dt\, S(t) e^{i2\pi ft}}{\int_{-\infty}^{+\infty} dt\, S(t)} = \frac{\left\| S(t) e^{i2\pi ft} \right\|}{\left\| S(t) \right\|} \tag{52}$$

where we have introduced the notation $\|S(t)\|$ to denote the integral of $S(t)$ over $]-\infty,+\infty[$:

$$\|S(t)\| = \int_{-\infty}^{+\infty} dt\, S(t) \tag{53}$$

Note that in most cases discussed in the following, the effective integration bounds are 0 and $+\infty$ due to the fact that the decays we will consider in this work are equal to zero for $t < 0$.

From the ratiometric nature of definition (52), it results that the phasor is invariant by dilation:

$$\left(\forall a \in \mathbb{R}^*\right), \quad z[aS] = z[S] \tag{54}$$

This phasor definition is related to the *Fourier transform* $\mathcal{F}[S]$ of *S*:

$$\mathcal{F}[S](f) = \int_{-\infty}^{+\infty} dt\, S(t) e^{-i2\pi ft} \tag{55}$$

defined for any value $f \geq 0$, by the following relation:

$$z[S](f) = \frac{\mathcal{F}^*[S](f)}{\mathcal{F}[S](0)} \tag{56}$$

where $x^*$ indicates the complex conjugate of $x$.

As it is obvious from the above definitions, if $\|S(t)\| = 1$ (*i.e.* if $S(t)$ is normalized), the relation between phasor and Fourier transform is further simplified into:

$$\|S(t)\| = 1 \Rightarrow z[S](f) = \mathcal{F}^*[S](f) \tag{57}$$

Because Eqs. (52) through (57) involve infinite integrals of non-periodic decay functions, the corresponding quantities are not directly accessible experimentally. It therefore appears important to show that this formalism can be replaced by an equivalent one involving finite integrals of periodic decays, which are experimentally accessible quantities.

### *3.1.2. Cyclic phasor and Fourier series*

Definition (52) of the phasor is in fact formally identical to an alternate definition involving *T*-periodic signals, which is more natural when dealing with experimental data. We will devote this section to establishing this connection.

When the phasor harmonic frequency *f* is a multiple of $1/T$:

$$f = f_n = \frac{n}{T},\ n \in \mathbb{N} \Rightarrow (\forall k \in \mathbb{Z})\ e^{i2\pi fkT} = 1 \tag{58}$$

the numerator of Eq. (52) can be rewritten:

$$\begin{aligned}
\int_{-\infty}^{+\infty} dt\, S(t) e^{i2\pi ft} &= \sum_{k=-\infty}^{+\infty} \int_{kT}^{kT+T} dt\, S(t) e^{i2\pi ft} = \sum_{k=-\infty}^{+\infty} \int_{kT}^{kT+T} dt\, S(t) e^{i2\pi f(t-kT)} \\
(u = t - kT) \quad &= \sum_{k=-\infty}^{+\infty} \int_{0}^{T} du\, S(u+kT) e^{i2\pi fu} \\
&= \int_0^T du \left( \sum_{k=-\infty}^{+\infty} S(u+kT) \right) e^{i2\pi fu} = \int_0^T dt\, S_T(t) e^{i2\pi ft}
\end{aligned} \tag{59}$$

in which we have introduced $S_T(t)$, the *T*-periodic summation of $S(t)$ (Section 2.1.1, Eq. (6)). As noted before, $S_T(t)$ is proportional to the signal measured in experiments, while the original signal $S(t)$ is generally not directly measurable.

Introducing the notation $\|S_T(t)\|_T$ for the integral of a *T*-periodic function $S_T(t)$ over a period *T* (we will occasionally use the simpler notation $\|S_T\|_T$ and omit the function's argument):

$$\|S_T(t)\|_T = \int_0^T dt\, S_T(t) \tag{60}$$

Eq. (59) can be rewritten:

$$\left\|S(t) e^{i2\pi ft}\right\| = \left\|S_T(t) e^{i2\pi ft}\right\|_T \tag{61}$$

Similarly, it is trivial to verify that:

$$\|S(t)\| = \|S_T(t)\|_T \tag{62}$$

thus establishing that, for phasor harmonic frequencies *f* equal to a multiple of $1/T$:

$$\underset{C}{z}\left[S_T\right](f) \triangleq \frac{\int_0^T dt\, S_T(t)\, e^{i2\pi ft}}{\int_0^T dt\, S_T(t)} = \frac{\left\|S_T(t)\, e^{i2\pi ft}\right\|_T}{\left\|S_T(t)\right\|_T} = \frac{\left\|S(t)\, e^{i2\pi ft}\right\|}{\left\|S(t)\right\|} = z\left[S\right](f) \tag{63}$$

While the two definitions Eq. (52) & Eq. (63) give identical results, it is again important to notice that one definition involves a non-periodic function ($S(t)$) while the other involves its *T*-periodic summation ($S_T(t)$).

The definition of the phasor of $S_T(t)$, Eq. (63), which we will call the *cyclic phasor* $\underset{C}{z}\left[S_T\right](f)$ (note the symbol 'C' underneath the 'z') because it involves a single period of the recorded periodic signal, connects it to the formalism of *Fourier series*, as discussed next. To simplify notations, we will omit the symbol 'C' below the phasor notation 'z' in the remainder of the discussion, as it should be obvious what definition is used based on the periodicity (or not) of the function involved.

The Fourier series of a *T*-periodic signal $S_T(t)$ is defined as:

$$S_T(t) = \frac{1}{2}a_0 + \sum_{n=1}^{+\infty} a_n \cos 2\pi f_n t + \sum_{n=1}^{+\infty} b_n \sin 2\pi f_n t \tag{64}$$

where the Fourier coefficients $(a_n, b_n)$ and harmonic frequencies $f_n$ are given by :

$$\begin{cases} a_n = \dfrac{2}{T}\displaystyle\int_0^T dt\;\; S_T(t)\cos(2\pi f_n t) \\ b_n = \dfrac{2}{T}\displaystyle\int_0^T dt\;\; S_T(t)\sin(2\pi f_n t) \\ f_n = \dfrac{n}{T}, \quad n \in \mathbb{N} \end{cases} \tag{65}$$

*n* (a positive integer) is the order of the Fourier harmonic. Note that contrary to the Fourier transform, defined for any frequency *f*, Fourier series only involve multiples of the fundamental frequency $f_0 = 1/T$ .

With these definitions, the cyclic phasor (Eq. (63)) can be rewritten:

$$z\left[S_T\right](f_n) = \frac{a_n + ib_n}{a_0} \tag{66}$$

If needed be, with the proper normalization of $S_T(t)$, we can obtain $a_0 = 1$, further simplifying the relation between cyclic phasor and Fourier series component.

*Notations*: in the remainder of this article, we will omit the mention of the phasor harmonic $f = f_n$ in the phasor notation when there is no ambiguity and write instead:

$$z\left[S_T\right](f) \equiv z\left[S_T\right] \tag{67}$$

The choice of the actual harmonic (or harmonics) to use in phasor analysis will not be discussed here, as it to some extent irrelevant to the topics addressed in this article. For some example of considerations involving harmonic(s) choices, see for instance refs. [18,31].

### 3.1.3. *Continuous vs discrete and ungated vs gated phasor*

All definitions so far, including Eq. (65), have assumed that the signal $S_T(t)$ was recorded for all values *t* in [0, *T*], which is an idealization. In practice, a signal is recorded experimentally only at a finite number of *t* values, in which case the integrations in Eq. (63) need to be replaced by summations. We will therefore distinguish in the following between 'continuous' and 'discrete' phasor definitions.

Moreover, most experimental data are effectively binned or time-gated. In other words, while data is tagged with precise time stamps $\{t_p\}_{1\leq p\leq N}$ measured with respect to the previous laser pulse, the recorded signal $S_T(t_p)$ corresponds effectively to the signal integrated over a period of time $[t_p, t_p + W]$, where *W* is the gate width (or bin duration). We will therefore also distinguish between 'ungated' (or instantaneous) and 'gated' (or 'binned') decay definitions, and by extension, speak of phasors of such experimental decays as 'continuous phasors' or 'discrete phasors'.

The next two sections will review the differences between continuous (Section 3.2) and discrete phasors (Section 3.3) of several classes of decays whose phasors can be easily computed analytically: (i) periodic single-exponential decays (PSEDs), (ii) PSEDs with single-exponential IRF, (iii) square-gated PSEDs and (iv) square-gated PSED with single-exponential IRF. Because they are useful in this type of calculations, properties of phasors of convolution products will be examined in both cases (continuous and discrete phasors).

## ***3.2. Phasor of continuous decays: continuous phasor***

### 3.2.1. *Continuous phasor of convolution products*

As discussed in Section 2.1, a recorded periodic decay $S_T(t)$ can generally be expressed as a cyclic convolution product (Eq. (14)):

$$S_T(t) = I_T \underset{T}{*} F_{0,T}(t) \tag{68}$$

where $I_T(t)$ is the *T*-periodic *instrument response function* and $F_{0,T}(t)$ is the *T*-periodic summation of the samples response $F_0(t)$ to a Dirac excitation. $F_{0,T}(t)$ is easily computed if the analytical form of $F_0(t)$ is known, while $I_T(t)$ is in principle measurable experimentally. However, the importance of Eq. (68) comes from the following property of the continuous cyclic phasor established in Appendix C ( '*continuous phasor convolution rule*', Eq. (C12)):

$$z\left[ f_T \underset{T}{*} g_T \right] = z[f_T]\, z[g_T] \tag{69}$$

where $f_T$ and $g_T$ are two *T*-periodic functions. This property simplifies the computation of the continuous phasor of experimental decays obtained as the cyclic convolution product of two or more functions, as encountered in Section 2 (Eqs. (10), (14), (38)).

### 3.2.2. *Continuous phasor of ungated PSEDs*

We will first review a few useful examples of phasors of ungated decays before presenting their counterpart in the presence of a square gate.

a. Ungated PSEDs with Dirac IRF

It is straightforward to verify that for the special case of an ungated PSED with lifetime $\tau$ ( $S_T(t) = \Lambda_{\tau,T}(t)$ defined by Eq. (17)), Eq. (63) reads:

$$z\left[\Lambda_{\tau,T}\right] = \frac{1}{1 - i2\pi f\tau} = \frac{1 + i2\pi f\tau}{1 + (2\pi f\tau)^2} \triangleq \zeta_f(\tau) \tag{70}$$

This expression, which we will refer to as the *canonical phasor* of a PSED with lifetime $\tau$, $\zeta_f(\tau)$, is of course identical to that of the phasor of infinite, non-periodic single-exponential decays generally encountered in the literature, with the following definitions of the phasor *components* ($g$, $s$) and phasor *modulus* $m$ and *phase* $\varphi$:

$$\begin{cases} \zeta_f(\tau) = g(\tau) + is(\tau) = m(\tau)e^{i\varphi(\tau)} \\ g(\tau) = \dfrac{1}{1 + (2\pi f\tau)^2} \\ s(\tau) = \dfrac{2\pi f\tau}{1 + (2\pi f\tau)^2} \\ m(\tau) = \dfrac{1}{\sqrt{1 + (2\pi f\tau)^2}} \\ \tan\varphi(\tau) = 2\pi f\tau \end{cases} \tag{71}$$

The locus ($g$, $s$) of the phasors $\zeta_f(\tau) = g(\tau) + is(\tau)$, $\tau \geq 0$ of continuous PSEDs is the so-called *universal semicircle* (sometimes called *universal circle*), noted $\mathcal{L}_\infty$ in the following, defined by:

$$\left(g - \frac{1}{2}\right)^2 + s^2 = \frac{1}{4}, \quad g, s \geq 0. \tag{72}$$

In particular $\zeta_f(0) = 1$ and $\zeta_f(\infty) = 0$.

b. Ungated PSEDs with single-exponential IRF

When the IRF is not a Dirac function, but a single-exponential with time constant $\tau_\times$, the phasor of the corresponding $T$-periodic signal, $\Psi_{\tau,\tau_\times,T}(t) = \Lambda_{\tau_\times,T} \underset{T}{*} \Lambda_{\tau,T}(t)$, the cyclic convolution of two PSEDs, is given by (Eq. (D14)):

$$z\left[\Psi_{\tau,\tau_\times,T}\right] = \zeta_f(\tau_\times)\zeta_f(\tau) \tag{73}$$

which describes a semicircle rotated by an angle $\varphi_\times$ and dilated by a factor $m_\times$ given by:

$$\begin{cases} \varphi_{\times} = \tan^{-1}(2\pi f \tau_{\times}) \\ m_{\times} = \left(1+(2\pi f \tau_{\times})^2\right)^{-\frac{1}{2}} \end{cases} \quad (74)$$

### 3.2.3. Continuous phasor of square-gated PSEDs

a. Square-gated PSEDs with Dirac IRF

Using Eq. (47) for the corresponding recorded decay $S_W(t) = \Lambda_{\tau,T,W}(t)$ and reporting it in Eq. (63), or alternatively, using the fact that $\Lambda_{\tau,T,W}(t) = \bar{\Pi}_{W,nT} \underset{T}{*} \Lambda_{\tau,T}(t)$ is the cyclic convolution of a PSED and a mirrored square gate, we obtain:

$$\begin{cases} z\left[\Lambda_{\tau,T,W}\right] = z\left[\bar{\Pi}_{W,nT}\right]\zeta_f(\tau) = M_W e^{-i\varphi_W}\zeta_f(\tau) \triangleq z_{[W]}\left[\Lambda_{\tau,T}\right] \\ \varphi_W = \pi f W \\ M_W = \dfrac{\sin\varphi_W}{\varphi_W} \end{cases} \quad (75)$$

where the expression of the canonical phasor $\zeta_f(\tau)$ is given by Eq. (70) and the phasor of a mirrored square gate is derived in Appendix C.5.1 (Eq. (C30)). Note the subscript '[W]' (*W* within square brackets) in the phasor notation, which indicates a square gate of width *W*.

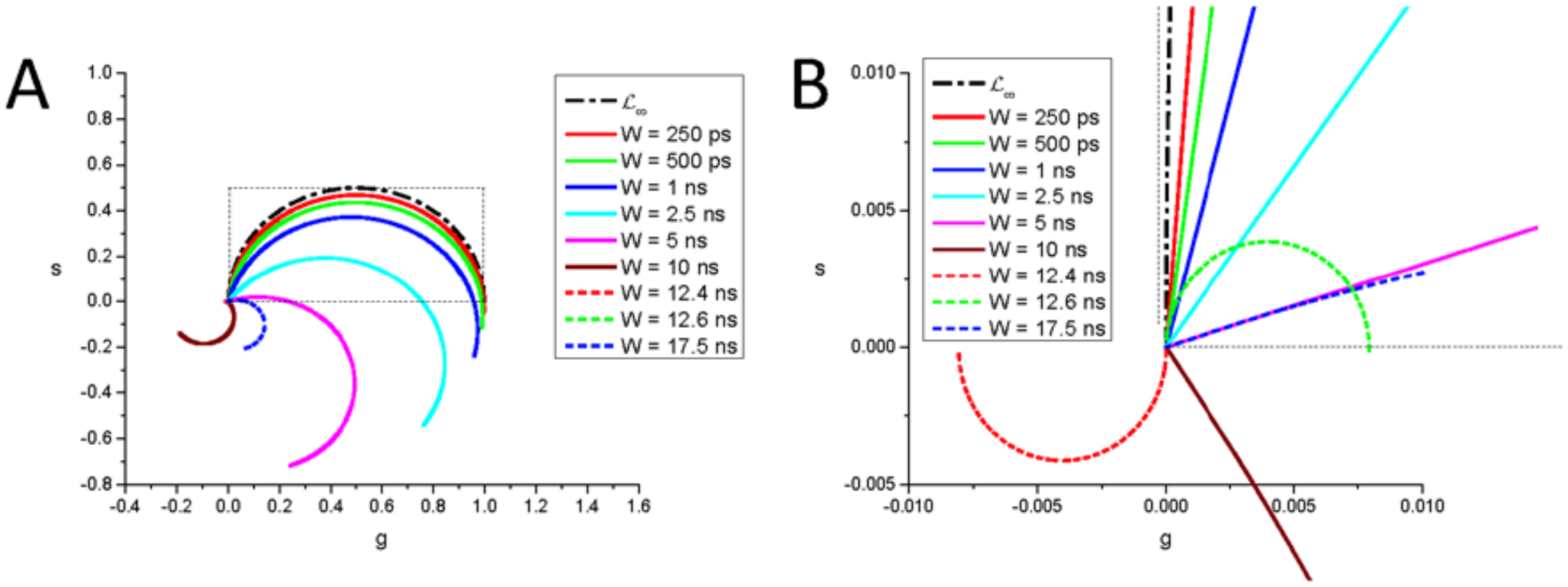

**Fig. 3**: Locus of continuous phasors of square-gated periodic single exponential decays for different value of the gate width W. T = 12.5 ns, $f_1$ = 80 MHz. A: overview, B: detail of the (0,0) region.

In other words, as illustrated in Fig. 3, the continuous phasor of a square-gated PSED is identical to that of the corresponding ungated PSED, up to a rotation by an angle $-\varphi_W = -\pi f W$ about the origin and a dilation by a factor $M_W$ given in Eq. (75), both of which are independent of $\tau$. The locus of continuous phasors of square-gated PSEDs is thus a rotated, dilated semicircle, which will refer to as the *SEPL for square-gated decays* and denote $\mathcal{L}_{[W]}$. Its equation is given by:

$$g^2 + s^2 = M_W \left( \cos\varphi_W g - \sin\varphi_W s \right) \tag{76}$$

which describes a circle whose center $\left(g_c, s_c\right)$ and radius *r* are given by:

$$\begin{cases} g_C = \dfrac{M_W}{2}\cos\varphi_W \\ s_C = -\dfrac{M_W}{2}\sin\varphi_W \\ r = \dfrac{M_W}{2} \end{cases} \tag{77}$$

The radius of this circle decreases inversely to *W*, and because of the sine function in the expression for $M_W$ (Eq. (75)), can occasionally be equal to zero. However, this only happens in trivial cases, when the gate period is a multiple of the laser period, in which case the square-gated decay is a constant, resulting in a constant phasor, no matter what lifetime is considered.

In all cases, $z_{[W]}\left[\Lambda_{0,T}\right] = M_W e^{i\varphi_W}$ and $z_{[W]}\left[\Lambda_{\infty,T}\right] = 0$.

b. Square-gated PSEDs with single-exponential IRF

The continuous phasor of a square-gated PSED convolved with a single-exponential excitation or IRF with time constant $\tau_\times$, $\Psi_{\tau,\tau_\times,T,W}(t) = \bar{\Pi}_{W,nT} \underset{T}{*} \Lambda_{\tau_\times,T} \underset{T}{*} \Lambda_{\tau,T}(t)$, is given by the product of three phasors (Eq. (D24)):

$$z\left[\Psi_{\tau,\tau_\times,T,W}\right] = M_W e^{-i\varphi_W} \zeta_f\left(\tau_\times\right)\zeta_f\left(\tau\right) = z_{[W]}\left[\Lambda_{\tau_\times,T}\right]\zeta_f\left(\tau\right) \triangleq z_{[W]}\left[\Psi_{\tau,\tau_\times,T}\right] \tag{78}$$

The locus of these phasors is thus a semicircle rotated by an angle $\varphi_* - \varphi_W = \tan^{-1}\left(2\pi f \tau_*\right) - \pi f W$ and dilated by a factor $m_* M_W$ (given in Eqs. (74) & (75)).

### 3.2.4. *Continuous phasor of arbitrary periodic decays*

a. Dirac IRF

By analogy with the discussion of Section 2.1.7, any *T*-periodic function $F_{0,T}(t)$ can be expressed in terms of a *T*-summation of a non-periodic function $F_0(t)$ (Eq. (11)), which in turns can be expressed in terms of a $\phi_0(\tau)$-weighted integral of normalized exponential functions $\Lambda_\tau(t)$, as:

$$F_0(t) = \int_0^\infty d\tau\, \phi_0(\tau)\Lambda_\tau(t) \tag{79}$$

It follows that $F_{0,T}(t)$ can be rewritten as:

$$F_{0,T}(t) = \int_0^\infty d\tau\, \phi_0(\tau)\Lambda_{\tau,T}(t) \tag{80}$$

The integral of $F_{0,T}(t)$ over [0, *T*] is given by:

$$\left\|F_{0,T}(t)\right\|_T = \int_0^T dt\, F_{0,T}(t) = \int_0^\infty d\tau\, \phi_0(\tau) = \int_0^\infty dt\, F_0(t) = \left\|F_0(t)\right\| \tag{81}$$

Inserting Eqs. (80)-(81) in Eq. (63) yields:

$$z\left[F_{0,T}\right] = \frac{\int_0^\infty d\tau\, \phi_0(\tau)\zeta_f(\tau)}{\left\|F_{0,T}(t)\right\|_T} = \frac{\int_0^\infty d\tau\, \phi_0(\tau)\zeta_f(\tau)}{\left\|F_0(t)\right\|} = z\left[F_0\right] \tag{82}$$

Due to the invariance of the phasor by dilation (Eq. (54)), this can also be written in terms of the $\| \ \|_T$-normalized decay $f_{0,T}(t)$ and $\| \ \|_T$-normalized weight function $\mu_0(\tau)$:

$$\begin{cases} f_{0,T}(t) = \dfrac{F_{0,T}(t)}{\left\|F_{0,T}(t)\right\|_T} \\ \mu_0(\tau) = \dfrac{\phi_0(\tau)}{\left\|F_{0,T}\right\|_T} \end{cases} \tag{83}$$

as:

$$\begin{cases} f_{0,T}(t) = \int_0^\infty d\tau\, \mu_0(\tau)\Lambda_{\tau,T}(t) \\ z\left[f_{0,T}\right] = \int_0^\infty d\tau\, \mu_0(\tau)\zeta_f(\tau) \end{cases} \tag{84}$$

Eq. (84) expresses the fact that the phasor of an arbitrary periodic function, expressed as a normalized weighted sum of PSEDs $\Lambda_{\tau,T}(t)$, is expressed as the same weighted sum but of the canonical phasors $\zeta_f(\tau)$. This formula provides a formal extension to arbitrary periodic function of the formalism discussed in this article, our discussion being mostly focused on PSEDs for simplicity.

An useful particular case of Eq. (79) is encountered when the samples emission can be written as a *sum of exponentials*:

$$\begin{cases} F_0(t) = \sum_{i=1}^n a_i e^{-t/\tau_i} = \sum_{i=1}^n a_i \tau_i \Lambda_{\tau_i}(t) = \int_0^\infty d\tau\, \phi_0(\tau)\Lambda_\tau(t) \\ \phi_0(\tau) = \sum_{i=1}^n a_i \tau_i \delta(\tau - \tau_i) \end{cases} \tag{85}$$

From this definition of $\phi_0(\tau)$, we obtain:

$$\begin{cases} \left\|F_0(t)\right\| = \sum_{i=1}^n a_i \tau_i = \left\|F_{0,T}(t)\right\|_T \\ \mu_0(t) = \sum_{i=1}^n \mu_i \delta(\tau - \tau_i); \ \ \mu_i = \dfrac{a_i \tau_i}{\sum_{j=1}^n a_j \tau_j} \end{cases} \tag{86}$$

Eq. (84) thus reads:

$$\begin{cases} f_{0,T}(t) = \sum_{i=1}^n \mu_i \Lambda_{\tau_i,T}(t) \\ z\left[f_{0,T}\right] = \sum_{i=1}^n \mu_i \zeta_f(\tau_i) \end{cases} \tag{87}$$

which expresses the fact that the phasor of a normalized weighted sum of normalized PSEDs $\Lambda_{\tau_i,T}(t)$ can be expressed with the same weighted sum of their individual phasors $\zeta_f(\tau_i)$.

b. Arbitrary IRF

The case of arbitrary IRFs will be discussed in the context of phasor calibration in Section 8.2.

### *3.3. Phasor of decays with discrete sampling: 'discrete' phasor*

#### *3.3.1. Definitions*

If a signal is only recorded at a finite number *N* of temporal locations $\{t_p\}_{1\le p\le N}$, separated by intervals $\{\theta_p\}_{1\le p\le N}$, $\theta_p = t_{p+1} - t_p$, a discrete version of the cyclic phasor definition (Eq. (63)) needs to be used:

$$\begin{cases} z_N[S_T](f) \triangleq \left\| S_T(t_p) e^{i2\pi f t_p} \right\|_N \Big/ \left\| S_T(t_p) \right\|_N \\ \left\| S_T(t_p) \right\|_N \triangleq \sum_{p=1}^{N} \theta_p S_T(t_p) \\ \left\| S_T(t_p) e^{i2\pi f t_p} \right\|_N \triangleq \sum_{p=1}^{N} \theta_p S_T(t_p) e^{i2\pi f t_p} \end{cases} \tag{88}$$

In the definition of the last value, $\theta_N$, the periodicity of the decay is used: $\theta_N = t_1 + T - t_{N-1}$. From this definition, it is obvious that the discrete phasor, like the continuous phasor, is invariant by dilation:

$$(\forall a), \quad z_N[aS_T] = z_N[S_T] \tag{89}$$

Definition (88) is *not* assuming that the *N* intervals cover the whole laser period. In other words, the record 'duration' *D*, defined by:

$$D = \sum_{p=1}^{N} \theta_p \tag{90}$$

might well be different from the decays period $T$, although $D = T$ is often the case experimentally. We will indicate when this assumption is used and dedicate a specific section to the cases where *D* < *T* (Section 5, The effect of decay truncation). As before, we will drop the mention of the phasor harmonic frequency when it is redundant and write:

$$z_N[S_T](f) \equiv z_N[S_T] \tag{91}$$

and use the shorthand notation $\|S_T\|_N \equiv \left\| S_T(t_p) \right\|_N$ when this does not create any ambiguity.

In typical cases where the recording locations are equidistant ($\theta_p = \theta,\ 1 \le p \le N;\ D = N\theta$):

$$\left\| S_T(t_p) \right\|_N = \theta \sum_{p=1}^{N} S_T(t_p) = \frac{D}{N} \sum_{p=1}^{N} S_T(t_p) \tag{92}$$

We will assume this condition to be met in the subsequent discussion.

From Eq. (92), it follows that:

$$\lim_{N\to\infty}\left\|S_T\left(t_p\right)\right\|_N = \int_0^T dt\, S_T\left(t\right) = \left\|S_T\left(t\right)\right\|_T \tag{93}$$

Some authors use a slightly different definition, where the argument of the complex exponential term in Eq. (88) & (92) is replaced by $2\pi f\left(t_p + W/2\right)$, *i.e.* the gates center $t_p + W/2$ is used instead of the gate beginning $t_p$ in the complex exponential argument [14]. While this choice is legitimate, it breaks the direct connection to the discrete Fourier transform (Eq. (66)). As we shall see, its only effect is to multiply the phasor as calculated in Eq. (92) by a constant term $e^{i\pi fW}$, *i.e.* it rotates the phasor by an angle $\pi fW$. This may have the undesirable effect to move the phasor of $\tau = 0$ away from its standard location $z\left[\Lambda_{0,T}\right] = 1$.

We will now examine some simple situations where the phasor of PSEDs can be expressed in compact form, as done for continuous phasors in the previous section.

### *3.3.2. Discrete phasor of convolution products*

When $D = T$, the discrete phasor as defined by Eq. (92) is related to the *discrete Fourier transform* (DFT) of the sequence of equidistant data points $\left\{S_T\left(t_k\right) = S_k\right\}, 1 \le k \le N$:

$$\mathcal{DF}\left[S_T\right](n) = \sum_{k=1}^{N} S_k e^{-i2\pi n\frac{k-1}{N}},\ 0 \le n \le N-1 \tag{94}$$

as can easily be seen from the following identities:

$$\begin{cases} \dfrac{k-1}{N} = \dfrac{(k-1)\theta}{N\theta} = \dfrac{t_k}{T} \\ f_n = \dfrac{n}{T} \end{cases} \tag{95}$$

Therefore:

$$z_N\left[S_T\right](f_n) = \frac{\mathcal{DF}^*\left[S_T\right](n)}{\mathcal{DF}^*\left[S_T\right](0)} \tag{96}$$

defined for all possible values of the phasor harmonic frequency $f_n = n/T$, $0 \le n \le N-1$. Note that this equivalence relies on a definition of the phasor harmonic frequency $f_n$ as a multiple of the inverse of the signal period $T$ and of the sampling times as $t_k = \left(k-1\right)\theta,\ 1 \le k \le N$ (Eq. (95)). As discussed in Appendix C.4, this connection to the DFT is not particularly useful, because convolution products involved in discrete phasor calculations are continuous convolutions, not discrete convolutions.

In fact, a *negative* 'discrete phasor convolution product rule' applies (Eq. (C17)):

$$z_N\left[F_T \underset{T}{*} G_T\right] \ne z_N\left[F_T\right] z_N\left[G_T\right] \tag{97}$$

which states that, in general, knowing the discrete phasors $z_N\left[F_T\right]$ and $z_N\left[G_T\right]$ of the components of a convolution product $F_T \underset{T}{*} G_T$ does not help with calculating the phasor $z_N\left[F_T \underset{T}{*} G_T\right]$ of the continuous convolution product. This has profound implications when dealing with discrete phasor calibration, as discussed in Section 8.

In some particular cases, a *weak version* of the discrete phasor convolution rule applies (Eq. (C21)):

$$z_N\left[I_T \underset{T}{*} F_{T,\lambda}\right] = \kappa\, z_N\left[I_T\right] z_N\left[F_{T,\lambda}\right] \tag{98}$$

where $\kappa$ is constant for a family of decays $\left\{F_{T,\lambda}(t)\right\}$, $\lambda \in \Omega$, where $\Omega$ is a subset of $\mathbb{R}$ and $I_T(t)$ represents the *T*-periodic instrument response function. This '*weak discrete phasor convolution rule*' allows a limited use of the standard phasor calibration approach for this specific family of decays, as will be discussed in Section 8.

### *3.3.3. Discrete phasor of ungated PSEDs*

a. Ungated PSEDs with Dirac IRF

For an ungated PSED, and assuming $\theta = T/N$ (*i.e.* gates covering the whole laser period, $D = T$ ) and $f = n/T$ , we obtain (Appendix B, Eq. (B3)):

$$\begin{cases} \zeta_{f,N}(\tau) \triangleq z_N\left[\Lambda_{\tau,T}\right] = \dfrac{1-x}{1-xe^{i\alpha}} \\ x(\tau) = e^{-\theta/\tau}; \alpha = 2\pi f \theta \end{cases} \tag{99}$$

We will refer to this function as the *canonical discrete phasor* of *T*-PSEDs, $\zeta_{f,N}(\tau)$.

In this case, the locus of phasors of discrete ungated PSEDs is a *circular arc* (see Fig. 3), whose properties are discussed in Appendix B. In particular, its equation, center and radius are given by (Eq. (B9)):

$$\begin{cases} \left(g-g_c\right)^2 + \left(s-s_c\right)^2 = r^2 \\ g_c = \dfrac{1}{2} \\ s_c = -\dfrac{1}{2}\tan(\alpha/2) \\ r = \dfrac{1}{2\left|\cos(\alpha/2)\right|} \end{cases} \tag{100}$$

The two extreme values, $z_N\left[\Lambda_{0,T}\right] = 1$ and $z_N\left[\Lambda_{\infty,T}\right] = 0$, remain identical to those of the continuous case. We will refer to this curve as the *SEPL for discrete phasors of ungated PSEDs* and denote it $\mathcal{L}_N$, the subscript '*N*' indicating the discrete nature of the decays (and the number of gates used to cover the whole laser period). Note that $\mathcal{L}_N$ also depends on the chosen harmonic *n* via the harmonic frequency *f* (compare Fig. 4A and 4B).

For large values of $N$, Eq. (99) tends to Eq. (70), as expected, and $\mathcal{L}_N$ tends to $\mathcal{L}_\infty$, the standard universal semicircle.

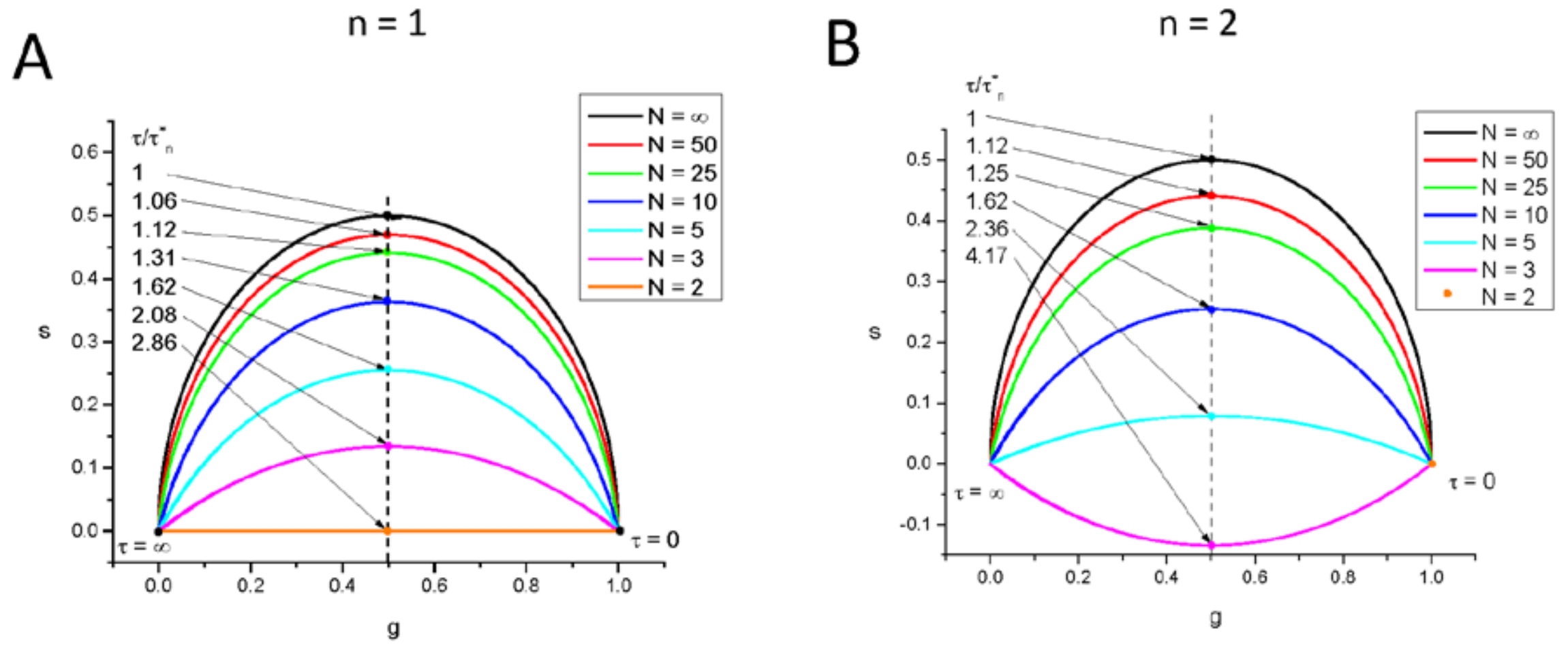

**Fig. 4**: Locus of discrete phasors of single-exponential decays for different choices of $N$, the number of sampling points. and $n$, the phasor harmonic. (A: $n$ = 1, B: $n$ = 2) The extremum of these curves (indicated by a dot of the same color) is located at $g = ½$ and is attained for different values of $\tau$ expressed in units of $\tau^*_n = T/2\pi n$. Notice that for $n$ = 1, $N$ = 2 results in a straight line, while for $n$ = 2, $N$ = 2 results in all phasors being located at 1. An example of a situation where $s < 0$ is provided in B ($n$ = 2, $N$ = 3).

b. Ungated PSEDs with single-exponential IRF

Using the definition of $\Psi_{\tau,\tau_\times,T}(t)$ (Eq. (20)) in Eq. (92) yields (Appendix D, Eq. (D31)):

$$\begin{cases} z_N\left[\Psi_{\tau,\tau_\times,T}\right] = \dfrac{1-x}{1-xe^{i\alpha}}\dfrac{1-x_\times}{1-x_\times e^{i\alpha}}e^{i\alpha} = e^{i\alpha}\zeta_{f,N}(\tau)\zeta_{f,N}(\tau_\times) \\ x(\tau) = e^{-\theta/\tau};\ x_\times = x(\tau_\times) = e^{-\theta/\tau_\times};\ \alpha = 2\pi f\theta \end{cases} \quad (101)$$

This can be rewritten:

$$z_N\left[\Psi_{\tau,\tau_\times,T}\right] = e^{i\alpha}z_N\left[\Lambda_{\tau_\times,T}\right]z_N\left[\Lambda_{\tau,T}\right] \quad (102)$$

which shows that the discrete phasor of single-exponential decays convolved with a $T$-periodic single-exponential IRF is a rotated and dilated version of the discrete phasor of PSEDs. We will return to this identity in Section 8, when discussing calibration. Note that Eq. (102) is an example of the weak discrete phasor convolution rule mentioned in Section 3.3.2, since $\Psi_{\tau,\tau_\times,T}(t) = \Lambda_{\tau_\times,T} \underset{T}{*} \Lambda_{\tau,T}(t)$ is a convolution product and the constant $\kappa = e^{i2\pi f\theta}$ does not depend on the lifetime $\tau$.

### *3.3.4. Discrete phasor of square-gated PSEDs*

a. Square-gated PSEDs with Dirac IRF

To obtain the phasor of a square-gated PSED recorded at discrete points, Eq. (92) is used with $S_T(t)$ given by $\Lambda_{\tau,T,W}(t)$ in Eq. (47). The result is (Appendix B, Eq. (B37)):

$$z_{N[W]}\left[\Lambda_{\tau,T}\right] \triangleq z_N\left[\Lambda_{\tau,T,W}\right] = \frac{-\dfrac{1-e^{ir\alpha}}{1-e^{i\alpha}} + \dfrac{1-uy^{-1}x^r e^{ir\alpha}}{1-xe^{i\alpha}}}{(k+1)N - r + \dfrac{1-uy^{-1}x^r}{1-x}} \tag{103}$$

where the following intermediate variables have been introduced:

$$\begin{cases} k = \lfloor W/T \rfloor \\ \omega = W[T] = W - kT \\ r = \left\lceil \dfrac{T-\omega}{\theta} \right\rceil \\ x(\tau) = e^{-\theta/\tau};\ y(\tau) = e^{-T/\tau};\ u(\tau) = e^{-\omega/\tau};\ \alpha = 2\pi f \theta \end{cases} \tag{104}$$

and the notation $\lceil x \rceil$ denotes the 'upper' integer part of *x* (or `ceil` – or ceiling - function in most programming languages):

$$\forall x \in \mathbb{R},\ x \in \left]n-1, n\right], n \in \mathbb{Z} \Rightarrow \lceil x \rceil = nx \tag{105}$$

As before, the gate duration *W* can take any positive value (including values larger than the laser period *T*, in which case, each gate in a frame will be separated by more than one laser period from the next); *r* is an index value used to determine which expression of $\Lambda_{\tau,T,W}(t)$ to use in Eq. (47); *x*, *y*, *u* and α are introduced to simplify notations.

The locus of the discrete phasors of PSEDs is, in general, a complex curve, which cannot be reduced to an algebraic equation due to the presence of the term $x^r$ in Eq. (103). We will refer to this curve as the *SEPL for discrete phasors of square-gated PSEDs* and denote it as $\mathcal{L}_{N[W]}$, subscript '*N* [*W*]' indicating that both discrete decays (*N* gates) and a square gate of duration *W* are considered. Eq. (103) introduces a similar notation, $z_{N[W]}\left[S_T\right]$, for the discrete phasor of square-gated periodic decay, $S_T$ .

In special cases where $T-\omega$ is proportional to the gate step *θ* (which, since $T = N\theta$ is assumed, is equivalent to the gate width *W* being proportional to *θ*), one obtains the following identity (Appendix B, Eq. (B40)):

$$W = q\theta \quad \Rightarrow \quad z_{N[W]}\left[\Lambda_{\tau,T}\right] = \frac{\sin q\dfrac{\alpha}{2}}{q\sin\dfrac{\alpha}{2}} e^{-i(q-1)\frac{\alpha}{2}}\ z_N\left[\Lambda_{\tau,T}\right] = \frac{\sin q\dfrac{\alpha}{2}}{q\sin\dfrac{\alpha}{2}} e^{-i(q-1)\frac{\alpha}{2}} \zeta_{f,N}(\tau) \tag{106}$$

Using the expression for the discrete phasor of the mirror square-gate function derived in Appendix C (Eq. (C37)), this equation can be rewritten:

$$W = q\theta \quad \Rightarrow \quad z_{N[W]}\left[\Lambda_{\tau,T}\right] = e^{i\alpha}\ z_N\left[\overline{\Pi}_{W,nT}\right] z_N\left[\Lambda_{\tau,T}\right] \tag{107}$$

In other words, it is an *arc of circle* rotated about 0 and with a diameter $d_{N[W]}$ given by:

$$\begin{cases} d_{N[W]} = \left| z_{N[W]}\left[\Lambda_{0,T}\right] \right| = \left| \dfrac{\sin q\dfrac{\alpha}{2}}{q\sin\dfrac{\alpha}{2}} \right| \\ T = N\theta,\ \ W = q\theta,\ \alpha = 2\pi f\theta \end{cases} \tag{108}$$

which decreases as $q$ (*i.e.* $W$) increases.

A particular case of interest is $W = \theta$, *i.e.* $q$ = 1 which corresponds to adjacent gates (contiguous and non-overlapping gates). We then have:

$$W = \theta \quad \Rightarrow \quad z_{N[W]}\left[\Lambda_{\tau,T}\right] = z_N\left[\Lambda_{\tau,T}\right], \tag{109}$$

*i.e.*, the discrete phasor of square-gated PSED with adjacent gates is equal to the discrete phasor of ungated PSED. This situation is that encountered with TCSPC data, where each 'bin' of a discrete decay is contiguous to the next one. In this case, the SEPL is therefore an arc of circle ($\mathcal{L}_N$) which only depends on the number of bins, not on their actual size.

In all other cases, $\mathcal{L}_{N[W]}$ is a complex curve passing through $z_{N[W]}\left[\Lambda_{\infty,T}\right] = 0$ and $z_{N[W]}\left[\Lambda_{0,T}\right] = z_{N[r(W)]}\left[\Lambda_{0,T}\right]$ where $r(W)$ is given by:

$$r(W) = \left\lceil \frac{T - W[T]}{\theta} \right\rceil \tag{110}$$

and the discrete phasor of a square-gated PSED with 0 lifetime is given by (Appendix B, Eqs. (B41) & (B42)):

$$z_{N[W]}\left[\Lambda_{0,T}\right] = \begin{cases} -\dfrac{1}{(k+1)N - r + 1} \dfrac{\sin(r-1)\dfrac{\alpha}{2}}{\sin\dfrac{\alpha}{2}} e^{ir\frac{\alpha}{2}}, & \dfrac{T-\omega}{\theta} \notin \mathbb{N} \\ -\dfrac{1}{(k+1)N - r} \dfrac{\sin r\dfrac{\alpha}{2}}{\sin\dfrac{\alpha}{2}} e^{i(r+1)\frac{\alpha}{2}}, & \dfrac{T-\omega}{\theta} \in \mathbb{N} \end{cases} \tag{111}$$

In other words, $\mathcal{L}_{N[W]}$ characterized by the same value of $r(W) = r$ (Eq. (110)), *i.e.* for which $\omega \in \left[(N-r)\theta, (N-r+1)\theta\right[$ share the same two points, 0 and $z_{N[W]}\left[\Lambda_{0,T}\right]$. This property is illustrated in Fig. 5A for $N$ =10, $T$ = 12.5 ns = 1/$f$, where curves characterized by the same $r$ value are represented with the same style, while colors indicate different values of $W$ within a given interval $\left[q\theta, (q+1)\theta\right[$. Because the curves at the boundaries of these intervals are different arcs of circle, the intermediate curves progressively 'interpolate' between those two regular curves, including curves that are best described qualitatively as a section of circular arc connected to the next circular arc by an almost straight 'stem'.

The only exception to this behaviour is the series of $\mathcal{L}_{N[W]}$ for $W \le \theta$, where the SEPL is identical to $\mathcal{L}_N$ as can be easily verified (Appendix B, Eq. (B46) and Figure 4A).

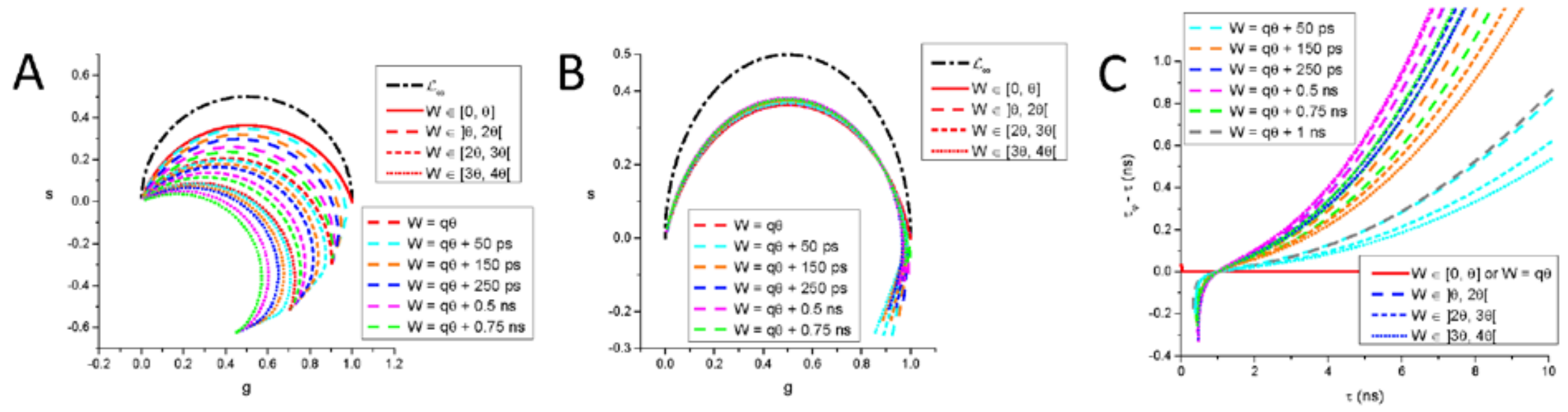

**Fig. 5**: SEPL for discrete square-gated phasors. $\mathcal{L}_{N[W]}$ for a constant number of gates $N$ = 10 (and gate step $\theta = T/N$, $T$ = 12.5 ns, $f = 1/T$) and varying gate width. A: Uncalibrated $\mathcal{L}_{N[W]}$, B: $\mathcal{L}_{N[W]}$ calibrated using $\tau_C$ = 1 ns. A: For $W \leq \theta$, $\mathcal{L}_{N[W]}$ is independent of W and equal to $\mathcal{L}_N$ (solid red circular arc). For $W \in ]\theta, 2\theta[$ (long dash curves), all $\mathcal{L}_{N[W]}$ share a common $z_{N[W]}\left[\Lambda_{0,T}\right]$ and $z_{N[W]}\left[\Lambda_{\infty,T}\right]$ but only $\mathcal{L}_{N[W]}$ for W = $\theta$ (solid red curve) is a circular arc. Similarly, for $W \in [2\theta, 3\theta[$ (short dash curves) and $W \in [3\theta, 4\theta[$ (dotted curves), the different $\mathcal{L}_{N[W]}$ share a common $z_{N[W]}\left[\Lambda_{0,T}\right]$ and $z_{N[W]}\left[\Lambda_{\infty,T}\right]$ in each group but only $\mathcal{L}_{N[W]}$ for $W = 2\theta$ (short red dash curve) and for $W = 3\theta$ (red dotted curve) are circular arcs. B: The fact that only one of the $\mathcal{L}_{N[W]}$ within each group is a circular arc is clearly visible after rotation bringing the phasor of $\tau$ = 1 ns back to $z_{N[W]}\left[\Lambda_{1,T}\right]$. All $\mathcal{L}_{N[W]}$ for $W = k\theta$, are mapped to $\mathcal{L}_N$ (circular arc) after calibration, while the others are clearly different from one another. C: Difference between the pseudo-phase lifetime and the real lifetime computed for the various calibrated $\mathcal{L}_{N[W]}$ curves shown in D. As expected due to the choice of $\tau_c$ = 1 ns as calibration lifetime, the difference is minimal around $\tau$ = 1 ns and increase around this value, demonstrating the limitations of phasor calibration in the general case.

It is easy to verify that in the limit $W \to 0$, one recovers the discrete phasor of an ungated signal (Eq. (99)). Similarly, in the limit $N \to \infty$, one recovers the continuous phasor of a square-gated signal (Eq. (75)).

b. Square-gated PSEDs with single-exponential IRF

The expression for the discrete phasor of a square-gated PSEDs with single-exponential IRF is derived in Appendix D.9 (Eq. (D47)) and does not correspond to any simple curve, even in the particular case where the gate width $W$ is equal to the gate step $\theta$ (contiguous gates):

$$\begin{cases} z_{N[W]}\left[\Psi_{\tau,\tau_\times,T}\right] \triangleq z_N\left[\Psi_{\tau,\tau_\times,T,W}\right] = \dfrac{-\dfrac{1-e^{ir\alpha}}{1-e^{i\alpha}} + \dfrac{1}{\tau-\tau_\times}\left(\tau\dfrac{1-\beta e^{ir\alpha}}{1-xe^{i\alpha}} - \tau_\times\dfrac{1-\beta_\times e^{ir\alpha}}{1-x_\times e^{i\alpha}}\right)}{(k+1)N - r + \dfrac{1}{\tau-\tau_\times}\left(\tau\dfrac{1-\beta}{1-x} - \tau_\times\dfrac{1-\beta_\times}{1-x_\times}\right)} \\ k = \left\lfloor \dfrac{W}{T} \right\rfloor;\ r = \left\lceil \dfrac{T-\omega}{\theta} \right\rceil \\ y(\tau) = e^{-T/\tau};\ y_\times(\tau) = e^{-T/\tau_\times};\ u(\tau) = e^{-\omega/\tau};\ u_\times(\tau) = e^{-\omega/\tau_\times} \\ \beta(\tau) = u\, x^r y^{-1};\quad \beta_\times(\tau) = u_\times x_\times^r y_\times^{-1} \end{cases} \tag{112}$$

This expression can however be used to explore the effect of different gate width/steps on the calibrated SEPL in a non-Dirac excitation case. As will be discussed in Section 8, the calibrated SEPL can in some cases be close to $\mathcal{L}_\infty$. Fig. 6 provides a simple illustration of the non-classical shape of the SEPL for binned TCSPC data for several bin numbers and IRF time constants, which shows that even for a number of bin as small as $N$ = 64 and an IRF time constant as large as 1/6 of the laser period, the SEPL is very similar to $\mathcal{L}_\infty$ (after calibration, see Section 8).

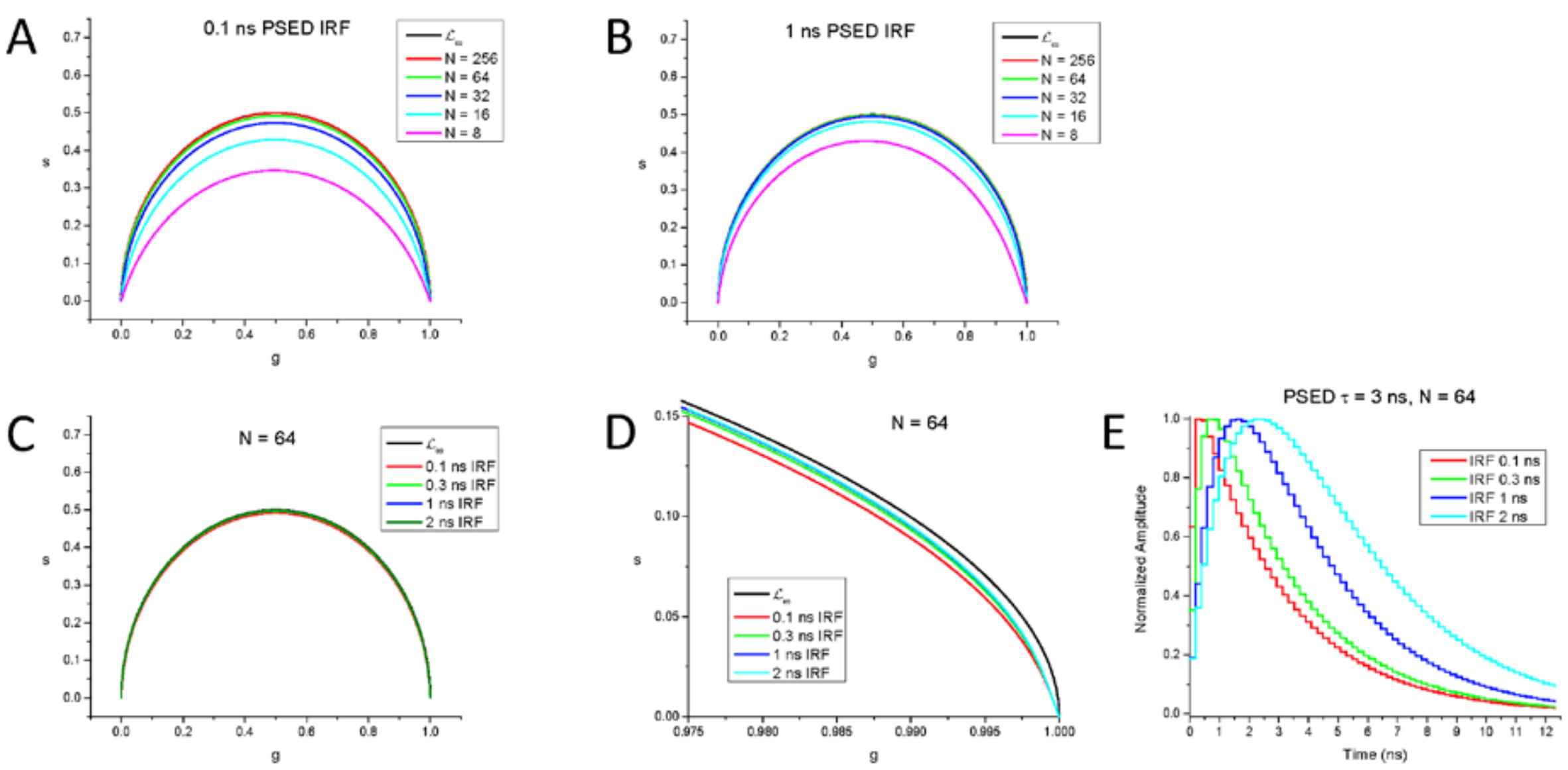

**Fig. 6**: Effect of binning and IRF width on the phasor of TCSPC data. This figure illustrates the discussion of § 3.3.4b on the phasor of PSED with a single-exponential IRF, binned with a finite number of bins $N$. A laser period $T$ = 12.5 ns was assumed, and an harmonic frequency $f = 1/T$ was used. A: Single-exponential IRF with time constant $\tau_0$ = 0.1 ns. B: Single-exponential IRF with time constant $\tau_0$ = 1 ns. Notice how the wider IRF brings the calibrated closer to $\mathcal{L}_\infty$. C, D: Effect of the IRF time constant ($\tau_0$ = 0.1 ns – 2 ns) on the SEPL at fixed bin number ($N$ = 64). All curves are very close to $\mathcal{L}_\infty$ as can be seen in the detail of the (1,0) region shown in D. E: Examples of binned $\tau$ = 3 ns PSED convolved with the single-exponential IRFs used in A-D.
Note that all SEPLs have been rotated/rescaled so that the phasor of the 0-lifetime decays are located at the point (1,0) [see Section 8: Phasor calibration].

### *3.3.5. Discrete phasor of arbitrary periodic decays*

a. Dirac IRF

A detailed discussion of this case can be found in Appendix B.3, whose results we will now summarize.

In the discrete case, a useful basis of functions to decompose a *T*-periodic decay in is the set of functions $\left\{\Lambda_{\tau,T,N}\left(t\right)\right\}_{\tau>0}$ proportional to the PSEDs defined in Eq. (17):

$$\Lambda_{\tau,T,N}\left(t\right)=\frac{\tau}{\theta}\left(1-e^{-\theta/\tau}\right)\Lambda_{\tau,T}\left(t\right) \tag{113}$$

where $\theta=T/N$ is the gate step. When $N\to\infty$, these functions are identical to the normalized PSEDs, $\left\{\Lambda_{\tau,T}\left(t\right)\right\}_{\tau>0}$.

Any *T*-periodic function $F_{0,T}\left(t\right)$ can be written, after normalization by $\left\|F_{0,T}\left(t_p\right)\right\|_N$ (*i.e.* $\|\ \|_N$-normalization) (Eq. (92)):

$$f_{0,T}\left(t\right)=\int_0^\infty d\tau\mu_0\left(\tau\right)\Lambda_{\tau,T,N}\left(t\right) \tag{114}$$

where the $\|\ \|_N$*-normalized* weight function $\mu_0\left(\tau\right)$ also appears in the expression for the phasor of $F_{0,T}\left(t\right)$:

$$z_N\left[f_{0,T}\right]=\int_0^\infty d\tau\mu_0\left(\tau\right)\zeta_{f,N}\left(\tau\right) \tag{115}$$

where the *canonical discrete phasors* of *T*-PSEDs, $\zeta_{f,N}\left(\tau\right)$ are defined in Eq. (99).

In particular, for a linear combination of PSEDs,

$$F_{0,T}\left(t\right)=\sum_{i=1}^{n}a_i\tau_i\Lambda_{\tau_i,T}\left(t\right) \tag{116}$$

the $\|\ \|_N$*-normalized* weight function $\mu_0\left(\tau\right)$ reads:

$$\begin{cases}\mu_0\left(t\right)=\sum_{i=1}^{n}\mu_i\delta\left(\tau-\tau_i\right)\\ \mu_i=\dfrac{a_i}{\left(1-e^{-\theta/\tau_i}\right)}\bigg/\sum_{j=1}^{n}\dfrac{a_j}{\left(1-e^{-\theta/\tau_j}\right)}\end{cases} \tag{117}$$

In other words, the discrete phasor of a linear combination of single-exponential decays can be expressed as a linear combination of phasors, but, in order for the same functional form to be preserved, the discrete decay needs to be expressed in the basis of $\left\{\Lambda_{\tau_i,T,N}\left(t\right)\right\}_{i=1\ldots n}$. The discrete phasor of the total decay is then expressed in the same functional form using their individual phasors, $\left\{\zeta_{f,N}\left(\tau_i\right)\right\}_{i=1\ldots n}$.

b. Arbitrary IRF

The case of arbitrary IRFs will be discussed in the context of phasor calibration in Section 8.3.

## 4. The effect of decay offset

So far, we have assumed that time 0 of the recording was identified with the IRF maximums location $t_0$, but this is not the case in general, for various experimental reasons. For instance, when using TCSPC hardware based on time-to-amplitude converter (TAC), it is customary to artificially delay the detector recording window with respect to the laser pulse signal, in order that photons emitted quasi-simultaneously with the excitation pulse are associated with a non-zero time-stamp: this allows avoiding the short and long time delay regions of the electronics, which are associated with the largest uncertainties or artefacts. In the general case, however, the experimentalist may simply choose to offset the location of the rising part of the recorded signal away from time 0 of the electronics, so that the rising part of the signal is clearly visible. Yet another reason why the time stamp corresponding to the IRF maximum might not be precisely known could be that some of the gates are discarded for one reason or another, or the timestamp assigned to the first gate is set to a non-zero value. While such a general offset is easily handled in fluorescence decay fitting approaches by incorporating an additional offset parameter, the result of phasor calculation according to Eq. (63) (continuous phasor) or Eq. (88) (discrete phasor) lead to phasor properties which depend on the precise value of the offset, and in general, differ from those discussed so far.

One trivial option to avoid these changes is to first subtract the offset from the recorded data timestamps:

$$t \mapsto t' = t - t_0 \tag{118}$$

and if the resulting timestamps are negative, use the periodicity $T$ of the decay to correct them:

$$t \mapsto t' = \left(t - t_0\right)[T] \tag{119}$$

where '[T]' indicates the modulo-$T$ operation. This operation, which amounts to a periodic shift of the decay, yields corrected timestamps with which formulas derived in the previous section can be used. However, this approach requires determining the exact value of the offset. This can in principle be done by recording the excitation (laser) signal as detected by the system but this procedure might not be always practical, for instance because the laser signal is efficiently rejected by the detection system and its signal can therefore not be recorded. In those cases, obtaining expressions corresponding to those derived in the previous section, modified by the presence of an offset, allows carrying out phasor analysis without prior knowledge of the offset (that is using Eq. (63) or Eq. (88)) and interpret the results in light of the formulas derived next.

### *4.1. Periodic decays with offset*

The effect of an offset on results derived in Section 2 is simply to replace any formula with a time argument $t$ by the same formula with the substitution defined by Eq. (119). For instance, the expression for a PSED with lifetime $\tau$ and period $T$ (Eq. (17)) is modified into:

$$\Lambda_{\tau,T|t_0}\left(t\right) \triangleq \frac{1}{\tau\left(1 - e^{-T/\tau}\right)} e^{-\left(t - t_0\right)[T]/\tau} = \Lambda_{\tau,T}\left(t - t_0\right) \tag{120}$$

with an integral of 1 over [0, $T$]. Similarly, the expression for a square-gated PSED (Eq. (47)) is replaced by:

$$\Lambda_{\tau,T,W|t_0}(t) \triangleq \begin{cases} (a)\ \dfrac{1-e^{-\omega/\tau}}{1-e^{-T/\tau}} e^{-(t-t_0)[T]/\tau} + k, & (t-t_0)[T] \in [0, T-\omega[ \\ (b)\ \dfrac{1-e^{-(\omega-T)/\tau}}{1-e^{-T/\tau}} e^{-(t-t_0)[T]/\tau} + k + 1, & (t-t_0)[T] \in [T-\omega, T[ \end{cases} \tag{121}$$

where $\begin{cases} k = \lfloor W/T \rfloor \\ \omega = W[T] = W - kT \end{cases}$

More generally, for an arbitrary $T$-periodic IRF $I_T(t)$ expressed in terms of the normalized $\Lambda_{\tau,T}(t)$ (Eq. (25)):

$$I_{T|t_0}(t) = \int_0^\infty d\tau \xi_0(\tau) \Lambda_{\tau,T|t_0}(t) \tag{122}$$

and for any PSED $\Lambda_{\tau_0,T}(t)$ convolved with such an IRF (see Appendix D.1.2, Eq. (D5)):

$$I_{T|t_0} \underset{T}{*} \Lambda_{\tau_0,T}(t) = \int_0^\infty d\tau \xi_0(\tau) \Lambda_{\tau,T|t_0} \underset{T}{*} \Lambda_{\tau_0,T}(t) = \int_0^\infty d\tau \xi_0(\tau) \Psi_{\tau,\tau_0,T|t_0}(t) \tag{123}$$

Likewise, for the square-gated version of such a decay (see Appendix D.1.2, Eq. (D21)):

$$I_{T|t_0} \underset{T}{*} \Lambda_{\tau_0,T,W}(t) = \int_0^\infty d\tau \xi_0(\tau) \Lambda_{\tau,T|t_0} \underset{T}{*} \Lambda_{\tau_0,T,W}(t) = \int_0^\infty d\tau \xi_0(\tau) \Psi_{\tau,\tau_0,T,W|t_0}(t) \tag{124}$$

For an arbitrary gate shape, the corresponding formal expression is:

$$I_{T|t_0} \underset{nT}{*} \overline{\Gamma}_{W,nT} \underset{T}{*} \Lambda_{\tau_0,T}(t) = \int_0^\infty d\tau \xi_0(\tau)\, \overline{\Gamma}_{W,nT} \underset{nT}{*} \Psi_{\tau,\tau_0,T,W|t_0}(t) \tag{125}$$

which may or may not be simplified.

Using these expressions, it is relatively simple to obtain the modified formulas for the phasor in the different situations explored in Section 3. The results are discussed next. As before, we will distinguish between continuous and discrete phasor, and present examples of ungated and square-gated PSEDs.

### *4.2. Continuous phasor of PSEDs with offset*

#### *4.2.1. Continuous phasor of ungated PSEDs*

a. PSEDs with Dirac IRF with offset

It is easy to verify that the continuous phasor of a PSED defined by Eq. (120) is given by (Appendix A):

$$z\left[\Lambda_{\tau,T|t_0}\right] = \frac{1}{1 - i2\pi f\tau} e^{i2\pi f t_0} = \zeta_f(\tau) e^{i2\pi f t_0} \tag{126}$$

In other words, it is equal to the phasor of a PSED without offset, $\zeta_f(\tau)$, rotated about 0 by an angle $2\pi f t_0$.

b. PSEDs with single-exponential IRF with offset

Details of the calculations are provided in Appendix D.5.2, resulting in the following phasor expression (Eqs. (D16)-(D17)):

$$\begin{aligned} z\left[\Psi_{\tau,\tau_\times,T|t_0}\right](f) &= \frac{1}{1-i2\pi f\tau}\frac{1}{1-i2\pi f\tau_\times}e^{i2\pi ft_0} = \zeta_f(\tau_\times)\zeta_f(\tau)e^{i2\pi ft_0} \\ &= z\left[\Lambda_{\tau_\times,T|t_0}\right]\zeta_f(\tau) \end{aligned} \tag{127}$$

where $\tau_\times$ is the time constant of the single-exponential IRF. Once again, the phasor is equal to the phasor of the same PSED with single-exponential IRF without offset, $\zeta_f(\tau_\times)\zeta_f(\tau)$, rotated about 0 by an angle $2\pi ft_0$.

### *4.2.2. Continuous phasor of square-gated PSEDs with offset*

Although the calculation is made a bit cumbersome, as two cases need to be distinguished ( $t_0 < \omega$ and $t_0 \geq \omega$, see Appendix A, derivation of Eq. (A10) for details in the case of a Dirac IRF and a square gate), the result is again simply a rotation by an angle $2\pi ft_0$ of the results in absence of offset (Eqs. (75) & (78)).

For instance, for a Dirac IRF (Eq. (A10)):

$$\begin{cases} z_{[W]}\left[\Lambda_{\tau,T|t_0}\right] \triangleq z\left[\Lambda_{\tau,T,W|t_0}\right] = M_W e^{-i\varphi_W}\zeta_f(\tau)e^{i2\pi ft_0} = z\left[\overline{\Pi}_{W,nT}\right]\zeta_f(\tau)e^{i2\pi ft_0} \\ M_W = \dfrac{\sin\varphi_W}{\pi fW};\ \varphi_W = \pi fW \end{cases} \tag{128}$$

and for a single-exponential IRF with time constant $\tau_\times$ (Eq. (D28)):

$$\begin{aligned} z_{[W]}\left[\Psi_{\tau,\tau_\times,T|t_0}\right] \triangleq z\left[\Psi_{\tau,\tau_\times,T,W|t_0}\right] &= M_W e^{-i\varphi_W}\zeta_f(\tau_\times)\zeta_f(\tau)e^{i2\pi ft_0} \\ &= z\left[\overline{\Pi}_{W,nT}\right]\zeta_f(\tau_\times)e^{i2\pi ft_0}\zeta_f(\tau) \end{aligned} \tag{129}$$

All these results are of the same general form:

$$z\left[I_{W,T|t_0} \underset{T}{*} \Lambda_{\tau,T}\right] = z\left[I_{T|t_0} \underset{nT}{*} \overline{\Gamma}_{W,nT} \underset{T}{*} \Lambda_{\tau,T}\right] = z\left[\overline{\Gamma}_{W,nT}\right]z\left[I_{T|t_0}\right]\zeta_f(\tau) \tag{130}$$

where the right hand side singles out the phasor of the (gated) instrument response function with offset. They also confirm that the effect of an offset $t_0$ on the continuous phasor is simply a rotation by an angle $2\pi ft_0$.

## ***4.3. Discrete phasor of PSEDs with offset***

### *4.3.1. Discrete phasor of ungated PSEDs with offset*

Analytical results for discrete phasors of decays with offset are a bit more complicated to compute, in particular in the presence of gating, and differ from those for continuous phasors, which are characterized by their simplicity and universality. We discuss only a few cases that can be fairly simply calculated analytically.

a. PSEDs with Dirac IRF with offset

In the case where the decay sampling points cover the whole period ($T = N\theta$), one obtains the following expression for the discrete phasor of an ungated PSED with a Dirac IRF (see Appendix B.1.2, Eq. (B26); an expression for the case where the decay samples do not cover the full period is also provided in Appendix B, Eq. (B27)):

$$\begin{cases} z_N\left[\Lambda_{\tau,T|t_0}\right] = \dfrac{1-x}{1-xe^{i\alpha}} e^{i\varphi_N(t_0)} = \zeta_{f,N}(\tau) e^{i\alpha\lceil t_0/\theta \rceil} \\ x(\tau) = e^{-\theta/\tau};\ \alpha = 2\pi f \theta \end{cases} \tag{131}$$

which is a rotated version of the expression $z_N\left[\Lambda_{\tau,T}\right]$ obtained in the absence of offset (Eq. (99)).

The argument of the constant exponential factor in the righthand side of Eq. (131) states that, if the offset $t_0 \in [0,T[$ is a multiple of the gate step size $\theta$, then the resulting phasor is simply a rotated version of the discrete phasor of the ungated PSED without offset (Eq. (99)), by an angle $2\pi f t_0$.

If however the offset $t_0$ is *not* a multiple of the gate step size $\theta$ , the resulting phasor is still a rotated version of the phasor without offset, but the angle is now given by a slightly different expression (see Appendix B, Eq. (B26) for a derivation). In particular, this expression predicts that, in some cases, the phasor of identical decays with *different* offsets (within $\theta$ of each other) will be equal. This is for instance the case for $T$ = 12.5 ns, $\theta$ = 1.25 ns ($N$ = 10) and $t_0 \in ]0,1.25]$ ns, as can be verified by a direct calculation using Eqs. (92) & (120).

b. PSEDs with single-exponential IRF with offset

The formula for the discrete phasor of an ungated PSED with a single-exponential IRF with time constant $\tau_*$ is derived in Appendix D (Eq. (D40)):

$$z_N\left[\Psi_{\tau,\tau_\times,T|t_0}\right] = \zeta_{f,N}(\tau)\zeta_{f,N}(\tau_\times)\Omega(\tau,\tau_\times,t_0) e^{i\lceil \frac{t_0}{\theta} \rceil \alpha} \tag{132}$$

where $\Omega(\tau,\tau_\times,t_0)$ is, in general, a complex function of $\tau$ , $\tau_\times$ and $t_0$ , and $\alpha$ is defined as before (*e.g.* Eq. (131)). This shows that the corresponding SEPL is not a simple curve.

If, however, the offset is commensurate with the gate step ($t_0 = q\theta$), this complex function $\Omega(\tau,\tau_\times,t_0)$ reduces to $e^{i\alpha}$ and the phasor can be written as the product of two phasors and a constant (Eq. (D42)):

$$t_0 = q\theta \Rightarrow z_N\left[\Psi_{\tau,\tau_\times,T|t_0}\right] = z_N\left[\Lambda_{\tau,T}\right] z_N\left[\Lambda_{\tau_\times,T|t_0}\right] e^{i\alpha} \tag{133}$$

The corresponding SEPL is therefore a rotated version of $\mathcal{L}_N$ .

### 4.3.2. *Discrete phasor of square-gated PSEDs with offset*

The calculations in this situation are a bit cumbersome and require the distinction of different cases depending on the respective values of the offset, period, gate width and gate step. We will look at the Dirac IRF case in some detail, limiting the discussion of the single-exponential IRF to a general formula, and skipping the case of arbitrary IRF altogether.

a. Square-gated PSEDs with Dirac IRF with offset

The expression for the discrete phasor of a square-gated PSED with Dirac IRF with offset is derived in Appendix B (Eq. (B60)):

$$\begin{cases} z_N\left[\Lambda_{\tau,T,W|t_0}\right] = \dfrac{\dfrac{e^{ir\alpha} - e^{iq\alpha}}{1-e^{i\alpha}} + \dfrac{x^q e^{iq\alpha} - u x^r e^{ir\alpha}}{1-xe^{i\alpha}} e^{t_0/\tau}}{kN + q - r + \dfrac{x^q - u x^r}{1-x} e^{t_0/\tau}} \\ q = \left\lceil \dfrac{t_0}{\theta} \right\rceil;\ r = \left\lceil \dfrac{t_0 - \omega}{\theta} \right\rceil \end{cases} \quad (134)$$

where $x(\tau) = e^{-\theta/\tau}$ , $u(\tau) = e^{-\omega/\tau}$ and $\alpha = 2\pi f \theta$ as before. This equation does not in general describe any simple algebraic curve, except in particular cases.

A special case of interest is encountered when the gates are adjacent ($W = \theta$). Two different situations may occur:

<u>Case $W = \theta$ , $t_0 = q\theta$</u>

When the offset is proportional to the gate step (*i.e.* the offset falls on one of the gates start), the discrete phasor reads:

$$z_{N[\theta]}\left[\Lambda_{\tau,T|t_0}\right] = \zeta_{f,N}(\tau) e^{i2\pi f t_0} = z_N\left[\Lambda_{\tau,T|t_0}\right] \quad (135)$$

In other words, the discrete phasor of the square-gated PSED is then equal to the discrete phasor of the *ungated* version of the PSED, with the same offset. Consequently, $\mathcal{L}_{N[\theta]}$ is a rotated arc of circle.

<u>Case $W = \theta$, $t_0 = q\theta - \theta_0$, $0 < \theta_0 < \theta$</u>

When the offset is not proportional to the gate step (*i.e.* when the offset is distinct from any of the gates start), the phasor reads:

$$z_{N[\theta]}\left[\Lambda_{\tau,T|t_0}\right] = z_N\left[\Lambda_{\tau,T,\theta|t_0}\right] = \frac{\left(1 - x + \left(1-e^{i\alpha}\right)\left(x - e^{-\theta_0/\tau}\right)\right)}{1 - xe^{i\alpha}} e^{i(q-1)\alpha} \quad (136)$$

This expression does not describe any simple curve, but can be studied numerically, as illustrated in Fig. 7 in which $q = 0$, and $t_0$ was incremented from 0 to $\theta$. As $t_0$ is increased, $\mathcal{L}_{N[\theta]}$ progressively deforms from the circular arc corresponding to $\mathcal{L}_N$ for $t_0 = 0$ into that corresponding to $t_0 = \theta$, with a quasi-linear 'stem' overlapping the segment connecting $z_N\left[\Lambda_{0,T}\right]$ to $z_N\left[\Lambda_{0,T|\theta}\right]$. This scenario is repeated with each additional increment of $\theta$ to $t_0$, with the circular arc ($\mathcal{L}_N$) corresponding to $t_0 = q\theta$ replacing the $\mathcal{L}_N$ corresponding to $t_0 = 0$.

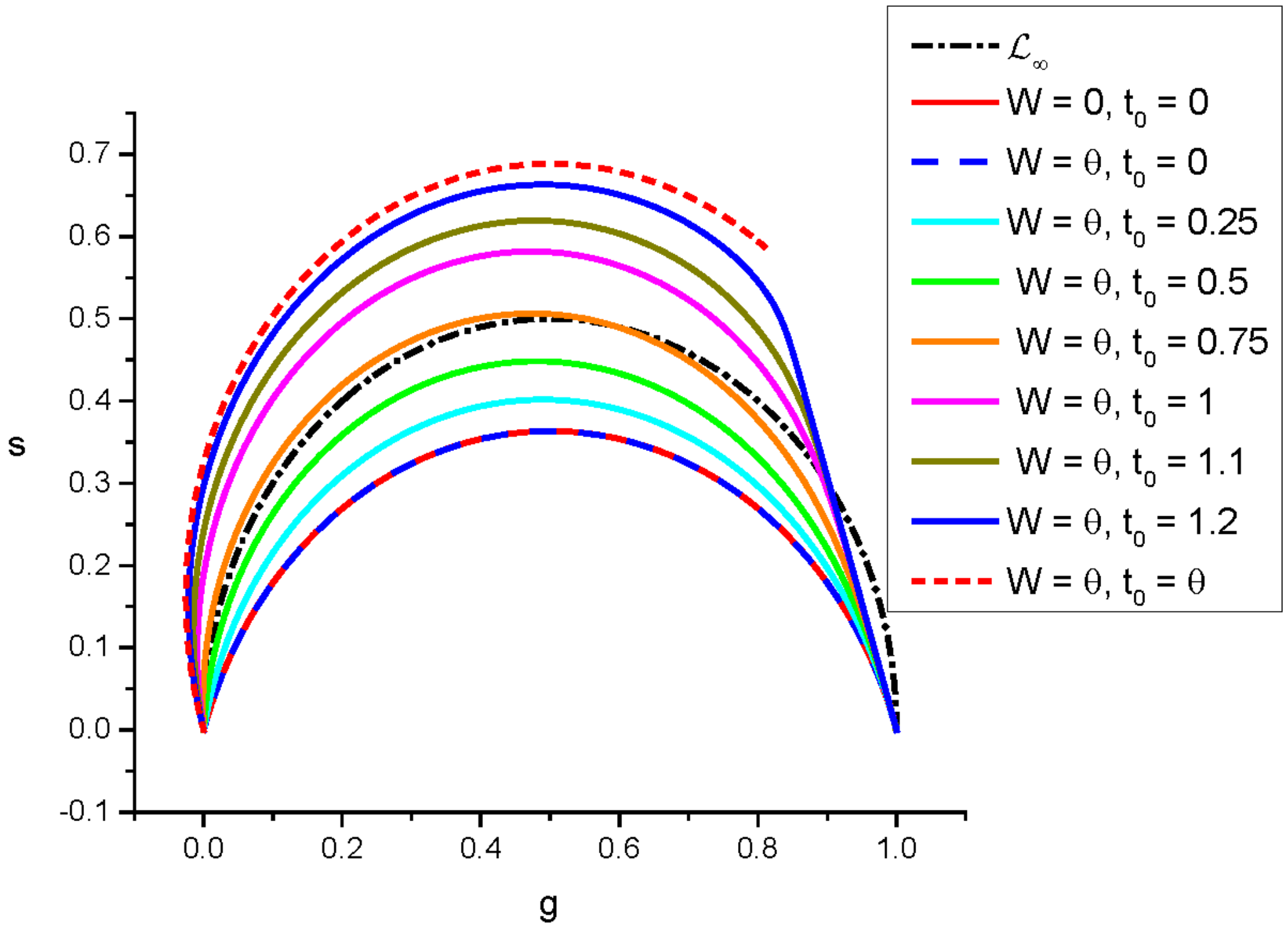

**Fig. 7**: Effect of a decay offset on the $\mathcal{L}_{N[W]}$ in the simple case where the gate width $W$ is equal to the gate step $\theta$. The calculations were made with $T$ = 12.5 ns, $N$ = 10 (i.e. $\theta$ = 1.25). The standard $\mathcal{L}_\infty$ (semi-circle) is represented as a black dot-dash curve. $\mathcal{L}_N$ (plain red) and $\mathcal{L}_{N[\theta]}$ (plain dark blue) are identical, yielding a dashed red/dark blue circular arc. As the decay offset $t_0$ increases (with $t_0 < \theta$), $\mathcal{L}_{N[W]}$ progressively rotates towards and is deformed into $\mathcal{L}_{N[\theta]}$, which is a circular arc (dashed red curve). Note that when $t_0 \sim \theta$, $\mathcal{L}_{N[\theta]}$ looks like a circular arc with a straight stem connecting to the 0-lifetime phasor $z_{N[W]}\left[\Lambda_{0,T}\right] = 0$.

This discussion shows that the SEPL can have rather odd shapes if the decay offset $t_0$ does not correspond to gate starts. This problem is minimized if the number of gates is large, since in this case, the difference between $\mathcal{L}_{N[\theta]}$s for $t_0 = q\theta$ and $t_0 = (q+1)\theta$ is a rotation of $2\pi f\theta = 2\pi \frac{n}{N}$, which is in general a small angle.

b. Square-gated PSEDs with single-exponential IRF with offset

The expression for the discrete phasor of a square-gated PSED with single-exponential IRF with offset is derived in Appendix D (Eq. (D53)). It does not correspond to a simple curve, but is useful to study the effect of the different parameters $(\theta, W, \tau_x, t_0)$ on the shape of the SEPL.

## 5. The effect of decay truncation

So far, we have assumed that the recorded decay comprises $N$ equidistant data points which cover the whole laser period $T$ ( $N\theta = T$ ). However, this might not always be the case experimentally, for various reasons. For instance, if the laser period is much longer than most of the lifetimes encountered in a study, and if each data point requires a long acquisition time, it might be advantageous to dispense with recording data for gates starting past a few times the largest lifetime after the laser pulse (*i.e.* offset $t_0$). Alternatively, the user may decide to skip the first few gates based on the argument that they are the most affected by the IRF, or do both (truncation on both sides of the laser period window).

While such a truncated decay may provide sufficient data for a good fit of the decay with a mono- or multi-exponential model, calculating its phasor based on Eq. (88) will in general result in a phasor that does not behave as described in the previous sections. In particular, the choice of the phasor harmonic frequency $f$ as a multiple of the fundamental frequency $T^{-1}$ turns out to be a poor choice in general.

As will soon become clear, an analytical expression of the phasor in the general truncated case is not particularly illuminating, and it is more efficient to analyze the effect of truncation numerically, starting from the easily calculated Eq. (88), which we will do in the examples discussed in Section 9. Here, we will limit ourselves to the case of the continuous and discrete phasors of ungated PSED with Dirac excitation, as they provide some insight on the different points discussed above. The situation of the discrete phasor of truncated square-gated PSED will be studied only in a single case where the phasor takes a simple form.

### *5.1. Continuous phasor of truncated ungated PSEDs*

We define the truncated decay by its initial recording position, $t_1 \geq 0$, and the total 'span' of the record, $D = N\theta$, such that $t_1 + D \leq T$. The definitions of $\left\| S_T(t) \right\|$ and the numerator of the corresponding 'truncated' phasor $\overleftrightarrow{z}\left[S_T\right]$ are given by a trivial modification of Eq. (63):

$$\begin{cases} \overleftrightarrow{z}\left[S_T\right]_{t_1,D} = \dfrac{\left\| S_T(t) e^{i2\pi ft} \right\|_{t_1,D}}{\left\| S_T(t) \right\|_{t_1,D}} \\ \left\| S_T(t) e^{i2\pi ft} \right\|_{t_1,D} = \displaystyle\int_{t_1}^{t_1+D} dt\ S_T(t) e^{i2\pi ft} \\ \left\| S_T(t) \right\|_{t_1,D} = \displaystyle\int_{t_1}^{t_1+D} dt\ S_T(t) \end{cases} \tag{137}$$

where we have used a double-ended arrow above the phasor symbol, $\overleftrightarrow{z}$, to indicate that the start ( $t_1$) and end ($t_1 + D$) of the integration are non-standard. Subscript '$t_1$, $D$' added to all quantities indicate the value of the first gate start and the records span. It is easy to verify that, for a PSED excited by a Dirac pulse (Eq. (17)):

$$\begin{cases} \vec{z}\left[\Lambda_{\tau,T}\right]_{t_1,D} = \dfrac{1-\eta e^{i2\pi fD}}{1-\eta} e^{i2\pi ft_1}\zeta_f(\tau) \\ \eta(\tau) = e^{-D/\tau} \end{cases} \tag{138}$$

This expression shows that the continuous phasor of a truncated PSED with Dirac excitation is equal to the canonical phasor $\zeta_f(\tau)$ rotated about the origin by a lifetime-dependent angle $2\pi f(t_1 - \beta)$ and scaled by a lifetime-dependent factor $\lambda(\tau)$ given by:

$$\begin{cases} \vec{z}\left[\Lambda_{\tau,T}\right] = \lambda e^{i2\pi f(t_1-\beta)}\zeta_f(\tau) \\ \beta(\tau) = \tan^{-1}\dfrac{\eta\sin(2\pi fD)}{1-\eta\cos(2\pi fD)} \\ \lambda(\tau) = \dfrac{\left(1-2\eta\cos(2\pi fD)+\eta^2\right)^{\frac{1}{2}}}{1-\eta} \end{cases} \tag{139}$$

The asymptotic behaviors of $\beta$ and $\lambda$ are:

- For $\tau \to 0$, $\eta(\tau) \to 0$ and therefore $\beta(\tau) \to 0$: the right hand side of Eq. (138) tends to $e^{i2\pi ft_1}\zeta_f(\tau)$, located on a rotated version of the $UC_\infty$.
- For $\tau \to \infty$, $\eta(\tau) \to 1$ and thus $\tan\beta(\tau) \to \cot\pi fD$, while $\lambda(\tau) \to 2\left|\sin\pi fD\right|\tau/D$. Because $\zeta_f(\tau) \xrightarrow[\tau\to\infty]{} e^{i\varphi(\tau)}/2\pi f\tau$ (see Eq. (71)), the result is:

$$\vec{z}\left[\Lambda_{\infty,T}\right]_{t_1,D} = \frac{\left|\sin(\pi fD)\right|}{\pi fD} e^{i\pi fD} e^{i2\pi ft_1}, \tag{140}$$

a value which is in general different from 0.

Fig. 8 illustrates, without loss of generality, these properties in the special case $t_1 = 0$, as $t_1 \neq 0$ simply adds a constant rotation (Eq. (138)). In that particular case, we truncated the decay down to $D = T/2$, for which Eq. (140) yields $\vec{z}\left[\Lambda_{\infty,T}\right]_{0,T/2} = i2/\pi$, which is located at the vertical of the locus of infinite lifetime in the standard $\mathcal{L}_\infty$ ($z = 0$).

It is also clear from Eq. (138), that if $f = n/D$, $n \in \mathbb{N}$, the first term $\dfrac{1-\eta e^{i2\pi fD}}{1-\eta}$ is equal to 1 and we are left with a rotated version of the continuous phasor of non-truncated ungated PSED, Eq. (99):

$$f = \frac{n}{D} \Rightarrow \vec{z}\left[\Lambda_{\tau,T}\right]_{t_1,D} = e^{i2\pi ft_1}\zeta_f(\tau) \tag{141}$$

This phasor frequency does not belong to the series of Fourier harmonics associated with $T$-periodic decays, $\{n/T\}_{n>0}$, but since it leads to a simpler functional form of $\vec{z}\left[\Lambda_{\tau,T}\right]_{t_1,D}$, and therefore a simpler interpretation of the calculated phasor, it is a natural choice to adopt.

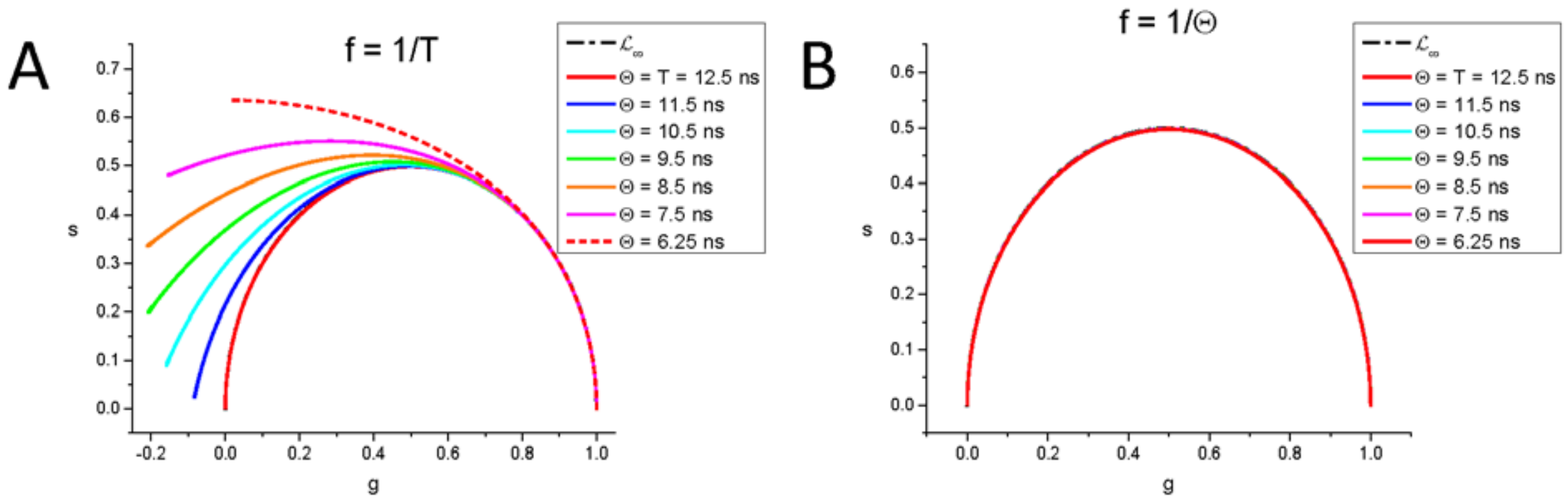

**Fig. 8**: Continuous phasor of truncated decays. To illustrate the effect of decay truncation, we use $T$ = 12.5 ns and compute the loci of continuous phasors of ungated PSED (SEPL) as a function of the observation duration $\Theta \leq T$ when $f = 1/T$ (A) or $f = 1/\Theta$ (B). In the latter case, all curves are identical to $\mathcal{L}_\infty$. However, if the phasor frequency is chosen to be the fundamental Fourier frequency $f = 1/T$, the SEPL increasingly departs from the $\mathcal{L}_\infty$ as the observation duration $\Theta$ decreases. For $\Theta = T/2$, the $z_{t_1,D}\left[\Lambda_{\infty,T}\right]$ ends up at the vertical of the 0 point (the location of $z\left[\Lambda_{\infty,T}\right]$ in $\mathcal{L}_\infty$).

### *5.2. Discrete phasor of truncated ungated PSEDs*

The discrete phasor of a truncated decay is defined by Eq. (88), in which $0 \leq t_1 < t_N \leq T$ (we will assume that the gates are equidistant: $\theta_p = \theta, \quad p \in [1, N]$) and they span $D = N\theta < T$ .

In the only case we will treat analytically, that of a PSED in the presence of a Dirac IRF, a straightforward calculation yields:

$$\tilde{z}_N\left[\Lambda_{\tau,T}\right]_{t_1,D} = \frac{1-x^N e^{iN\alpha}}{1-x^N}\frac{1-x}{1-xe^{i\alpha}}e^{i2\pi f t_1} = \frac{1-x^N e^{iN\alpha}}{1-x^N}e^{i2\pi f t_1}\zeta_{f,N}\left(\tau\right) \tag{142}$$

where we have used the previous notations $x(\tau) = e^{-\theta/T}$, $\alpha = 2\pi f\theta$ and $\zeta_{f,N}(\tau)$ is the discrete phasor of a non-truncated, ungated PSED, Eq. (99). The prefactor in front of the term $e^{i2\pi f t_1}\zeta_{f,N}(\tau)$ in Eq. (142) depends on $\tau$ and therefore shows that the SEPL is in general complex, unless $N\alpha = 2n\pi$, i.e. $f = n/D$, $n \in \mathbb{N}$.

We can therefore distinguish two situations:

#### *5.2.1. $f = n/D$, $n \in \mathbb{N}$*

As in the continuous case, if $f = n/D$, $n \in \mathbb{N}$, the fractional prefactor in Eq. (142) is equal to 1 and we are left with a rotated version of the discrete phasor of non-truncated ungated PSED with Dirac excitation, Eq. (99), which is an arc of circle rotated about the origin:

$$f = \frac{n}{D} \Rightarrow \tilde{z}_N\left[\Lambda_{\tau,T}\right]_{t_1,D} = e^{i2\pi f t_1} z_N\left[\Lambda_{\tau,T}\right] \tag{143}$$

### 5.2.2. General case, $f \neq n/D$

In the general case, the fractional prefactor in Eq. (142) can be rewritten $\lambda_N e^{-i\beta_N}$ with:

$$\begin{cases} \lambda_N(\tau) = \dfrac{\left(1-2x^N \cos N\alpha + x^{2N}\right)^{\frac{1}{2}}}{1-x^N} \\ \beta_N(\tau) = \tan^{-1} \dfrac{x^N \sin N\alpha}{1-x^N \cos N\alpha} \end{cases} \tag{144}$$

This prefactor depends on $\tau$, and its asymptotic behavior when $\tau \to 0$ and $\tau \longrightarrow \infty$ is easy to compute:

- when $\tau \to 0$, $\lambda_N \to 1$ and $\beta_N \to 0$, therefore the phasor $\vec{z}_N \left[\Lambda_{\tau,T}\right]_{t_1,D}$ tends to the expression of Eq. (143), which defines a rotated circular arc.
- when $\tau \longrightarrow \infty$, $x \sim 1-\theta/\tau$ and $x^N \sim 1-N\theta/\tau$, which leads to the asymptotic expression:

$$\vec{z}_N \left[\Lambda_{\tau,T}\right]_{t_1,D} \xrightarrow[\tau\to\infty]{} \frac{1}{N}\frac{1-e^{iN\alpha}}{1-e^{i\alpha}} e^{i2\pi f t_1} \tag{145}$$

This value is different from 0 when $f \neq n/D$. It is easy to verify that in the limit $N \to \infty$, we recover Eq. (140).

Overall, we see that the curve described by Eq. (142) is close to a rotated arc of circle (Eq. (143)) when $\tau \to 0$ and ends up on a point $\vec{z}_N \left[\Lambda_{\infty,T}\right]_{t_1,D}$ which is in general different from the origin.

## 5.3. Discrete phasor of truncated square-gated PSEDs with offset

This situation is the most complicated, but also the most general, and, as in the previous discussion, can be simplified with an adequate choice of starting gate position and phasor frequency.

As shown in Appendix B, Section C.2.4.b, in the case of a positive offset ($t_0 \in \left]0,T\right[$) and first gate chosen to start at the decay offset, the phasor reads, assuming a phasor harmonic frequency proportional to $D^{-1}$:

$$\left.\begin{matrix} t_1 = t_0,\ t_N < T-\omega \\ f = \dfrac{n}{D} \end{matrix}\right\} \Rightarrow \begin{cases} \vec{z}_{N[W]} \left[\Lambda_{\tau,T}\right]_{t_0,D} = \dfrac{1-x}{1-xe^{i\alpha}} e^{i2\pi f t_0} = e^{i2\pi f t_0} \zeta_{f,N}(\tau) \\ x(\tau) = e^{-\theta/\tau};\quad \alpha = 2\pi f \theta \end{cases} \tag{146}$$

This is the same equation as for the discrete phasor of an ungated PSED and describes a circular arc rotated about the origin. Note that this simple formula is valid only if the phasor frequency is a multiple of $D^{-1}$ (it is not correct if $f$ is chosen to be a multiple of $T^{-1}$ instead, see Section C.2.4.b).

In other words, as shown in Fig. 9, in the case of a discrete square-gated PSED with offset, it might be advantageous to make sure that the IRF location $t_0$ coincides with the start of a gate, chosen as the starting gate ($t_1 = t_0$), and truncate the recording at gate $N$ such that:

$$N \leq \left\lfloor \frac{T-\omega-t_0}{\theta} \right\rfloor \tag{147}$$

Using a phasor frequency $f = n/N\theta = n/D$ will result in a phasor:

$$z_{N[W]}\left[\Lambda_{\tau,T}\right] = \frac{1-x}{1-xe^{i\alpha}} e^{i2\pi ft_0} = e^{i2\pi ft_0} \zeta_{f,N}(\tau) \tag{148}$$

that is, an arc of circle rotated about the origin.

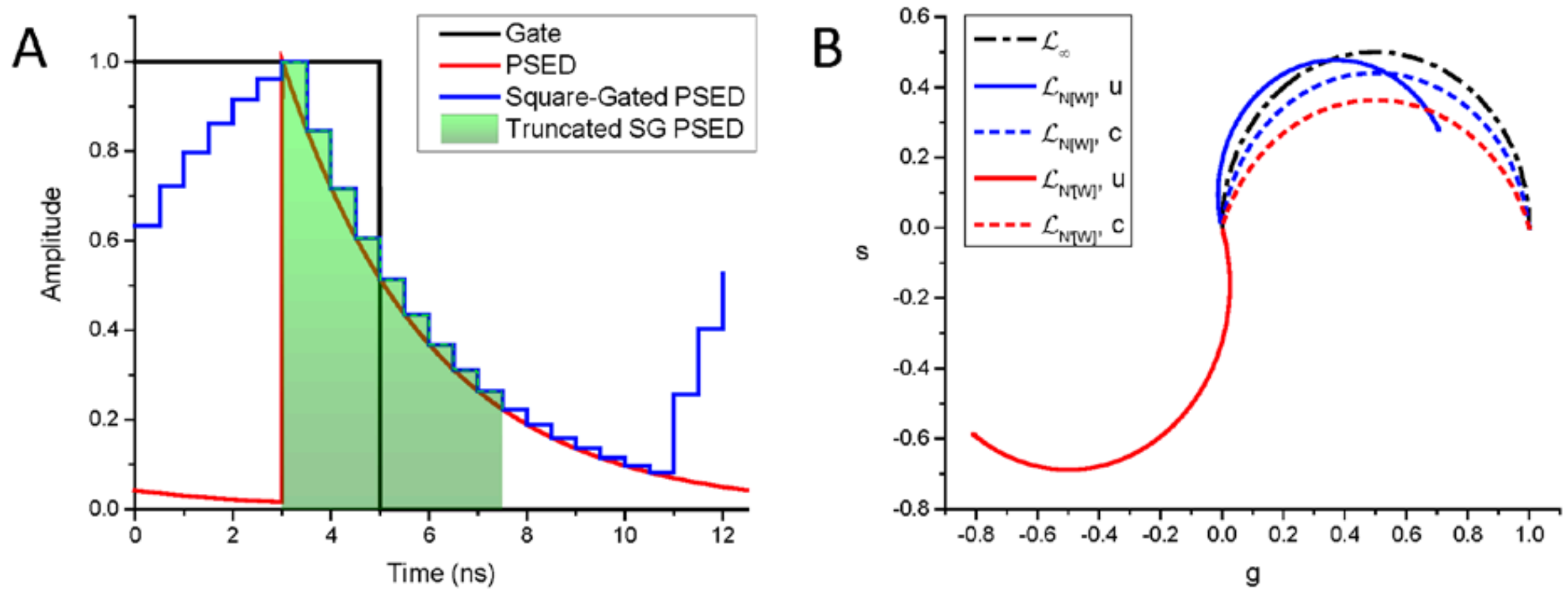

**Fig. 9**: Discrete phasor of an offset and truncated square-gated PSED. A: Square Gate ($W$ = 5 ns, black), periodic single-exponential decay (PSED, $\tau$ = 3 ns, red), gated PSED sampled every $\Theta$ = 0.5 ns ($N$ = 25, blue) and truncated version ($N$ = 10, green) starting at $t_1$ = 3 ns, the position of the laser IRF, and ending at $t_N = T - W$. The laser period is T = 12.5 ns. The 'duration' of the truncated decay is D = T – W – $t_0$ + Θ = 5 ns. B: SEPLs for the PSED shown in A. blue: full PSED, $f$ = $1/T$, red: truncated PSED, $f$ = $1/D$. Uncalibrated SEPLs are shown as solid curves, while calibrated ($\tau_C$ = 0 ns) SEPLs are shown as dashed curves.

## 6. The effect of gate shape

In previous sections, we have used square gates as an example of gate shape that can be easily treated analytically, at least in the case of simple excitation pulse shapes. This has allowed us to study the effect of gate width on the phasors of PSEDs. This model is adapted to data acquired with photon-counting detectors followed by electronics that effectively bin photons, or to study the effect of binning on such data. For instruments whose response is actually electronically gated (turned on and off), the resulting detection efficiency temporal profile is rarely rectangular (or 'square'). In practice, the finite temporal resolution of the response leads to smooth instead of sharp edges, and potentially to ringing or irregular rather than flat top. Examples of such departure from ideality can be found in the literature (e.g. [28,29,32]). In other cases, the electronics 'gating' profile might simply not be rectangular at all, but instead triangular, or ramp-like or even sinusoidal, among many possible examples.

Note that it also possible to modulate the phase of the modulation of the detector response in some frequency modulation approaches [33,34]. While this may give rise to interesting modulation shapes, these are not directly relevant to the topic of this article, concerned exclusively with fixed phase (or offset) gate functions.

### *6.1. Effect of gate shape on continuous phasors*

In the case of continuous phasors, the effect of gating on the phasor is indeed trivial, due to the fact that:

(i) a gates effect on a decay can be described as an additional term in a convolution product (Eq. (42)),

(ii) the convolution rule (Eq. (69)) shows that this gate term amounts to multiplying the phasor by a constant term.

In other words, two experiments differing only by theirs gate shapes will result in phasors that differ only by a constant complex scaling factor, *i.e.* a dilation and a rotation of the universal circle ($\mathcal{L}_\infty$). As discussed in Section 8, this difference is taken care of by phasor calibration.

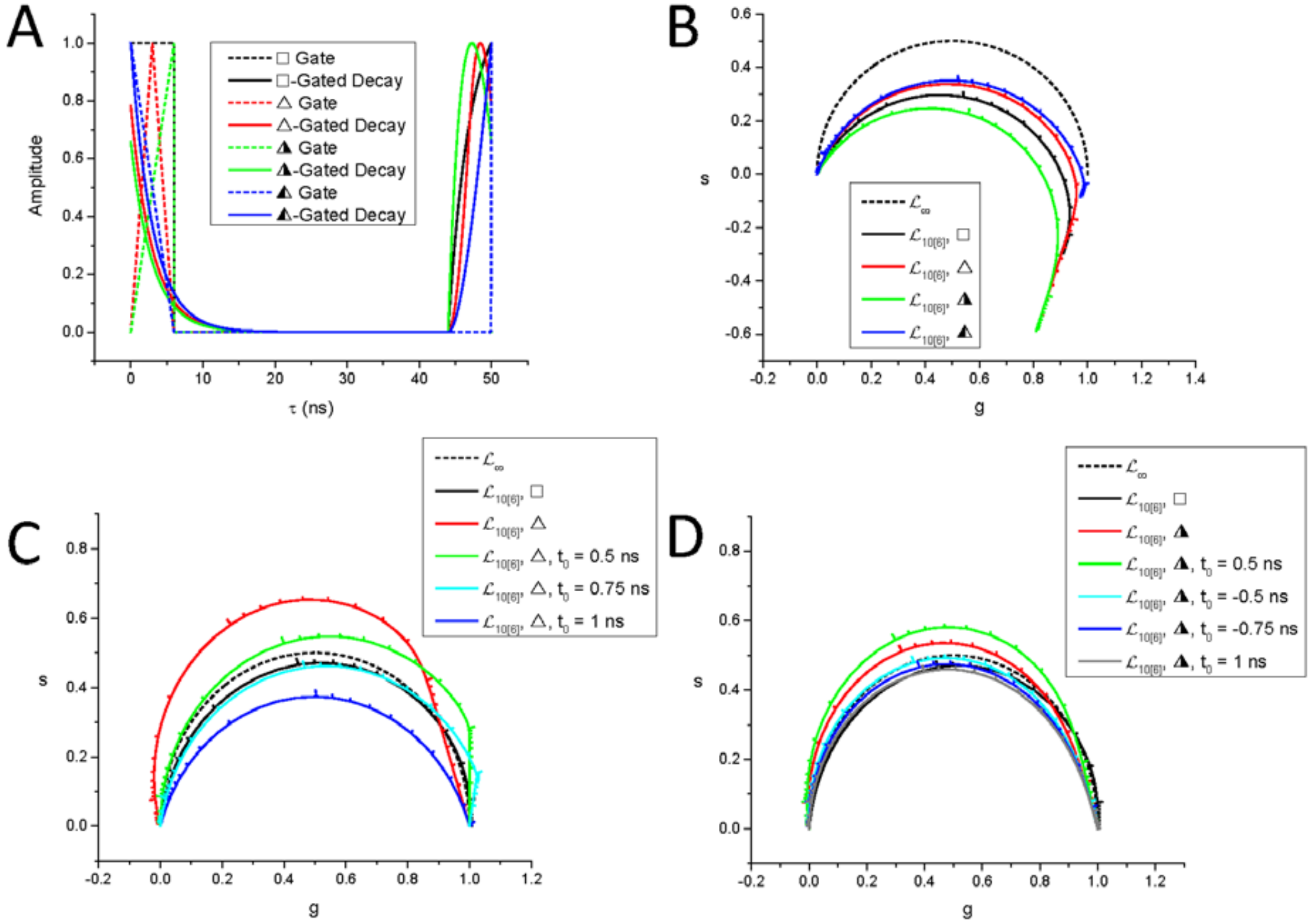

**Fig. 10**: Effect of gate shapes on the discrete phasors. A: 4 gates of width $W$ = 6 ns (square, triangle, sawtooth and reversed sawtooth, dashed curves) are shown starting at $t$ = 0 within a period of duration $T$ = 50 ns ($f$ = 20 MHz). The corresponding gated-decays for a PSED with $\tau$ = 3 ns are shown as plain curves. Due to the different locations of the maximum of each gate, the corresponding $T$-periodic gated decays exhibit different maxima locations as well as different shapes. This effect is similar for all PSEDs and therefore results in different SEPL shown in B. B: universal semicircle ($\mathcal{L}_\infty$, dashed curve) and the 4 SEPL with gate width $W$ = 6 ns for the same number of equidistant gate locations $N$ = 10. While the SEPLs look fairly similar, noticeable differences exist. Ticks indicate the locations of PSEDs with lifetime 0.1 – 1 in steps of 0.1, 1-10 in steps of 1, etc., with ticks corresponding to 1, 10, and 100 drawn slightly longer. C: By shifting the triangle gate or equivalently, introducing a positive IRF offset t0 (indicated in the legend), the difference between the calibrated SEPL and that corresponding to the square gate (black curve) can be minimized if not completely eliminated. D: A similar but negative adjustment of the sawtooth gate offset achieves a similar better similarity between the SEPL corresponding to the square gate (black curve) and the sawtooth one (offset indicated in the legend).

### *6.2. Effect of gate shape on discrete phasors*

The difference between SEPLs corresponding to different gate shapes is more subtle for discrete phasors. Indeed, we have seen in Section 3.3.4, that even in the case of a simple square gate and Dirac excitation, the shape of the SEPL can change significantly by a mere change of gate width. This effect will be more noticeable the smaller the number of gates.

A numerical comparison of the SEPLs obtained for a few gate shapes can be instructive, and is presented in Fig. 10. Fig. 10A shows 4 examples of $W = 6$ ns-wide gate shapes (square, triangle, sawtooth and reversed sawtooth) and their effect on a $\tau = 3$ ns PSED. All broaden the decay but also shift and deform it in different ways. This results in different discrete SEPLs computed for $N = 10$ gates for the 4 gate shapes, as shown in Fig. 10B. These differences are not overly surprising, considering that the gates may effectively shift the IRF differently. Fig. 10C&D illustrate this point, by showing that by adding or subtracting some IRF offset, the SEPL of triangle-gated decays (Fig. 10C) or of sawtooth-gated decays (Fig. 10D) can be somewhat (but not perfectly) made to look closer after proper rotation and rescaling so that the phasor of the 0-lifetime PSEDs are all located at 1 (i.e. after proper phasor calibration, as discussed in detail in Section 8).

While these observations are merely qualitative, they suggest that, in general, gates of similar duration (*i.e.* support size *W*) result in similar SEPL shapes, provided the proper IRF shift is implemented.

More quantitative estimates of the effects of small gate shape variations on the calculated phasors can of course be obtained in case an analytical expression for the phasor is available, such as those provided in this work. We give an example of such an analysis in Section 9.3.2 when discussing the effect of gate width variations on the phase and modulus lifetime. The necessary expressions are derived in the next section.

## 7. Phase and modulus lifetimes

### *7.1.General considerations*

Equipped with this better understanding of the differences between continuous versus discrete phasor of ungated and square-gated PSED, the effect of complete or truncated recording, IRF offset and various combinations thereof, we can now look into ways to use the computed phasors to gain information on the recorded decays.

In 'standard' phasor analysis, by which term we mean analysis of *continuous* phasors of *ungated* decays excited by a Dirac IRF, definition (70) of the phasor $\zeta_f(\tau)$ of a PSED with lifetime $\tau$ can be rewritten $\zeta_f(\tau) = me^{i\varphi}$ as in Eq. (71), where:

$$\begin{cases} m(\tau) = \dfrac{1}{\sqrt{1+(2\pi f \tau)^2}}; \\ \varphi(\tau) = \tan^{-1}(2\pi f \tau) \end{cases} \tag{149}$$

defining the phasor phase $\varphi$ and modulus $m$. For PSED, this leads to two equivalent expressions for the lifetime $\tau$: the *phase lifetime* $\tau_\varphi$ and the *modulus lifetime* $\tau_m$, given by [13,35]:

$$\begin{cases} \tau_\varphi = \dfrac{1}{2\pi f}\tan\varphi \\ \tau_m = \dfrac{1}{2\pi fm}\sqrt{1-m^2} \end{cases} \quad (150)$$

These expressions can be used formally with the modulus and phase of the phasor of non-single-exponential decays as well, but in that case, the two values $\tau_\varphi$ and $\tau_m$ are likely to (i) be different from one another, and (ii) their interpretation will be ambiguous at best. In particular, if the phasor of a decay $S_T(t)$ is outside $\mathcal{L}_\infty$, *i.e.* $m > 1$, $\tau_m$ given by Eq. (150) will be imaginary (due to the presence of the square root of a negative number). No such problem exists for $\tau_\varphi$, but the result should be interpreted cautiously, as any decay whose phasor is not located on the universal circle is obviously not a PSED. In particular, $\tau_\varphi$ is different from the average lifetime:

$$\begin{cases} \tau_\varphi \neq \langle \tau \rangle = \dfrac{\int_0^\infty dt\ t\,F_0(t)}{\int_0^\infty dt\,F_0(t)} = \dfrac{\left\| \mathcal{F}_{0,T}(t) \right\|_T}{\left\| F_{0,T}(t) \right\|_T} \\ \mathcal{F}_0(t) = -\int du\,F_0(u) \end{cases} \quad (151)$$

where $F_0(t)$ is the signal emitted upon a Dirac excitation, $F_{0,T}(t)$ its $T$-periodic summation, $-\mathcal{F}_0(t)$ its primitive and $\mathcal{F}_{0,T}(t)$ the $T$-periodic summation of $\mathcal{F}_0(t)$.

In the case of the different examples of phasor expressions studied previously:

(i) continuous phasor of ungated PSEDs with single-exponential IRF (Eq. (73)),
(ii) continuous phasor of square-gated PSEDs with Dirac IRF (Eq. (75)),
(iii) continuous phasor of square-gated PSED with single-exponential IRF (Eq. (78)),
(iv) discrete phasor of ungated PSEDs with Dirac IRF (Eq. (99)),
(v) discrete phasor of ungated PSEDs with single-exponential IRF (Eq. (101)),
(vi) discrete phasor of square-gated PSEDs with Dirac IRF (Eq. (103)),

it is still possible to define a modulus $m$ and phase $\varphi$ of the calculated continuous or discrete phasor $z = me^{i\varphi}$, where $m = m(\tau)$ and $\varphi = \varphi(\tau)$ are different functions of $\tau$ than those of Eq. (149) and the lifetimes $\tau_\varphi$ and $\tau_m$ defined as:

$$\begin{cases} \tau_\varphi = \varphi^{-1}(\arg(z)) \\ \tau_m = m^{-1}(|z|) \end{cases} \quad (152)$$

are not given by Eq. (150) anymore, because the phasor expression is different from $\zeta_f(\tau)$ (Eq. (70)).

It is possible to obtain analytical formulas for the phasor modulus and phase in the continuous square-gated decay case and the discrete ungated decay case. However, in general, only implicit formulas can be obtained in the discrete square-gated decay case, due to the complexity of Eq. (103), although, in a few special cases, analytical formulas can be obtained.

We will look at these cases in the next sections.

### *7.2. Phase and modulus lifetime of continuous phasors*

#### *7.2.1. Phase and modulus lifetime of ungated PSEDs in the presence of a single-exponential IRF*

Equation (73) for the phasor in this case can be rewritten:

$$\begin{cases} z\left[\Lambda_{\tau,T}\right] = m(\tau)e^{i\varphi(\tau)} \\ m(\tau) = \dfrac{1}{\sqrt{1+\left(2\pi f \tau_*\right)^2}} \dfrac{1}{\sqrt{1+\left(2\pi f \tau\right)^2}} \\ \varphi(\tau) = \tan^{-1}\left(2\pi f \tau_*\right) + \tan^{-1}\left(2\pi f \tau\right) \end{cases} \tag{153}$$

from which a phase and modulus lifetime can be defined by:

$$\begin{cases} \tau_\varphi = \tan\left(\varphi - \varphi_*\right) \\ \tau_m = \dfrac{1}{2\pi f}\sqrt{\left(m_*/m\right)^2 - 1} \end{cases} \tag{154}$$

where angle $\varphi_*$ and modulus $m_*$ were defined in Eq. (74). These equations are the same as Eq. (150) after rotation of the phasor by $\varphi_*$ and dilation by a factor $1/m_*$. Obviously, for this calculation to be possible, $m_*$ and $\varphi_*$, the modulus and phase of the excitation pulse need to be known. A simpler alternative is provided by phasor calibration, as discussed in Section 8.

#### *7.2.2. Phase and modulus lifetime of continuous square-gated PSEDs with Dirac IRF*

Writing Eq. (75) for the continuous phasor of an ungated PSED with Dirac IRF as:

$$z_{[W]}\left[\Lambda_{\tau,T}\right] = m(\tau)e^{i\varphi(\tau)} \tag{155}$$

we obtain the following equations for the phase and modulus lifetimes of square-gated PSEDs:

$$\begin{cases} \tau_\varphi = \dfrac{1}{2\pi f}\tan\left(\varphi + \varphi_W\right) \\ \tau_m = \dfrac{1}{2\pi f}\sqrt{\left(M_W/m\right)^2 - 1} \end{cases} \tag{156}$$

where $\varphi_W$ and $M_W$ are defined in Eq. (75). These equations simply express the fact that $\mathcal{L}_{[W]}$ is a rotated and dilated version of $\mathcal{L}_\infty$ and thus the modulus and phase of the phasor need to be respectively rescaled and corrected before using the formulas valid for $\mathcal{L}_\infty$ (Eq. (150).

#### *7.2.3. Phase and modulus lifetime of continuous square-gated PSEDs with single-exponential IRF*

Writing Eq. (78) for the continuous phasor of a square-gated PSED with single-exponential IRF as:

$$z_{[W]}\left[\Psi_{\tau,\tau_*,T}\right] = m(\tau)e^{i\varphi(\tau)} \tag{157}$$

we obtain the following equations for the phase and modulus lifetimes:

$$\begin{cases} \tau_\varphi = \dfrac{1}{2\pi f} \tan\left(\varphi + \varphi_W - \varphi_*\right) \\ \tau_m = \dfrac{1}{2\pi f} \sqrt{\left(m_* M_W / m\right)^2 - 1} \end{cases} \tag{158}$$

### *7.3. Phase and modulus lifetime of discrete phasors*

#### *7.3.1. Phase and modulus lifetime of discrete ungated PSEDs with Dirac IRF*

Writing $z_N\left[\Lambda_{\tau,T}\right]$ in Eq. (99) as:

$$z_N\left[\Lambda_{\tau,T}\right] = m(\tau) e^{i\varphi(\tau)} \tag{159}$$

we obtain the following equations for the phase and modulus lifetimes of discrete ungated PSEDs with Dirac IRF, after some straightforward algebra:

$$\left.\begin{matrix} m \leq 1 \Rightarrow \tau_m = \tau_m^- \\ 1 < m \leq \left|\cos\dfrac{\alpha}{2}\right|^{-1} \Rightarrow \tau_m = \tau_m^+ \end{matrix}\right\} \rightarrow \begin{cases} \tau_\varphi = \dfrac{\theta}{\ln\left[\cos\alpha\left(1 + \dfrac{\tan\alpha}{\tan\varphi}\right)\right]} \\ \tau_m^\pm = \dfrac{\theta}{\ln\left(\dfrac{1-m^2}{\left(\sqrt{1-m^2\cos^2\dfrac{\alpha}{2}} \pm m\left|\sin\dfrac{\alpha}{2}\right|\right)^2}\right)} \end{cases} \tag{160}$$

where we have used the notation $\alpha = 2\pi f\theta$ introduced in Eq. (99). Notice that the term $\left|\cos\dfrac{\alpha}{2}\right|^{-1}$ in Eq. (160) is the diameter of the circle of which $\mathcal{L}_N$ is a part of and therefore the condition expressed in Eq. (160) simply states that the phasor needs to be inside $\mathcal{L}_N$ for the modulus lifetime to be defined, as expected.

Notice also that there is no simple connection between these expressions and those valid in the simpler case of ungated decays, Eq. (150).

#### *7.3.2. Phase and modulus lifetime of discrete ungated PSEDs with single-exponential IRF*

Writing $z_N\left[\Psi_{\tau,\tau_*,T}\right]$ in Eq. (101) as:

$$z_N\left[\Psi_{\tau,\tau_*,T}\right] = m(\tau) e^{i\varphi(\tau)} \tag{161}$$

we obtain the same equations as those obtained for the phase and modulus lifetimes of discrete ungated PSEDs with Dirac IRF (Eq. (160)), with the simple replacements:

$$
\begin{cases}
m \mapsto \dfrac{m}{\left(\dfrac{x_* \sin\alpha}{1 - x_* \cos\alpha}\right)} \\
\varphi \mapsto \varphi - \alpha - \tan^{-1}\left(\dfrac{x_* \sin\alpha}{1 - x_* \cos\alpha}\right)
\end{cases}
\tag{162}
$$

### 7.3.3. *Phase and modulus lifetime of discrete square-gated PSEDs with Dirac IRF*

As mentioned previously, Eq. (103) for $z$ does not, in general, lead to any simple relation between the phase and modulus of $z$ and the lifetime $\tau$. In fact, as is visible in Fig. 11, for some choices of gate width $W$ and gate step $\theta$, different PSEDs can be associated with the same modulus, showing that, in some cases, looking for an unambiguous modulus lifetime is impossible using the

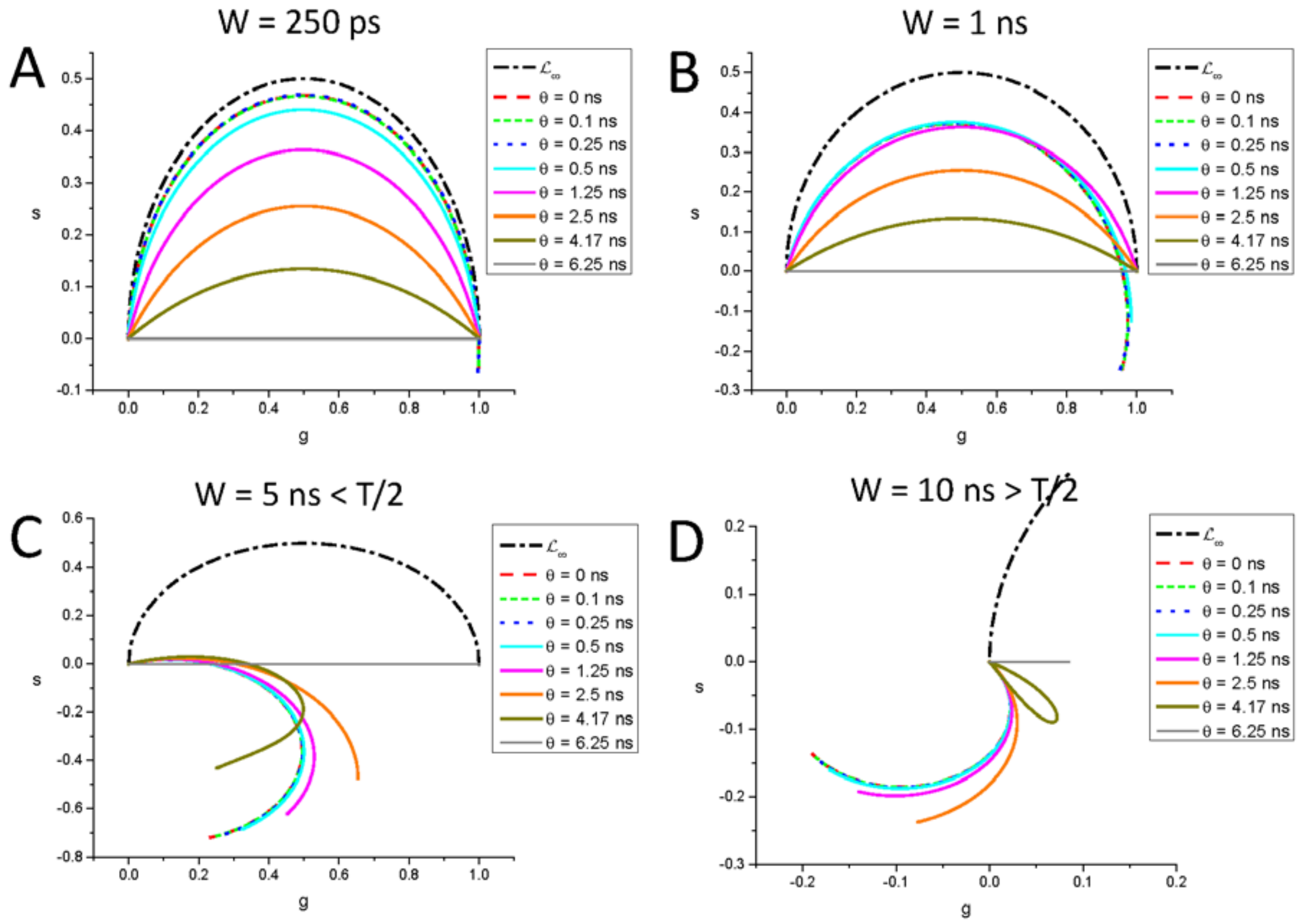

**Fig. 11**: Locus of the discrete phasors of periodic square-gated single-exponential decays, $\mathcal{L}_{N[W]}$ for different $N$ and $W$. The curves were calculated for a laser period $T$ = 12.5 ns and the corresponding first harmonic frequency $f_1$ = 80 MHz. Each panel corresponds to a different gate width, each curve corresponding to a different gate step choice $\theta$ (the number of gates is $N = T/\theta$). A: $W$ = 250 ps, B: $W$ = 1 ns, C: $W$ = 5 ns, D: $W$ = 10 ns. Note in particular the locus of phasors for $N$ = 2 ($\theta$ = 6.25 ns), which covers the whole [0,1] segment, except when $W > T/2$. Of particular interest is the case $N$ = 3 and $W > T/2$, for which the $\mathcal{L}_{N[W]}$ forms a closed loop.

implicit formula:

$$
\left| z_{N[W]}\left[\Lambda_{\tau,T}\right] \right| = m(\tau)
\tag{163}
$$

Fortunately, this is not the case for the phase $\varphi$, which appears to be uniquely defined for $\tau \geq 0$. The implicit relation between the phase $\varphi(\tau)$ and lifetime $\tau$ is given by:

$$\frac{\operatorname{Im}\left(z_{N[W]}\left[\Lambda_{\tau,T}\right]\right)}{\operatorname{Re}\left(z_{N[W]}\left[\Lambda_{\tau,T}\right]\right)} = \tan\varphi(\tau) \tag{164}$$

Numerical solutions of these equations can be obtained using standard iterative procedures and are implemented in the *Phasor Explorer* software provided with this article (see Supplementary Material and Data Availability sections below).

Note that all the results above are only valid in the case where there is no decay offset (the laser pulse corresponds to $t_0 = 0$ in the decay recording coordinates) and the decay is not truncated ($T = N\theta$). In practice, the decay might well be complete, but unless care is taken to circularly shift the decay such that time 0 in the recording corresponds to the IRF peak, the locus of phasors of PSEDs will be different from one of the tractable situations described above making it impractical to obtain modified equations for the phase and modulus lifetime.

While this seems to imply that the prospect of interpreting phasor data in terms of phase (or modulus) lifetime in the most general case is compromised, it turns out that a simple approach based on the concept of phasor calibration can provide useful quantitative results in most practical situations, as discussed next.

## 8. Phasor calibration and pseudo-calibration

Phasor calibration is a central concept in *continuous* phasor analysis, abstracting all experimental details into a simple algebraic operation, in order to bring back the phasors of PSEDs to the universal semicircle $\mathcal{L}_\infty$. In Section 8.1, we first qualitatively discuss how this is modified in the non-standard cases examined in this article, before briefly reviewing the case of continuous phasors (Section 8.2) and examining discrete phasors quantitatively (Section 8.3).

### *8.1. Reference Single-Exponential Phasor Loci*

The results of the previous sections have shown that different data recording situations may result in different loci of the phasor of PSEDs (a curve dubbed SEPL) in the phasor plot. Does this mean that phasor data needs to interpreted differently each time the SEPL changes, *i.e.* each time some modification of the experimental conditions happens? In the case of continuous phasors of complete decays, the answer is no, provided *phasor calibration* is used, as discussed in the next sub-section. We will first qualitatively discuss this familiar situation, and contrast it with non-standard cases discussed in previous sections, in order to extend this notion to these more complex cases. To help with the discussion, the different cases addressed in this section are schematically illustrated in Fig. 12.

As discussed in Sections 3 & 4, for *continuous phasors of non-truncated decays*, the effects of excitation pulse profile, instrument integration (gating) or offset, etc., can all be described as convolution products of various decay-independent functions with the 'pure' luminescence decays of interest (*i.e.* the signal emitted by the sample upon excitation by a Dirac comb). Due to the *continuous phasor convolution rule* (Section 3.2.1, Eq. (69)), this means that the resulting phasors are merely multiplied by a constant complex number, compared to the ideal situation where the IRF is a Dirac function (and therefore no detector gating or binning takes place). In other words, the experimental continuous phasors only differ by a dilation and a rotation from the phasors of the pure single-exponential decays, whatever the nature of the excitation function and electronic response function (including gating and offset) are. The corresponding SEPL is thus a rescaled and rotated version of the canonical $\mathcal{L}_\infty$ obtained for continuous, offset-free, ungated PSEDs.

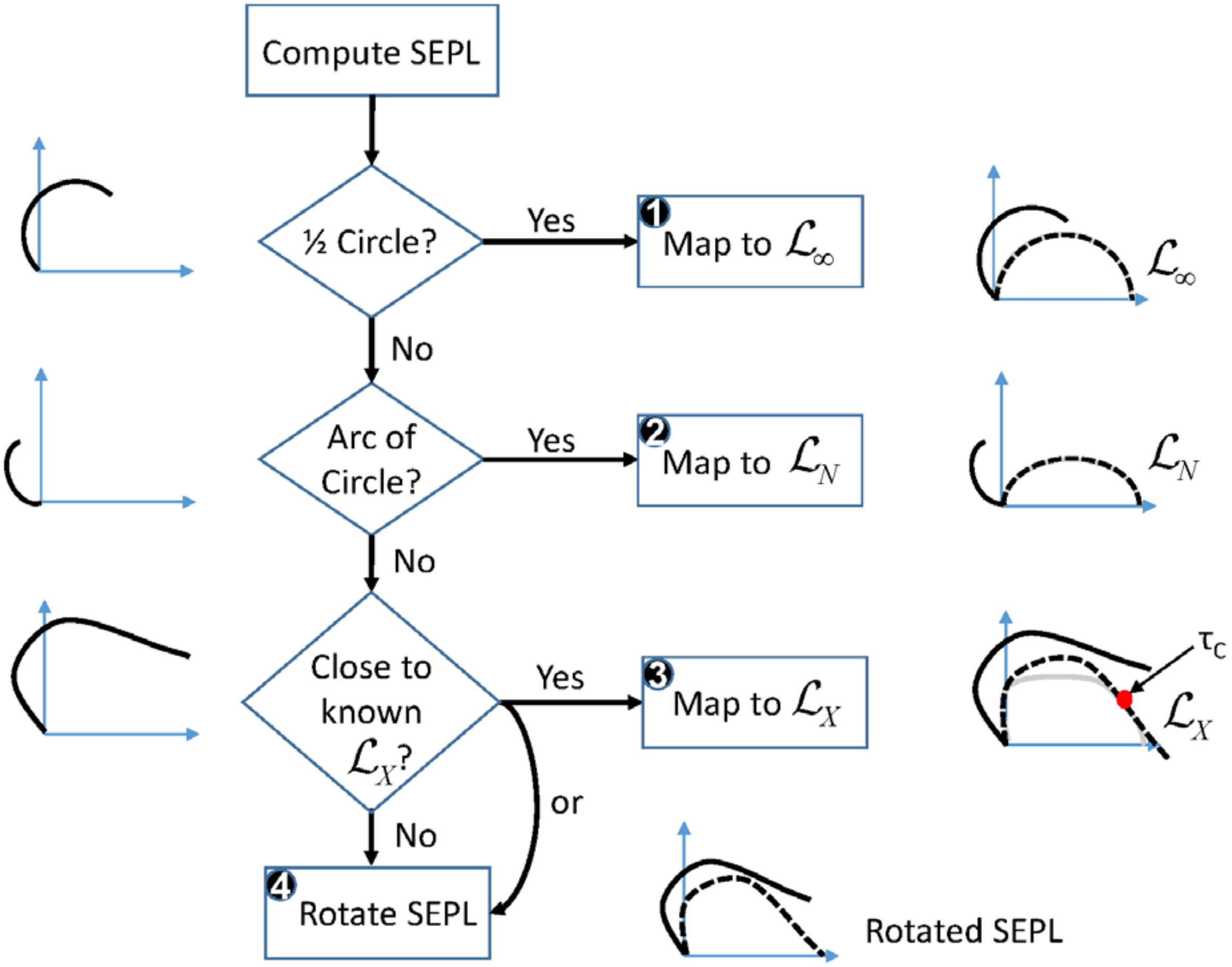

**Fig. 12**: Phasor calibration workflow. The Single-Exponential Phasor Loci (SEPL) can be computed if the IRF shape (including the gate shape effect) is known. The resulting curve (shown on the left side of the figure) can be compared to the simple cases studied here. Case 1: if the SEPL is a rotated, scaled half circle, it is natural to use the universal semicircle, $\mathcal{L}_\infty$, as reference. In that case, the calibrated phasors are mapped exactly to the corresponding phasors on the universal semicircle. Case 2: if the SEPL is a rotated, scaled arc of circle, it is natural to use $\mathcal{L}_N$ as reference. In that case, the calibrated phasors are mapped exactly to the corresponding phasors on $\mathcal{L}_N$. Case 3: if the SEPL, after rotation and scaling is identical to (resp. close to) a simple SEPL (generically referred to as $\mathcal{L}_X$), it is natural to it as a reference. In that case, the calibrated phasors are mapped exactly to (resp. close to) the corresponding phasors on $\mathcal{L}_X$. Case 4: if there is no good match of the rotated/rescaled SEPL to any known reference SEPL, it is simpler to rotate and scale the SEPL so that the phasor of a PSED with zero lifetime is mapped to 1.

In these simple cases, an opposite rotation and dilation will bring the experimental *SEPL* back to the reference $\mathcal{L}_\infty$, as illustrated in Fig. 12①. Data interpretation and analysis are thus simplified, being performed within the same familiar framework, where the phasor of a zero lifetime PSED is equal to 1, the phasor of an infinite lifetime PSED is equal to 0, and the phasor of PSEDs with finite lifetimes are located at predictable locations on the universal semi-circle $\mathcal{L}_\infty$.

In all other cases, however, the difference is more profound, such as for instance in the case of discrete phasors (discussed in Sections 3.3 and 4.2.3 & 4.2.4), or for continuous phasors of truncated decays (Section 5). Indeed, the locus of discrete phasors of *ungated*, *offset-free* PSEDs with Dirac IRF, $\mathcal{L}_N$ (Eq. (99)), is a circular *arc* instead of a full semicircle, whose radius *r* depends on the number of gates *N* according to Eqs. (99)-(100). In the presence of a single-exponential IRF (Eq. (101)) or an IRF offset $t_0$ (Eq. (131)), the corresponding SEPL is still a circular arc, but the arc is rotated by an angle which depends in a non-trivial manner on the IRF or its offset, and additionally, dilated in the single-exponential IRF case. In those cases, it would seem desirable to use $\mathcal{L}_N$ as the reference SEPL, since trying to map those SEPLs back to $\mathcal{L}_\infty$ (a half-circle) by inverse rotation and dilation, will clearly not succeed, as illustrated in Fig. 12②. The advantage of such a remapping to $\mathcal{L}_N$ would be that, in all these cases, the phasor of a PSED with zero lifetime would always be located at 1 and the phasor of a PSED with infinite lifetime at 0, with the phasor of PSEDs with intermediate lifetimes in-between on the $\mathcal{L}_N$ at predictable locations.

The fact that things are otherwise more complex is obvious because, for an arbitrary IRF, the discrete phasor of an ungated PSED does not verify any weak rule for the discrete phasors of convolution products.

In the case of *gated* PSEDs, the situation is obscured by the fact that the analytical expression for the phasor is relatively complex, even in the simple case of a *square gate* and a Dirac IRF (Eq. (103)). In some cases, however, the corresponding SEPL, $\mathcal{L}_{N[W]}$, is ‘close to’ a circular arc ($\mathcal{L}_N$) as can be seen in Fig. 11, which suggests once again that, in favorable situations, it might be possible to bring the SEPL ‘close to’ a familiar curve ($\mathcal{L}_N$ or $\mathcal{L}_\infty$) by inverse rotation and dilation, in order to fall back ‘approximately’ to a familiar situation, as illustrated in Fig. 12③.

In some other cases, however, some examples of which can be seen in Fig. 11, even such an ‘approximate mapping’ is impossible, because the shape of the *SEPL* departs too much from an arc of circle. In these cases, it will be up to the practitioner to decide whether to try and remap the SEPL ‘partially’ to one of the reference SEPLs identified so far ($\mathcal{L}_N$ or $\mathcal{L}_\infty$), or instead use a rotated/dilated $\mathcal{L}_{N[W]}$ as reference SEPL, or even a rotated/dilated version of the actual SEPL bringing a specific phasor to a particular point in the complex plane (for instance, the phasor of the PSED with zero lifetime to 1), as illustrated on Fig. 12④.

The remainder of this section will clarify and examine the validity and usefulness of these general statements, in the continuous and discrete cases.

### *8.2. Calibration of continuous phasor of periodic decays*

According to the continuous phasor convolution rule (Eq. (69)), the continuous phasor of an arbitrary *T*-periodic signal, $S_T(t) = I_T \underset{T}{*} F_{0,T}(t)$, is the product of two phasors: that of the IRF used to acquire the signal, and that of the decay obtained (hypothetically) with a Dirac excitation only:

$$z[S_T] = z[I_T]\,z[F_{0,T}] \tag{165}$$

To keep the discussion general, we do not specify the *T*-periodic instrument response function $I_T(t)$, which could be characterized by an arbitrary gate profile $\Gamma_{s,W}(t)$ and offset $t_0$. This is of course true for a PSED, for which the phasor reads:

$$z\left[I_T \underset{T}{*} \Lambda_{\tau,T}\right] = z[I_T]\,\zeta_f(\tau) \tag{166}$$

Both equations use the same IRF phasor $z[I_T]$, which can therefore be computed using any reference sample such as one characterized by single-exponential decay with lifetime $\tau_C$. Eq. (166) for that reference yields:

$$z[I_T] = \frac{z\left[I_T \underset{T}{*} \Lambda_{\tau_C,T}\right]}{\zeta_f(\tau_C)} \tag{167}$$

In that equation, $z\left[I_T \underset{T}{*} \Lambda_{\tau_C,T}\right]$ is the *calibration phasor*, or phasor of the *reference decay before calibration* obtained from Eq. (63) (a quantity that can be computed from the data) and $\zeta_f(\tau_C)$ is the *calibrated phasor of the reference sample*, given by the simple analytical formula of Eq. (70). Their ratio, Eq. (167), or *calibration factor*, is equal to the phasor of the IRF.

This relation is true for any reference lifetime $\tau_C$, including $\tau_C = 0$, for which $\zeta_f(0) = 1$, yielding:

$$z[I_T] = z\left[I_T \underset{T}{*} \Lambda_{0,T}\right] \tag{168}$$

Eq. (168) might not always be very useful in practice, as it may not be simple to measure the decay of a sample of lifetime 0 (or close enough to 0), *i.e.* the IRF.

In any case, with the help of such a *calibration factor*, it is possible to obtain the *calibrated phasor* $\tilde{z}[S_T]$ of any measured decay $S_T(t)$ as:

$$\tilde{z}[S_T] \triangleq \frac{z[S_T]}{z[I_T]} = z[F_{0,T}] \tag{169}$$

We will use a tilde sign above the phasor symbol ($\tilde{z}$) to indicate a calibrated phasor in the remainder of this article. If the calibration phasor is correct (*i.e.* it is acquired with the same IRF as the samples of interest) and the reference lifetime is accurately known, the calibrated phasor computed by the formula on the right of the $\triangleq$ symbol in Eq. (169) is identical to that of the underlying decay hypothetically excited by a Dirac pulse, $z[F_{0,T}]$.

While the calibration phasor is intended to be computed based on experimental data, it can be computed analytically in the simple cases studied before. As an example, for a *T*-periodic single-exponential excitation function with time constant $\tau_*$, a square gate of width *W* and offset $t_0$, the calibration phasor $z\left[I_T\right]$is formally given by (Eq. (129)):

$$z\left[I_{T,W|t_0}\right]=M_W e^{-i\varphi_W}\zeta_f\left(\tau_*\right)e^{i2\pi f t_0}=\frac{\sin\left(\pi fW\right)/\pi fW}{1-i2\pi f\tau_*}e^{i\left(2\pi f t_0-\pi fW\right)} \tag{170}$$

Comparison of the analytical result of Eq. (170) with the numerical result of Eq. (167) might be used for instance to estimate the width *W* of a square gate, or the IRF offset $t_0$.

In the case of truncated decays, however, the continuous phasor convolution rule does not apply in general (see Section 5). Therefore phasor calibration using a reference PSED as described above (that is, by division by a constant term) will only be useful for PSEDs, and only in the case $f = n/D$, in the absence of gating, and for an instantaneous instrument response function (see Section 5, Eq. (141)). In this case, phasor calibration using a PSED reference will bring phasors of PSEDs back to the universal circle $\mathcal{L}_\infty$, but the calibrated phasors of other arbitrary decays $S_T\left(t\right)=I_T \underset{T}{*} F_{0,T}\left(t\right)$ will not, in general, be identical to the phasor of $F_{0,T}\left(t\right)$, the original *T*-periodic emission due to a Dirac comb excitation.

In all other cases, $f \neq n/D$, *ad hoc* choices will have to be made regarding which phasor calibration to apply (if any) to bring, say, part of the calibrated SEPL 'close to' a region of $\mathcal{L}_\infty$ within which minor differences between some characteristic of the decays (for instance the lifetime and the phase lifetime extracted using $\mathcal{L}_\infty$) will exist. We will examine an example of such a situation in Section 9.2.

### *8.3. Calibration of discrete phasor of periodic decays*

In the case of decay functions sampled at a finite number of time points, the continuous phasor is replaced by a discrete phasor (Eq. (92)) and the corresponding discrete phasor convolution rule states that, in general (see Section 3.3.2) (Eq. (97)):

$$z_N\left[f_T \underset{T}{*} g_T\right]\neq z_N\left[f_T\right]z_N\left[g_T\right] \tag{171}$$

When $f_T$ is the IRF $I_T\left(t\right)$ (gated or ungated, with or without offset) and $g_T$ is the *T*-periodic decay $F_{0,T}\left(t\right)$ resulting from the hypothetical excitation of a sample with a Dirac comb, Eq. (171) says that the discrete case analogue of the continuous phasor calibration, *i.e.* division of the calculated (gated or ungated) phasor by the IRFs phasor (Eq. (169)), will *not* provide any direct useful information (in particular, it will not provide $z_N\left[F_{0,T}\right]$). We shall examine some special cases in the next sub-sections.

However, the situation may sometimes be more favorable, and a weak discrete phasor convolution rule apply. In this case, a similar form of phasor calibration as in the continuous case can be implement, as we shall discuss first.

*8.3.1. Weak discrete phasor convolution rule cases*

In some special cases, the inequality above is replaced by a *weak* discrete phasor convolution rule valid for some families of functions only (Eq. (98)):

$$z_N\left[f_T \underset{T}{*} g_T\right] = \kappa\, z_N\left[f_T\right] z_N\left[g_T\right] \tag{172}$$

where $\kappa$ is a constant.

In cases where Eq. (172) applies to the convolution of PSEDs and the instrument response function $I_T$:

$$z_N\left[I_T \underset{T}{*} \Lambda_{\tau,T}\right] = \kappa z_N\left[I_T\right] z_N\left[\Lambda_{\tau,T}\right] \tag{173}$$

the following modified discrete phasor calibration equation can be used:

$$\tilde{z}_N\left[I_T \underset{T}{*} \Lambda_{\tau,T}\right] \triangleq \frac{z_N\left[I_T \underset{T}{*} \Lambda_{\tau,T}\right]}{\kappa z_N\left[I_T\right]} = z_N\left[\Lambda_{\tau,T}\right] \tag{174}$$

As in the continuous phasor case. the *calibration factor* $\kappa z_N\left[I_T\right]$ can be obtained with the help of a single-exponential decay with known lifetime $\tau_C$ by:

$$\kappa z_N\left[I_T\right] = \frac{z_N\left[I_T \underset{T}{*} \Lambda_{\tau_C,T}\right]}{z_N\left[\Lambda_{\tau_C,T}\right]} \tag{175}$$

where $z_N\left[I_T \underset{T}{*} \Lambda_{\tau_C,T}\right]$ is the phasor of the *reference decay before calibration* obtained from Eq. (88), using the measured gated values $S_T\left(t_p\right)$ ($p$ = 1… $N$). $z_N\left[\Lambda_{\tau_C,T}\right]$ is given by the same Eq. (88) computed using the analytical formula for $S_T\left(t\right) = \Lambda_{\tau_C,T}\left(t\right)$ and takes the simple form $\zeta_{f,N}\left(\tau\right)$ given by Eq. (99). As in the continuous case, the calibration factor could in principle be obtained with a reference sample of lifetime 0 (*i.e.* the IRF) for which $z_N\left[\Lambda_{0,T}\right] = 1$:

$$\kappa z_N\left[I_T\right] = z_N\left[I_T \underset{T}{*} \Lambda_{0,T}\right] \tag{176}$$

although this could be a challenging measurement to perform. It is generally easier to use a sample with known finite lifetime $\tau_C$.

As we have seen in Section 3.3, a relation such as Eq. (173) can be obtained only in a few special cases such as for ungated decays with single-exponential IRF (Eq. (102)) and square-gated decays with Dirac IRF (when the gate width is proportional to gate step, Eq. (107)).

In those cases, since the discrete phasor $\zeta_{f,N}\left(\tau\right)$ given by Eq. (99) is located on a circular arc, $\mathcal{L}_N$, described in Section 3.3.3, Eq. (174) states that the calibrated discrete phasor of such IRF-convolved PSEDs are mapped back to $\mathcal{L}_N$, which therefore takes the role which $\mathcal{L}_\infty$ played for continuous phasors.

The usefulness of such calibration is not limited to PSEDs. The weak discrete phasor convolution rule for PSEDs (Eq. (174)) results in a similar relation for any arbitrary recorded $T$-periodic decay $S_T(t)$. Indeed, we can introduce the $\| \|_N$-normalized recorded decay $\sigma_T(t)$:

$$\begin{cases} \sigma_T(t) = \dfrac{S_T(t)}{\|S_T\|_N} = \displaystyle\int_0^\infty d\tau \mu_0(\tau) \dfrac{I_T \underset{T}{*} \Lambda_{\tau,T}(t)}{\left\| I_T \underset{T}{*} \Lambda_{\tau,T} \right\|_N} \\ \displaystyle\int_0^\infty d\tau \mu_0(\tau) = 1 \end{cases} \tag{177}$$

where $\mu_0(\tau)$ is a weight function in the $\| \|_N$-normalized base $\left\{ I_T \underset{T}{*} \Lambda_{\tau,T}(t) \middle/ \left\| I_T \underset{T}{*} \Lambda_{\tau,T} \right\|_N \right\}_{\tau \geq 0}$. It follows from Eq. (177) that:

$$\begin{aligned} z_N[S_T] = z_N[\sigma_T] &= \int_0^\infty d\tau \mu_0(\tau) \frac{\left\| I_T \underset{T}{*} \Lambda_{\tau,T}(t_p) e^{i2\pi f t_p} \right\|_N}{\left\| I_T \underset{T}{*} \Lambda_{\tau,T} \right\|_N} = \int_0^\infty d\tau \mu_0(\tau) z_N\left[ I_T \underset{T}{*} \Lambda_{\tau,T} \right] \\ &= \kappa z_N[I_T] \int_0^\infty d\tau \mu_0(\tau) \zeta_{f,N}(\tau) \end{aligned} \tag{178}$$

where we have used the definition of $z_N\left[\Lambda_{\tau,T}\right]$ given in Eq. (99).

In other words, the weak discrete phasor convolution rule allows the phasor of an arbitrary recorded decay $S_T(t)$, expressed in the basis of $\| \|_N$-normalized IRF-convolved PSEDs, to be written as the product of $\kappa z_N[I_T]$ with the phasor of the same weighted sum of PSEDs, $\left\{\Lambda_{\tau,T}(t)\right\}_{\tau>0}$. This also means that, after calibration with the calibration factor given by Eq. (175), the calibrated phasor of $S_T(t)$ (and $\sigma_T(t)$) is given by a similar linear relation:

$$\tilde{z}_N[S_T] = \tilde{z}_N[\sigma_T] = \int_0^\infty d\tau \mu_0(\tau) \zeta_{f,N}(\tau) \tag{179}$$

In particular, for a Dirac comb-excited decay equal to a linear combination of exponentials PSEDs:

$$F_{0,T}(t) = \sum_{i=1}^{n} a_i \tau_i \Lambda_{\tau_i,T}(t) \tag{180}$$

The sum in Eq. (180) can be rewritten in terms of the $\| \|_N$-normalized PSEDs, $\left\{\Lambda_{\tau,T,N}(t)\right\}_{\tau>0}$ using the results of Appendix B.3:

$$\begin{cases} f_{0,T}(t) = \dfrac{F_{0,T}(t)}{\|F_{0,T}\|_N} = \displaystyle\sum_{i=1}^{n} \mu_i \Lambda_{\tau_i,T,N}(t) \\ \mu_i = \dfrac{a_i}{\left(1 - e^{-\theta/\tau_i}\right)} \Bigg/ \displaystyle\sum_{j=1}^{n} \dfrac{a_j}{\left(1 - e^{-\theta/\tau_j}\right)} \end{cases} \tag{181}$$

This gives the phasor of the IRF-convolved decay $S_T(t)$ as:

$$z_N\left[S_T\right]=\kappa z_N\left[I_T\right]\int_0^\infty d\tau\mu_0(\tau)\,z_N\left[\Lambda_{\tau,T,N}\right]=\kappa z_N\left[I_T\right]\sum_{i=1}^{n}\mu_i\zeta_{f,N}(\tau_i) \tag{182}$$

After calibration with (*i.e.* division by) $\kappa z_N\left[I_T\right]$, the calibrated phasor is thus:

$$\tilde{z}_N\left[S_T\right]=\sum_{i=1}^{n}\mu_i z_N\left[\Lambda_{\tau_i,T}\right] \tag{183}$$

The *calibrated* discrete phasor of a *T*-periodic decay is therefore equal to the same linear combination of the PSED phasors as the normalized 'pure decay' (obtained with a Dirac comb IRF) is of the normalized PSEDs (Eq. (181)).

In summary, whenever a weak discrete phasor convolution rule applies for PSEDs and the instrument response function $I_T$, phasor calibration with the calibration factor $\kappa z_N\left[I_T\right]$ obtained using a reference sample (Eq. (175)) maps any phasor to an easily interpretable phasor.

We shall now look at the two special cases discussed previously.

### 8.3.2. Special case 1: ungated PSEDs with single-exponential IRF

As seen in Section 3.3.3, the discrete phasor of ungated PSEDs with single-exponential IRF with time constant $\tau_\times$ (Eq. (102)) can be rewritten in the form of Eq. (174) with $\kappa=e^{i\alpha}=e^{i2\pi f\theta}$, since the IRF $I_T$ is equal to the excitation function $\Lambda_{\tau_\times,T}$:

$$z_N\left[\Psi_{\tau,\tau_\times,T}\right]=e^{i\alpha}z_N\left[I_T\right]z_N\left[\Lambda_{\tau,T}\right] \tag{184}$$

The calibration factor $e^{i\alpha}\,z_N\left[I_T\right]$ can thus be obtained using any reference PSED with lifetime $\tau_C$:

$$e^{i\alpha}\,z_N\left[I_T\right]=\frac{z_N\left[\Psi_{\tau_C,\tau_\times,T}\right]}{z_N\left[\Lambda_{\tau_C,T}\right]}=e^{i\alpha}z_N\left[\Lambda_{\tau_\times,T}\right]=e^{i\alpha}\,\frac{1-x_\times}{1-x_\times e^{i\alpha}} \tag{185}$$

### 8.3.3. Special case 2: square-gated PSEDs with Dirac IRF, gate width W is proportional to the gate step θ

As discussed in Section 3.3.4, in the special cases where the gate width *W* is proportional to the gate step *θ*, $W=q\theta$, the discrete phasor of square-gated PSEDs reads (Eq. (107)):

$$W=q\theta\quad\Rightarrow\quad z_{N[W]}\left[\Lambda_{\tau,T}\right]=e^{i\alpha}\,z_N\left[I_{T,W}\right]z_N\left[\Lambda_{\tau,T}\right] \tag{186}$$

where we have used the fact that the gated IRF, $I_{T,W}$, is identical to the gate function $\Pi_{W,nT}$ in that specific case. This relation is of the form of Eq. (174), with $\kappa=e^{i\alpha}=e^{i2\pi f\theta}$, and the calibration factor $e^{i\alpha}\,z_N\left[I_{T,W}\right]$:

$$e^{i\alpha}\,z_N\left[I_{T,W}\right]=\frac{z_{N[W]}\left[\Lambda_{\tau_C,T}\right]}{z_N\left[\Lambda_{\tau_C,T}\right]}=e^{-i(q-3)\alpha/2}\,\frac{\sin q\dfrac{\alpha}{2}}{q\sin\dfrac{\alpha}{2}} \tag{187}$$

where the last expression comes from Eq. (B40) in Appendix B.

#### *8.3.4. Cases where discrete phasor calibration does not map the SEPL to a known SEPL: pseudo-calibration*

Even in the simple case of a Dirac excitation and a square gate, as soon as the gate width is not proportional to the gate step, the discrete phasor of square-gated PSEDs takes a complex form (Eq. (103)), and no weak discrete phasor convolution rule applies.

Another way of describing this property is to note that the loci of discrete phasor of square-gated PSEDs describe a SEPL ( $\mathcal{L}_{N[W]}$) distinct from an arc of circle (see Fig. 5A and its discussion in Section 3.3.6) and cannot be mapped to another, simpler SEPL (for instance $\mathcal{L}_N$ or $\mathcal{L}_\infty$, which are both arcs of circle).

Of course, this does not mean that the discrete phasors cannot be 'pseudo-calibrated' by choosing a particular reference PSED with lifetime $\tau_C$ (or any other decay with known analytical form) and defining a *pseudo-calibrated phasor* $\tilde{z}_N\left[S_T\right]$ by an equation of the form Eq. (174):

$$\tilde{z}_N\left[S_T\right] \triangleq \frac{z_N\left[S_T\right]}{z_N\left[\Lambda_{\tau_C,T}\right] / \zeta_f\left(\tau_C\right)} \neq z_N\left[F_{0,T}\right] \tag{188}$$

However, as indicated by the $\neq$ symbol, the resulting pseudo-calibrated phasor of the measured decay $S_T$ will in general be different from that of the underlying decay $f_{0,T}$ obtained in the presence of a Dirac excitation function. We will note the corresponding pseudo-calibrated SEPL with a tilde as well: $\tilde{\mathcal{L}}$.

In some cases, however, the SEPL does not differ much from an arc of circle, and therefore, such an attempt to map it back to another one (for instance $\mathcal{L}_N$ or $\mathcal{L}_\infty$) might be justified, if it simplifies phasor data interpretation. The Phasor Explorer software discussed in Appendix F allows fitting a calculated SEPL to an arc of circle, thereby enabling to quantify the similarity of the SEPL to a circle (for instance by mean square error of the fit or a graphical comparison of both).

More generally, the SEPL might be reasonably close to an arc of circle for a range of lifetimes of interest $\left[\tau_{\min}, \tau_{\max}\right]$, which means that using a relation such as Eq. (188), where the reference lifetime $\tau_C$ is chosen in the interval $\left[\tau_{\min}, \tau_{\max}\right]$, will bring the discrete phasors of PSEDs with lifetime in this interval close to their 'standard' locations on the chosen target SEPL (for instance $\mathcal{L}_N$ or $\mathcal{L}_\infty$). For lifetimes outside that range, the pseudo-calibrated discrete phasors of PSEDs (and obviously of arbitrary decays) will depart from the discrete canonical discrete phasors $\zeta_{f,N}\left(\tau\right)$.

Because of the variety of experimental situations encountered in terms of excitation function, gate shape, width and separation, it is impossible to provide quantitative or even qualitative rules to determine when such a pseudo-calibration may be useful or which reference lifetime might be appropriate. However, a simple metric reporting on the proximity (or lack thereof) of the resulting tentative mapping consists in comparing the phase lifetime of pseudo-calibrated PSED phasors (calculated using the appropriate formula for the phase lifetime, see Section 7) to their known PSED lifetimes. Small differences will be indicative of a reasonable calibration, while significant departure will indicate a poor approximation. In general, however, the best strategy is described in the next section.

### 8.3.5. Pseudo-calibration of the discrete phasor in the general case

Whenever no such relation as Eq. (173) exists, the loci of PSED phasors (SEPL) is a curve which, in general will be complex (*i.e.* not an arc of circle) and dilated/rotated about the origin, in the sense that the phasor of the PSED with lifetime 0 will be different from 1 and its norm generally will generally be different from 1 too. However, except in the case of truncated decays discussed in Section 5, for which the phasor of a PSED with infinite lifetime does not coincide with the origin when the phasor harmonic is not well-matched with the record duration *D*, in all other cases, the SEPL is anchored at the origin and it is possible to rotate and dilate it such that the phasor of a PSED with zero lifetime is mapped to $z = 1$ (as illustrated in Fig. 12④), using the following definition of the *pseudo-calibrated phasor*:

$$\tilde{z}_N\left[I_T \underset{T}{*} \Lambda_{\tau,T}\right] \triangleq \frac{z_N\left[I_T \underset{T}{*} \Lambda_{\tau,T}\right]}{z_N\left[I_T \underset{T}{*} \Lambda_{0,T}\right]} \tag{189}$$

The pseudo-calibration factor used in this equation, $z_N\left[I_T \underset{T}{*} \Lambda_{0,T}\right]$, corresponds to the phasor of a PSED with 0 lifetime, *i.e.* the IRF. Although it might not always be simple to acquire such an IRF signal in the same conditions as other samples of interest, there is a strong motivation in attempting to do so: the uncalibrated phasor of the IRF can then be used as pseudo-calibration factor (Eq. (189)) and the IRF decay itself, $I_T \underset{T}{*} \Lambda_{0,T}\left(t_p\right) = I_T\left(t_p\right),\ p = 1, \dots, N$ can be used to compute the SEPL for this experiment, at least approximately (see Appendix E for details), providing a convenient reference curve for phasor interpretation.

After this pseudo-calibration operation, the pseudo-calibrated SEPL, $\tilde{\mathcal{L}}$, will be a curve anchored at both points 0 (locus of pseudo-calibrated phasors with infinite lifetime) and 1 (locus of pseudo-calibrated phasors with zero lifetime), and can be labelled with indicator marking the location of PSEDs with specific lifetimes (*e.g.* 0.1 – 0.9 ns, 1 – 9 ns, etc., as shown for instance in Fig. 10).

## 9. Application to experimental data

### *9.1. 4-gate confocal FLIM*

In a pioneering work examining the use of phasor analysis with time-gated fluorescence data, Fereidouni *et al.* used a slightly different formalism than that used here and limit their analysis to the case of adjacent gates ($W = \theta$) [14]. As we have seen (Section 3.3.4), in this case the discrete phasor of a square-gated PSED is identical to that of an ungated PSED (Eq. (109)). The purpose of this section is to connect both works and clarify some differences.

In ref. [14]'s theoretical part, a phasor harmonic $f_n = n/D,\ D = N\theta$ is used, but *D*, the total span of the *N* adjacent gates (width *W* = gate separation $\theta$) is not assumed to cover the whole laser period $T$ (the decay is truncated, as defined in Section 5). Indeed, while they define $T = N\theta$, in their notation, $T$ can be different from the laser period – and in fact is, in the subsequent experimental section. We will replace it with the notation used throughout this article, Eq. (88) & (92), *i.e.* $D = N\theta$ to avoid confusion, and reserve the notation $T$ for the laser period.

Additionally the first gate is assumed to start at $t_1 = 0$ and the gate centers are used as arguments of the complex exponentials in Eq. (92), rather than the gate beginnings, $t_p$:

$$e^{i2\pi t_p} \mapsto e^{i2\pi(t_p + W/2)}, \quad t_p = (p-1)\theta,\ p \in [1, N] \tag{190}$$

Finally, while ref. [14] deals nominally with square-gated PSEDs, which would call for the use of $\Lambda_{\tau,T,W}(t_p)$ (defined by Eq. (47)) in Eq. (92), it is instead replaced by the simpler form, appropriate for ungated periodic signals:

$$\Lambda_{\tau,T,W}(t_p) \underset{W=\theta}{\mapsto} \int_{t_p}^{t_p+W} e^{-t/\tau} dt = \tau\left(1 - e^{-\theta/\tau}\right) e^{-t_p/\tau} \tag{191}$$

While this replacement ignores the $T$-periodicity of the decay, in cases where the time argument $t_p$ is smaller than $T$ - $W$, this is an acceptable form (compare Eq. (17) and Eq. (47)). This clearly requires that *truncated* decays are considered.

As discussed in Section 3.3, using the gate centers rather than their beginning, results in a mere rotation of the SEPL calculated using the beginning of the gates (the choice made in this work) by an angle $\pi f \theta$, where $f$ is the phasor harmonic.

Using the replacement of Eq. (191) amounts to using Eq. (142) with $t_1 = \theta/2$ for the phasor, *i.e.* the discrete phasor of a truncated ungated PSED with a first gate starting at $t_1 = \theta/2$. The result can be rewritten, in the case $D = N\theta$ and $f = n/D$ as:

$$\tilde{z}_N\left[\Lambda_{\tau,D}\right] = \frac{\sinh \theta/2\tau}{\sinh\left(\theta/2\tau - i\,\alpha/2\right)} \tag{192}$$

with $\alpha = 2\pi f \theta$, which is ref. [14]'s Eq. (10).

The solid curves in Fig. 13A, representing the locus of Eq. (192) for different number of gates $N$ are identical to those shown in Fig. 2 of ref. [14], which assumes $D = T$. The $UC_{N[W]}$ calculated using the convention used in this work, that is with an exponential argument in Eq. (92) equal to $t_p$, the location of the beginning of the $p^{\text{th}}$ gate rather than its center, are shown as dashed curves on the same graph.

In the subsequent experimental section of ref. [14], the authors use the discrete phasor to study fluorescent samples characterized by distinct lifetimes. The information provided is: $T$ = 20 ns, $W$ = $\theta$ = 2 ns, $N$ = 4 (imposed by the hardware), and that the first gate starts at $t_1$ = 0.5 ns (= $\theta$/4). Consistent with the definition of the phasor harmonic in terms of the total recording span, $f = 1/D = 1/N\theta$ = 125 MHz is chosen. In this case, because the gates do not cover the full laser period, this frequency does not belong to the series of Fourier harmonic frequencies for a $T$-periodic decay ($f = 50$ MHz, 100 MHz, etc.). Using Eq. (143) and accounting for the additional rotation of $\pi fW$ due to the choice of the gate centers as argument of the complex exponential term (see Section 3.3), we expect:

$$z = e^{i2\pi f(t_1+W/2)} z_N\left[\Lambda_{\tau,T}\right] = e^{i\frac{3\pi}{8}} z_N\left[\Lambda_{\tau,T}\right] \tag{193}$$

where we have used $t_1 = \theta/4, W = \theta, f = 1/4\theta$ , as defined above. Eq. (193) is identical to Eq. (192), but has the advantage to be easier to interpret, since the locus of $z_N\left[\Lambda_{\tau,T}\right]$ is an arc of circle, namely $\mathcal{L}_N$, and the prefactor amounts to rotation angle of 3π/8. The corresponding curve is indicated in red in Fig. 13B. This is a rotated version (by an angle π/8) of the curve obtained for adjacent gates covering the whole period (Fig. 13B, blue curve), which is also shown in Fig. 13A as '$N = 4$,C' (solid light blue). Since the authors reported their calculated experimental phasors as falling on that latter curve, we have to conclude that, in practice, they used the replacement $t_p \mapsto t_p' = (p-1)\theta$ in their phasor calculations. Assuming that the IRF location was $t_0 = 0$ in the original time frame, this implies that the IRF was now located at $t_0' = -\theta/4$. Computation of $\mathcal{L}_{N[W]}$ with this parameter leads to a curve identical to that shown in the cited article. Note that this same curve is actually obtained for a large range of possible $t_0'$ values, which makes the exact location of the IRF relatively irrelevant.

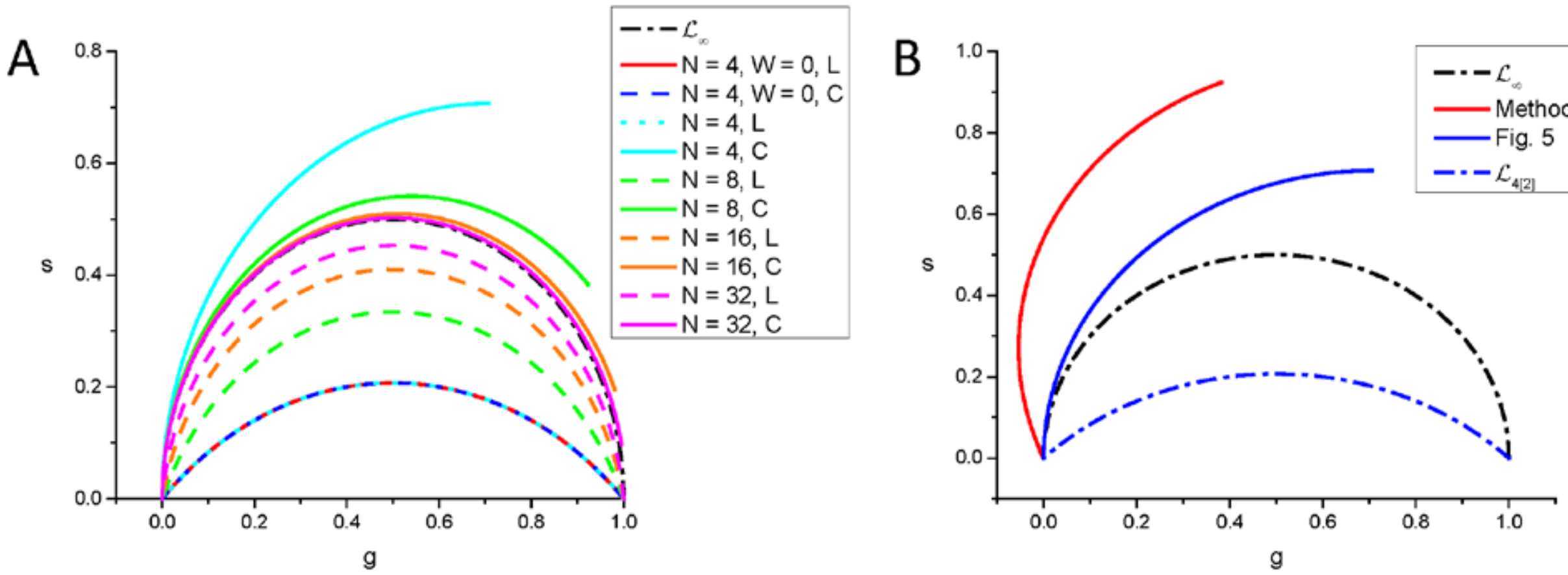

**Fig. 13**. A: Comparison of $\mathcal{L}_{N[W]}$ computed for $T = N\theta$, $W = \theta$, $f = 1/T$ using the argument of the complex exponential terms equal to the beginning of each gate (L, dashed curves) or the center of each gate (C, plain curves) Both differ only by a rotation of $\pi f\theta = \pi/N$. Note that when $W = 0$ (instead of $\theta$), there is no difference between the two conventions (compare red and dashed blue curves). The plain curves are identical to those presented in Fig. 2 of ref. [14]. B: $\mathcal{L}_{N[W]}$ computed as described in the experimental methods of ref. [14], and as presented in Fig. 5 of that article (this latter curve is identical to the curve $N = 4$, C of panel A). Also shown is $\mathcal{L}_{N[W]}$ corresponding to $N = 4$ and $W = 2$ ns without offset (dot-dashed blue curve).

### *9.2. Time-gated ICCD*

In an article using a different type of detector (time-gated ICCD) [15], Chen *et al.* reported time-gated phasor analysis of NIR dye fluorescence with short lifetime ($\tau \leq 1$ ns) using overlapping gates (nominally $\theta$ = 40 ps, $W$ = 300 ps) and offset and truncated decays ($t_0$ ~ 1.5 ns, $T$ = 12.5 ns, $D$ ~ 6 ns). Some of the calculations were also performed with non-overlapping gates ($\theta$ = 320 ps).

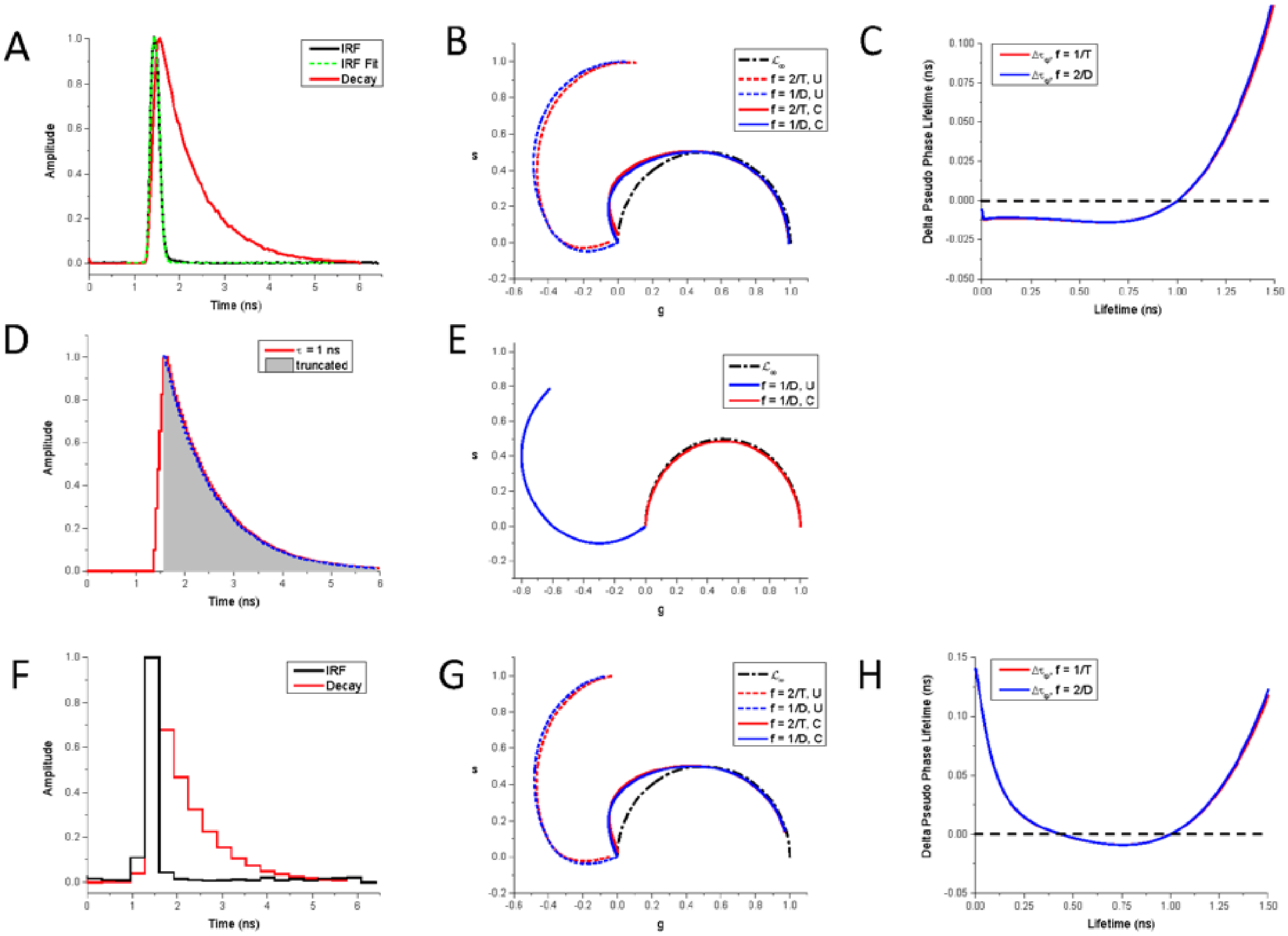

**Fig. 14**: Time-gated ICCD imaging of NIR dyes (data from ref. [15]) studied using overlapping gates (A-C: $N$ = 150, $\theta$ = 40 ps) or non-overapping gates (F-H: $N$ = 19. $\theta$ = 320 ps). A, F: Example of IRF and fluorescence decay. The laser period is $T$ = 12.5 ns, but data was recorded only for the first 6 ns. The IRF is not square, and instead is well approximated by a logistic-edge gate model with FWHM = 230 ps. B, G: $\mathcal{L}_{N[W]}$ computed for $W$ = 230 ps and $t_0$ = 1.57 ns are shown for two choices of phasor frequency: $f = 2/T$ = 160 MHz (red) and $f = 1/D$ = 167 MHz (blue) without (dashed curves) or with calibration (solid curves) using $\tau$ = 1 ns as reference. C, H: Difference between pseudo-phase lifetime and true lifetime for calibrated phasors as in B (red: $f = 2/T$, blue: $f = 1/D$) for $\tau \leq 1.5$ ns. D: Overlay of the 1 ns lifetime decay shown in A (red) and its truncated version (gray area) at $t_0$ = 1.57 ns (duration D = 4.4 ns). E: $\mathcal{L}_{N[W]}$ for decays truncated as in D, harmonic frequency $f = 1/D$ = 227 MHz (blue: uncalibrated, red: calibrated using $\mathcal{L}_N$). The phase lifetime of the calibrated phasor is obviously identical to the original lifetime, since the calibrated SEPL is identical to $\mathcal{L}_N$.

The phasor harmonic used was $f = 2/T$ = 160 MHz.

The first noteworthy feature of this work is that the gates used were not square, due to their brevity and the finite rise and fall time of the gating electronics and microchannel plate (MCP) response: as shown in Fig. 14A, the gated-IRF profile is well fitted by a 0-width logistic edge gate, *i.e.* the gates profile is dominated by the rising ($\sigma_R$ = 21 ps) and falling ($\sigma_F$ = 37 ps) times of the gating electronics plus MCP response (FWHM ~ 230 ps). Because the gates are so short, this shape discrepancy is expected to have negligible influence. The second noteworthy characteristic is that the decays are truncated (only the first ~6 ns of the complete laser period are available). Finally, the IRF is offset, its peak being located at $t_0$ ~ 1.57 ns. Combined with the previous property, this means that only ~ 4.5 ns of the actual decay part is available. The work used standard phasor calibration with a sample with known reference lifetime, $\tau_C$ = 1 ns.

Fig. 14B shows the corresponding PSED phasor loci (dashed curves: uncalibrated $\mathcal{L}_{N[W]}$; solid curves: calibrated or pseudo SEPL, $\tilde{\mathcal{L}}_{N[W]}$) for $f_1 = 2/T$ = 160 MHz (red), the phasor frequency used in ref. [15], and for $f_2 = 1/D$ = 166.7 MHz, the suggested phasor frequency for a truncated decay. Due to the minor difference between these two frequencies, the results are unsurprisingly very similar. The most noteworthy feature of Fig. 14B is the increasing departure of $\tilde{\mathcal{L}}_{N[W]}$ from $\mathcal{L}_\infty$ for $\tau > \tau_C$. On the other hand, the difference with $\mathcal{L}_\infty$ for $\tau \le \tau_C$ is minimal.

Fig. 14C represents the corresponding pseudo-phase lifetime $\tilde{\tau}_\varphi$ as a function of $\tau$. While unsurprisingly this difference increases for $\tau > \tau_C$ (and in fact diverges for large lifetimes), the difference remains below 14 ps for $\tau \le \tau_C$. Since that work was concerned with lifetimes shorter than 1 ns, this demonstrates that using a pseudo phasor was appropriate.

While this analysis (already presented in an abridged form in the Supporting Material of ref. [15]) justifies the use of a standard phasor calibration approach in this situation, a better solution can be found, based on the present work. Namely, lets consider the decay shown in Fig. 14A: truncating it by setting the first gate $t_1 = t_0$, and keeping all the other subsequent gates, we obtain a new recording span $D$ ~ 4.43 ns comprised of $N$ = 110 gates (Fig. 14D). Using $f = 1/D$ as the phasor frequency, we find ourselves in the situation of the special case discussed in Section 5.3, for which we have shown that the locus of the phasors of PSEDs is a rotated arc of circle (in fact it is identical to $\mathcal{L}_{N,t_0}$, the SEPL for discrete, ungated PSEDs with offset $t_0$). Using $\mathcal{L}_N$ as the reference SEPL would produce calculated phase lifetimes closer to the real lifetimes (for PSEDs) than obtained in ref. [15]. However, as shown above, the discrepancy in the original work was minimal. Note also that, because the number of gates, $N$ = 110, is rather large, $\mathcal{L}_N$ is actually very similar to $\mathcal{L}_\infty$, making it a valid reference SEPL as well for further analysis (Fig. 14E).

Chen *et al.* conclude by presenting results using a decimated subset of the original gates, using only 1 every 8 gates ($G$ = 19, $\theta$ = 320 ps). The corresponding IRF and decays, illustrated in Fig. 14F are poorly resolved, yet the corresponding PSED phasors calculated with these new parameters (Fig. 14G) are quite similar to those obtained previously (compare with Fig. 14B). The most notable difference is the fact that for $\tau \to 0$, the calibrated $\tilde{\mathcal{L}}_{N[W]}$ stops short of the locus of $\tau$ = 0 on the $\mathcal{L}_\infty$, namely the point $z$ = 1. This results in a pseudo-phase lifetime $\tilde{\tau}_\varphi$ presenting a discrepancy as high as 150 ps for $\tau$ = 0, but below 30 ps for $\tau \in [0.166, 1.217]$ ns (Fig. 14H), a range corresponding to that studied in ref. [15].

In summary, the choice of $f$ = 160 MHz and the use of standard phasor concepts and phasor calibration was justified in this work, but acquiring complete rather than truncated decays would have simplified everything.

### *9.3. Wide-field time-gated single-photon avalanche diode array*

Ulku *et al*. have reported several examples of time-gated phasor analysis of visible dye fluorescence using a SPAD array imager (SwissSPAD 2) [16,29]characterized by long gates ($W$ > 10 ns). Systematic studies of the influence of various parameters ($W$, $N$) on the calculated phase lifetime was presented in ref. [16]. The acquisition settings corresponded to discrete, non-truncated ( $D = T = N\theta = f^{-1}$ ) and offset ($t_0$ = 15 ns) decays.

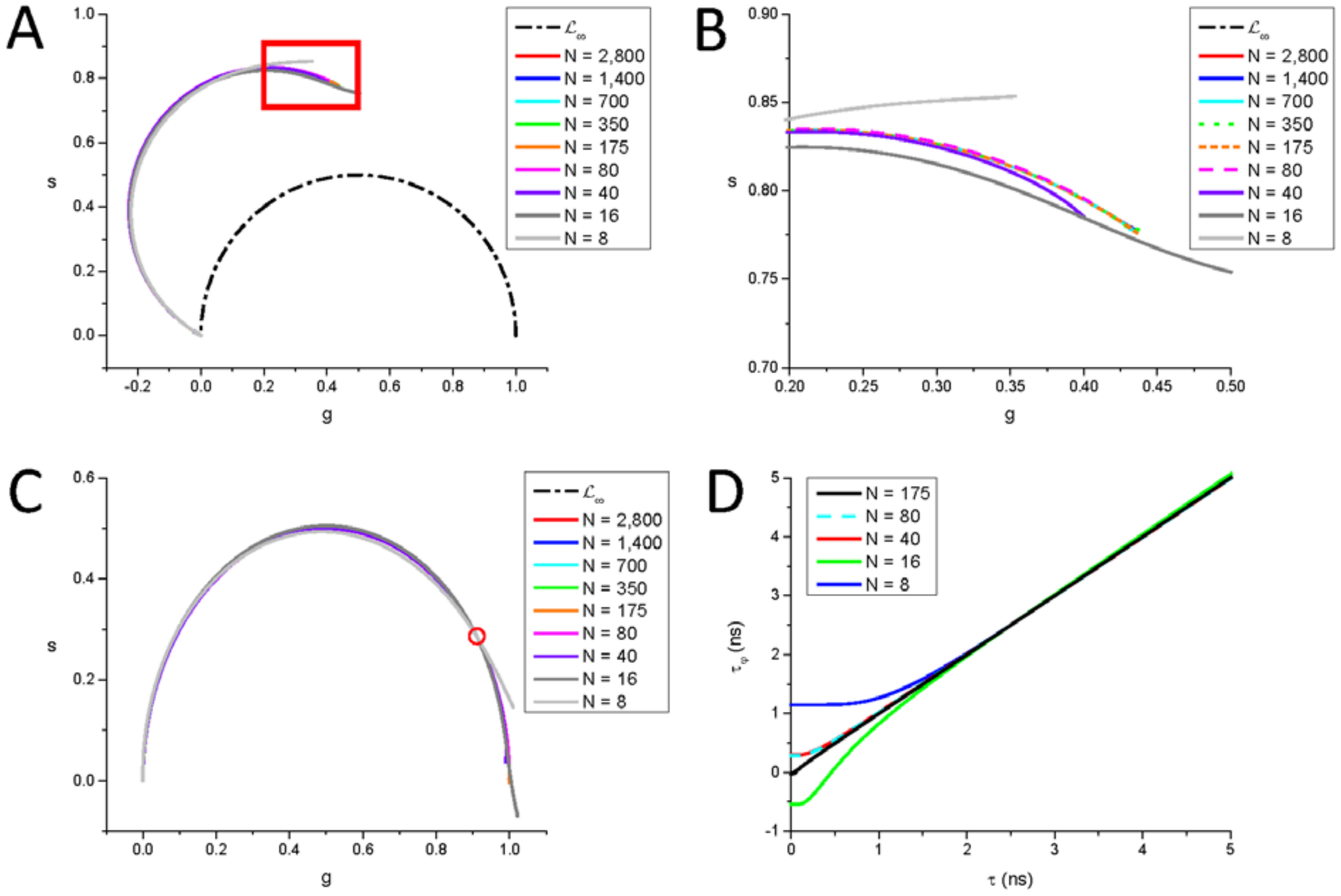

**Fig. 15**: Discussion of Fig. 7 in ref. [16]. A: uncalibrated $\mathcal{L}_{N[W]}$ for varying number of gates $N$. Laser period, T = 50 ns; phasor harmonic frequency, $f = 1/T$ = 20 MHz; gate width, $W$ = 13.1 ns; gate step $\theta = T/N$; IRF offset: $t_0$ = 0. B: Zoom of A in the boxed region. There are clear differences between the different curves for short lifetimes. C: Calibrated $\mathcal{L}_{N[W]}$ of panel A, using $\tau_C$ = 2.5 ns. The curves are close to $\mathcal{L}_\infty$ , except for those computed with few gates ($N$ < 40), which differ from it for short lifetimes. The red circle indicates the location of $\tau = \tau_C$ on $\mathcal{L}_\infty$ . D: Pseudo phase lifetimes $\tau_\varphi$ computed from the curves shown in panel C as a function of the known lifetime $\tau$. The only major discrepancies occur for $N$ < 40 and $\tau$ < 2 ns.

### 9.3.1. *Effect of gate number*

Fig. 15A shows $\mathcal{L}_{N[W]}$ computed with these parameters for different numbers $N$ of gates (as used in Fig. 7 of ref. [16]). Due to the decay offset, the curves are rotated away from the $\mathcal{L}_\infty$, and as shown in the zoomed region represented in Fig. 15B, some discrepancies with the $\mathcal{L}_\infty$ are noticeable in the region of short lifetimes for $N < 40$.

After calibration using $\tau_C = 2.5$ ns, as used in ref. [16], these curves are rotated back towards $\mathcal{L}_\infty$, but these discrepancies remain for short lifetimes. This can be quantified by plotting the pseudo-phase lifetime for these curves (Fig. 15D), which does indeed show that pseudo-phase lifetimes extracted for phasors on the calibrated $\mathcal{L}_{N[W]}$ ($N < 40$) are significantly different from the real lifetimes. Fortunately, this effect is only significant for lifetimes $\tau < 2$ ns, which are below the range of lifetimes considered in ref. [16]. Studies of shorter lifetimes would require using either a different calibration lifetime, or a sufficient number of gates to avoid large discrepancies.

### 9.3.2. *Effect of gate parameters non-uniformity*

In both refs. [36] and [16], a local phasor calibration was used, the reason invoked being the non-uniformity of the detectors response. This non-uniformity is detailed in ref. [29], where the gates rising edge and falling edge positions, as well as the gate width, are shown to be multimodal, and depend on the location of each SPAD within the array.

To analyse the effect of this non-uniformity of gate parameters on the calculated phasors, we can use the results derived here. While these works used a finite number of gates, their large number (from ~100 in ref. [36] to 2,800 in ref. [16]) implies that the calculated phasors are close to the continuous phasor discussed in this article. Ignoring for a moment the effect of decay offset, we can use Eq. (75) for the continuous phasor of square-gated PSEDs to understand the effect of a non-uniform gate width on the calculated phasors.

The changes in modulus prefactor $M_W$ and in phase $\varphi_W$ upon a small change $\delta W$ in width are given by:

$$\begin{cases} \delta M_W = M_W \left(\pi fW \cot\left(\pi fW\right) - 1\right) \dfrac{\delta W}{W} \\ \delta \varphi_W = \varphi_W \dfrac{\delta W}{W} \end{cases} \tag{194}$$

The effect on these variations on the calculated phase and modulus lifetimes can be computed from Eq. (156):

$$\begin{cases} \dfrac{\delta \tau_\varphi}{\tau_\varphi} = \dfrac{2\varphi_W}{\sin 2\left(\varphi + \varphi_W\right)} \dfrac{\delta W}{W} \\ \dfrac{\delta \tau_m}{\tau_m} = \dfrac{\left(M_W/m\right)^2}{\left(M_W/m\right)^2 - 1} \left(\pi fW \cot\left(\pi fW\right) - 1\right) \dfrac{\delta W}{W} \end{cases} \tag{195}$$

These expressions show that phase and modulus lifetime relative changes are proportional to the relative gate width variation $\delta W/W$, with a prefactor depending on the phase $\varphi$ or modulus $m$ of the phasor under consideration. For instance, for $W = 13.1$ ns and $f = 20$ MHz (data of Fig. 7 in ref. [16]), we obtain: $\varphi_W$ = 0.823 and $M_W$ = 0.891. Lets look at the influence of a $\delta W$ = 250 ps width variation on a width $W$ = 13.1 ns ($\delta W/W = 0.019$) on both $\tau_\varphi$ and $\tau_m$ for a lifetime $\tau$ = 2.5 ns. Using Eq. (156), we obtain $\varphi + \varphi_W = \tan^{-1}(2\pi f \tau) =$ 0.304 rad and $M_W/m = \sqrt{1+(2\pi f \tau)^2} = 1.05$, from which we get:

$$\begin{aligned} \frac{\delta\tau_\varphi}{\tau_\varphi} &= 2.88\frac{\delta W}{W} \sim 0.055 \\ \frac{\delta\tau_m}{\tau_m} &= -2.55\frac{\delta W}{W} \sim -0.049 \end{aligned} \tag{196}$$

*i.e.* approximately a 5% variation in phase and modulus lifetime (or ~125 ps). This corresponds to the upper end of phase lifetime standard deviations observed in Fig. 7b of ref. [16]. In other words, without accounting for this systematic dispersion of phase lifetimes due to gate width variation by using local phasor calibration, smaller effects such as that of shot noise studied in ref. [16] would have remained undetectable.

A similar analysis shows that relative variation of phase and modulus lifetimes of a similar magnitude are induced by small gate offsets as described in ref. [29]. Local phasor calibration solves this problem as well.

## 10. Conclusion

In this article, we have examined the effect of several experimental parameters on the calculated phasor of periodic single-exponential decays (PSEDs), notably in cases where the traditional notion of the "universal circle" loses some of its relevance. In particular, extending the work of ref. [14], we have provided analytical formulas for a number of practical cases encountered experimentally, not only in the analysis of data acquired with novel time-resolved approaches, but also in the more traditional case of TCSPC data. Indeed, when such data is decimated or binned down to a small number of bins, one is formally in the case of discrete ungated decays and square-gated decays discussed in the text. Likewise, truncated or offset decays are encountered with all types of instrumentation and cause their own specific issues.

This study has shown that, in some cases, the resulting loci of the phasor of these PSEDs (the *single-exponential phasors loci* curve or *SEPL,* as referred to in this article) is a simple analytical curve, namely an arc of circle, which can be mapped back to, either the standard universal semicircle (noted $\mathcal{L}_\infty$) or one of the discrete circular arcs, $\mathcal{L}_N$, by a simple *phasor calibration* using a sample with known single lifetime. These cases are encountered when the gate width $W$ is proportional to the gate step $\theta$, which suggests that such a relation between width and step should be strived for whenever possible. In particular, as noted above, this situation applies to binned TCSPC data, which consists of contiguous bins (*i.e.* $W = \theta$). The results presented in this work should therefore facilitate interpretation of such binned data, which have the advantage of requiring much less storage space and accordingly, much less processing time, opening the prospect of extremely fast phasor analysis.

Even when the SEPL is not a simple curve and therefore no exact mapping to one of the simple SEPLs described in this work is possible, we have shown that calibration can be convenient to use nonetheless, provided precautions are taken to interpret data in specific regions of the phasor plot.

Such calibration, or *pseudo-calibration* in the general case, allows using conventional knowledge of the universal circle and the 'canonical' phasor plot for phasor interpretation. Finally, in the general case, and when the IRF used to acquire the data can be measured, we have shown how a pseudo-calibrated SEPL can be used as a convenient reference to interpret discrete phasors.

## Supplementary Material

Supplementary material to this article consists of 6 appendixes (A-F) referred to in the text and 3 supplementary figures (S1-S3).

## Acknowledgments

I want to thank my collaborators for their patience during times when many aspects of phasor analysis applied to time-gated data remained unchartered. I am grateful for the constant support of Shimon Weiss and the Weiss lab at UCLA for listening to the various stages of development of the results presented here.

This work was funded in part by Human Frontier Science Program Grant RGP0061/2015, US National Institute of Health Grants R01 GM095904 & R01 CA250636, University of California CRCC Grant CRR-18-523872, US Department of Energy Grant DE-SC0020338, the European Research Council under European Union's Horizon 2020 research and innovation program under grant agreement No. 669941 and by the Partner University Fund, a program of the French American Culture Exchange.

## Data Availability

The data that support the finding of this study is available in a free online repository on Figshare at https://doi.org/10.6084/m9.figshare.11653182, reference number [37] and the software on Github at https://doi.org/10.5281/zenodo.3884101, reference number [38].

## References

[1] K. Suhling, P. M. W. French, and D. Phillips, Photochemical & Photobiological Sciences **4,** 13 (2005).

[2] E. B. van Munster and T. W. J. Gadella, Advances in Biochemical Engineering/Biotechnology **95,** 143 (2005).

[3] H. Wallrabe and A. Periasamy, Current Opinion in Biotechnology **16,** 19 (2005).

[4] J. R. Lakowicz, *Principles of Fluorescence Spectroscopy*, 3 ed. (Springer US, New York, 2006).

[5] W. Becker, A. Bergmann, M. A. Hink, K. König, K. Benndorf, and C. Biskup, Microscopy Research and Technique **63,** 58 (2004).

[6] H. E. Grecco, P. Roda-Navarro, and P. J. Verveer, Optics Express **17,** 6493 (2009).

[7] M. A. Digman, V. R. Caiolfa, M. Zamai, and E. Gratton, Biophysical Journal **94,** L14 (2008).

[8] A. H. A. Clayton, Q. S. Hanley, and P. J. Verveer, Journal of Microscopy **213,** 1 (2004).

[9] G. I. Redford and R. M. Clegg, Journal of Fluorescence **15,** 805 (2005).

[10] D. M. Jameson, E. Gratton, and R. D. Hall, Applied Spectroscopy Reviews **20,** 55 (1984).

[11] D. V. O'Connor and D. Philips, *Time-Correlated Single Photon Counting* (Academic Press, 1984).

[12] W. Becker, *Advanced time-correlated single photon couting techniques* (Springer, Berlin, 2005).

[13] R. A. Colyer, C. Lee, and E. Gratton, Microscopy Research and Technique **71,** 201 (2008).

[14] F. Fereidouni, A. Esposito, G. A. Blab, and H. C. Gerritsen, Journal of Microscopy **244,** 248 (2011).

[15] S.-J. Chen, N. Sinsuebphon, A. Rudskouskaya, M. Barroso, X. Intes, and X. Michalet, Journal of Biophotonics **12,** e201800185 (2019).

[16] A. Ulku, A. Ardelean, M. Antolovic, S. Weiss, E. Charbon, C. Bruschini, and X. Michalet, Methods and Applications in Fluorescence **8,** 024002 (2020).

[17] J. T. Smith, R. Yao, N. Sinsuebphon, A. Rudkouskaya, N. Un, J. Mazurkiewicz, M. Barroso, P. Yan, and X. Intes, Proceedings of the National Academy of Sciences USA **116,** 201912707 (2019).

[18] J. P. Eichorst, K. W. Teng, and R. M. Clegg, in *Fluorescence Spectroscopy and Microscopy: Methods and Protocols*, edited by Y. Engelborghs and V. A. J.W.G. (Springer Science+Business Media, 2014), p. 97.

[19] S. Ranjit, L. Malacrida, D. M. Jameson, and E. Grafton, Nature Protocols **13,** 1979 (2018).

[20] U. P. Wild, A. R. Holzwarth, and H. P. Good, Review of Scientific Instruments **48,** 1621 (1977).

[21] W. R. Ware, L. J. Doemeny, and T. L. Nemzek, Journal of Physical Chemistry **77,** 2038 (1973).

[22] M. N. Berberan-Santos, Chemical Physics **449,** 23 (2015).

[23] V. Venugopal, J. Chen, and X. Intes, Biomedical Optics Express **1,** 143 (2010).

[24] S. Cova, M. Ghioni, A. Lotito, I. Rech, and F. Zappa, J. Mod. Opt. **51,** 1267 (2004).

[25] V. Golovin and V. Saveliev, Nuclear Instruments & Methods in Physics Research Section A **518,** 560 (2004).

26 M. Suyama, Y. Kawai, S. Kimura, N. Asakura, K. Hirano, Y. Hasegawa, T. Saito, T. Morita, M. Muramatsu, and K. Yamamoto, IEEE Transactions on Nuclear Science **44,** 985 (1997).
27 R. A. Colyer, G. Scalia, F. Villa, F. A. Guerrieri, S. Tisa, F. Zappa, S. Cova, S. Weiss, and X. Michalet, Proceedings of SPIE **7905,** 790503 (2011).
28 S. Burri, Y. Maruyama, X. Michalet, F. Regazzoni, C. Bruschini, and E. Charbon, Optics Express **22,** 17573 (2014).
29 A. C. Ulku, C. Bruschini, I. M. Antolovic, Y. Kuo, R. Ankri, S. Weiss, X. Michalet, and E. Charbon, Journal of Selected Topics in Quantum Electronics **25,** 6801212 (2019).
30 K. A. O'Donnell, Journal of the Optical Society of America A **3,** 113 (1986).
31 A. Vallmitjana, A. Dvornikov, B. Torrado, D. M. Jameson, S. Ranjit, and E. Gratton, Methods and Applications in Fluorescence **8** (2020).
32 S.-J. Chen, N. Sinsuebphon, and X. Intes, Photonics **2,** 1027 (2015).
33 A. D. Elder, C. F. Kaminski, and J. H. Frank, Optics Express **17,** 23181 (2009).
34 E. B. van Munster and T. W. J. Gadella, Journal of Microscopy **213,** 29 (2004).
35 G. Weber, Journal of Physical Chemistry **85,** 949 (1981).
36 R. Ankri, A. Basu, A. C. Ulku, C. Bruschini, E. Charbon, S. Weiss, and X. Michalet, Acs Photonics **7,** 68 (2020).
37 X. Michalet (2020) "Data for Continuous and discrete phasor analysis of binned or time-gated periodic decays," Figshare, Raw data, analysis scripts & results. https://doi.org/10.6084/m9.figshare.11653182
38 X. Michalet (2020) "Phasor Explorer 0.2 source code," GitHub, Source Code. https://doi.org/10.5281/zenodo.3884101

## Tables

**Table I**: Different loci of phasors of PSEDs (SEPL) discussed in this article. Index notations: $\infty$ indicates a continuous phasor, $N$ a discrete phasor, $[W]$ a square-gated phasor.

| **SEPL** | **Discussed in Section** | **Equation Number** | **Type of curve** |
|---|---|---|---|
| $\mathcal{L}_{\infty}$ | 3.2.2 | (70) | Universal semicircle |
| $\mathcal{L}_{[W]}$ | 3.2.3 | (75) | Rotated, dilated semicircle |
| $\mathcal{L}_{N}$ | 3.3.3 | (99) | Circular arc |
| $\mathcal{L}_{N[W]}$ | 3.3.4 | (103) | Complex curve unless $W = q\theta$ |

**Table II**: Notations

| Symbol | Description | Defined in Eqs |
|---|---|---|
| $\lfloor x \rfloor$ | Floor function | (4) |
| $\lceil x \rceil$ | Ceiling function | (105) |
| $x[T]$ | Modulo operation | (18) |
| $H(t)$ | Heaviside function | (8) |
| $f_T \underset{T}{*} g_T(t)$ | Cyclic convolution product of two *T*-periodic functions | (9) |
| $I_T(t)$ | *T*-periodic instrument response function (IRF) | (15) |
| $\Lambda_\tau(t)$ | Normalized exponential function | (16) |
| $\Lambda_{\tau,T}(t)$ | Normalized *T*-periodic exponential function (PSED) | (17) |
| $\Lambda_{\tau,T\vert t_0}(t)$ | Normalized PSED with offset (Dirac excitation) | (120) |
| $\Psi_{\tau,\tau_\times,T}(t)$ | Convolution of 2 *T*-PSEDs | (20), (27) |
| $\overline{\Gamma}_{W,nT}(t)$ | Mirrored *nT*-periodic gate function of width *W* | (35) |
| $\overline{\Pi}_{W,nT}(t)$ | Mirrored *nT*-periodic square-gate function of width *W* | (37) |
| $I_{T,W}(t)$ | Square-gated (width *W*) *T*-periodic IRF | (43) |
| $\Lambda_{\tau,T,W}(t)$ | Square-gated (*W*) *T*-PSED (Dirac excitation) | (47) |
| $\Lambda_{\tau,T,W\vert t_0}(t)$ | Square-gated (*W*) *T*-PSED with offset (Dirac excitation) | (121) |
| $\Psi_{\tau,\tau_\times,T,W}(t)$ | Square-gated (*W*) *T*-PSED (single-exponential excitation) | (48) |
| $z[S_T]$ | Cyclic phasor of *T*-periodic function $S_T$ | (66), (67) |
| $\zeta_f(\tau)$ | Continuous phasor of ungated PSED (Dirac excitation) | (70) |
| $z_{[W]}[\Lambda_{\tau,T}]$ | Continuous phasor of square-gated PSED (Dirac excitation) | (75) |
| $z_N[S_T]$ | Discrete cyclic phasor of *T*-periodic function $S_T$ | (88), (91), (92) |
| $z_{N[W]}[\Lambda_{\tau,T}]$ | Discrete phasor of square-gated PSED (Dirac excitation) | (103) |
| $\Vert S(t)\Vert, \Vert S\Vert$ | Integral over ]- ∞,∞[ of function *S* | (53) |
| $\Vert S_T(t)\Vert_T, \Vert S_T\Vert_T$ | Integral over [0,*T*] of *T*-periodic function $S_T$ | (60) |
| $\Vert S_T(t_p)\Vert_N, \Vert S_T\Vert_N$ | Discrete version of $\Vert S_T(t)\Vert_D$ over [0,*D*], D: record duration | (88), (92) |

Supplementary Material for "**Continuous and discrete phasor analysis of binned or time-gated periodic decays**"

Xavier Michalet

*Department of Chemistry & Biochemistry, 607 Charles E. Young Drive E., Los Angeles, CA 90095, USA*

michalet@chem.ucla.edu

## Appendix A: Continuous phasor of PSEDs with Dirac IRF with offset

The continuous version of the phasor definition (Eq. (63)**Error! Reference source not found.**) assumes a periodic signal $S_T(t)$. The expressions derived here assume a decay offset $t_0 \in \left]0,T\right[$ (a negative offset $t_0 \in \left]-T,0\right]$ is equivalent to the positive offset $t_0' = t_0 + T \in \left]0,T\right[$).

### *A.1. Continuous phasor of ungated PSEDs with Dirac IRF with offset*

Using Eq. (120) for $S_T(t)$, we obtain, after splitting the integral into two parts ($t \in \left[0,t_0\right[$ and $t \in \left[t_0,T\right[$) to account for the different form taken by Eq. (119):

$$\begin{cases} \left\| \Lambda_{\tau,T|t_0}(t) \right\|_T = 1 \\ \left\| \Lambda_{\tau,T|t_0}(t) e^{i2\pi ft} \right\|_T = \int_0^T dt \Lambda_{\tau,T|t_0}(t) e^{i2\pi ft} = \dfrac{e^{i2\pi ft_0}}{1-i2\pi f\tau} \end{cases} \quad \text{(A1)}$$

from which Eq. (126) follows:

$$z\left[\Lambda_{\tau,T|t_0}\right] = \frac{1}{1-i2\pi f\tau} e^{2\pi ift_0} = \zeta_f(\tau) e^{2\pi ift_0} \quad \text{(A2)}$$

### *A.2. Continuous phasor of square-gated PSEDs with Dirac IRF with offset*

Using Eq. (121) for $S_T(t)$, we need to distinguish between the different possible forms taken by Eq. (121), depending on the value of $t' = t - t_0 - \lfloor (t-t_0)/T \rfloor T$:

$$\begin{aligned} &(i) \quad t' \in \left[0, T-\omega\right[ \\ &(ii) \quad t' \in \left[T-\omega, T\right[ \end{aligned} \quad \text{(A3)}$$

Within these two cases, the value of *t* depends on the location of *t* with respect to $t_0$:

$$\begin{aligned} &(a) \quad t \in \left[0,t_0\right[ : \ \lfloor (t-t_0)/T \rfloor = 0 \\ &(b) \quad t \in \left[t_0,T\right[ : \ \lfloor (t-t_0)/T \rfloor = 1 \end{aligned} \quad \text{(A4)}$$

It is convenient to distinguish between two cases:

*A.2.1. Case $t_0 < \omega$:*

$$\begin{aligned} \left\| \Lambda_{\tau,T,W|t_0}(t) \right\|_T &= \int_0^{t_0} dt \left( \frac{1-uy^{-1}}{1-y} e^{-(t-t_0+T)/\tau} + k + 1 \right) + \\ &+ \int_{t_0}^{t_0+T-\omega} dt \left( \frac{1-u}{1-y} e^{-(t-t_0)/\tau} + k \right) + \int_{t_0+T-\omega}^{T} dt \left( \frac{1-uy^{-1}}{1-y} e^{-(t-t_0)/\tau} + k + 1 \right) \\ &= W \end{aligned} \quad \text{(A5)}$$

where we have introduced the notations:

$$u = e^{-\omega/\tau} \qquad y = e^{-T/\tau} \tag{A6}$$

Likewise, the numerator of the phasor expression is given by:

$$\begin{aligned} \left\| \Lambda_{\tau,T,W|t_0}(t) e^{i2\pi ft} \right\|_T &= \int_0^{t_0} dt \left( \frac{1-uy^{-1}}{1-y} e^{-(t-t_0+T)/\tau} + k + 1 \right) e^{i2\pi ft} + \\ &+ \int_{t_0}^{t_0+T-\omega} dt \left( \frac{1-u}{1-y} e^{-(t-t_0)/\tau} + k \right) e^{i2\pi ft} + \int_{t_0+T-\omega}^{T} dt \left( \frac{1-uy^{-1}}{1-y} e^{-(t-t_0)/\tau} + k + 1 \right) e^{i2\pi ft} \end{aligned} \tag{A7}$$

*A.2.2. Case* $t_0 \geq \omega$:

$$\begin{aligned} \left\| \Lambda_{\tau,T,W|t_0}(t) \right\|_T &= \int_0^{t_0-\omega} dt \left( \frac{1-uy}{1-y} e^{-(t-t_0+T)/\tau} + k \right) + \\ &+ \int_{t_0-\omega}^{t_0} dt \left( \frac{1-uy^{-1}}{1-y} e^{-(t-t_0+T)/\tau} + k + 1 \right) + \int_{t_0}^{T} dt \left( \frac{1-u}{1-y} e^{-(t-t_0)/\tau} + k \right) \\ &= W \end{aligned} \tag{A8}$$

and:

$$\begin{aligned} \left\| \Lambda_{\tau,T,W|t_0}(t) e^{i2\pi ft} \right\|_T &= \left\{ \int_0^{t_0-\omega} dt \left( \frac{1-u}{1-y} e^{-(t-t_0+T)/\tau} + k \right) e^{i2\pi ft} + \right. \\ &\left. + \int_{t_0-\omega}^{t_0} dt \left( \frac{1-uy^{-1}}{1-y} e^{-(t-t_0+T)/\tau} + k + 1 \right) e^{i2\pi ft} + \int_{t_0}^{T} dt \left( \frac{1-u}{1-y} e^{-(t-t_0)/\tau} + k \right) e^{i2\pi ft} \right\} \end{aligned} \tag{A9}$$

After some basic calculations, we obtain, *in both cases*:

$$\begin{aligned} z\left[\Lambda_{\tau,T,W|t_0}\right] &= \frac{\sin \pi f \omega}{\pi f W} \frac{e^{-i\pi f \omega}}{1 - i2\pi f \tau} e^{i2\pi f t_0} \\ &= z_{[W]}\left[\Lambda_{\tau,T}\right] e^{i2\pi f t_0} = z_{[W]}\left[\Lambda_{\tau,T|t_0}\right] \end{aligned} \tag{A10}$$

which is the same as Eq. (75) for the continuous phasor of a square-gated PSED without offset, rotated by an angle $2\pi f t_0$.

## Appendix B: Discrete phasor of *T*-periodic decays

In this Appendix, ***w***e will not make the distinction between original decay (hypothetically resulting from a Dirac excitation), IRF and recorded decay, and only consider the recorded decay. In this sense, it can either be viewed as a discussion of decays recorded with a Dirac IRF, or a discussion of recorded decays with arbitrary IRF, but with particular functional form (for instance, T-periodic single-exponential decays). The point of the Appendix is simply to derive useful mathematical relations used in different parts of this work.

We will first assume that the recorded decay $\{S_T(t_p)\}, 1 \le p \le N$ covers the whole period $T$ ( $N\theta = T$ ) and that phasor harmonic frequencies are multiple of the fundamental frequency $f_1 = 2\pi T^{-1}$. We will briefly discuss cases where it may makes sense to use different frequencies when the $\{S_T(t_p)\}_{1\le p\le N}$ do not cover the whole period (Section B.2.3.b).

### *B.1. Discrete phasor of ungated PSEDs*

We will first treat the case without decay offset ($t_0$ = 0) in some detail and only sketch the derivation in the case with decay offset in the interest of length.

#### *B.1.1. No offset*

For an ungated PSED with lifetime $\tau$, $\Lambda_{\tau,T}(t)$ (Eq. (17)), we obtain:

$$\begin{cases} \left\| \Lambda_{\tau,T}(t_p) \right\|_N = \dfrac{\theta}{\tau\left(1-e^{-T/\tau}\right)} \displaystyle\sum_{p=1}^{N} e^{-(p-1)\theta/\tau} = \dfrac{\theta}{\tau\left(1-e^{-\theta/\tau}\right)} \\ \left\| \Lambda_{\tau,T}(t_p) e^{i2\pi f t_p} \right\|_N = \dfrac{\theta}{\tau\left(1-e^{-T/\tau}\right)} \displaystyle\sum_{p=1}^{N} e^{(p-1)\theta(-1/\tau + i2\pi f)} \\ \qquad = \dfrac{\theta}{\tau\left(1-e^{(-1/\tau+i2\pi f)\theta}\right)} \end{cases} \tag{B1}$$

Introducing:

$$\begin{cases} x(\tau) = e^{-\theta/\tau} \\ \alpha = 2\pi f\theta = 2\pi fT/N \end{cases} \tag{B2}$$

we get:

$$\zeta_{f,N}(\tau) \triangleq z_N\left[\Lambda_{\tau,T}\right] = \frac{1-x}{1-xe^{i\alpha}} = \frac{(1-x)\left(1-xe^{-i\alpha}\right)}{1-2x\cos\alpha + x^2} \tag{B3}$$

The functions $x(t)$ and discrete phasor of PSED $\zeta_{f,N}(\tau)$ will be used throughout this work.

Eq. (B3) can also be rewritten:

$$\begin{cases} z_N\left[\Lambda_{\tau,T}\right] = g(\tau) + is(\tau) = M_N(\tau)e^{i\Phi_N(\tau)} \\ M_N(\tau) = \dfrac{1-x}{\sqrt{1-2x\cos\alpha + x^2}} \\ \tan\Phi_N(\tau) = \dfrac{x\sin\alpha}{1-x\cos\alpha} \end{cases} \tag{B4}$$

For sin $\alpha$ = 0, (*i.e.* $\dfrac{n}{N} = \dfrac{q}{2}$, $q \in \mathbb{N}$ where $f = \dfrac{n}{T}$), Eq. (B4) implies that $s(\tau) = 0$ for all $\tau$. Specifically, if cos $\alpha$ = 1 (*i.e.* if $q$ is even), the locus of discrete phasors of PSEDs is the single point $z$ =

1. If cos $\alpha$ = -1 (*i.e.* if $q$ is odd), the locus of discrete phasors of PSEDs is a line connecting 0 ($\tau$ = ∞) and 1 ($\tau$ = 0) with:

$$g(\tau) = \tanh\frac{\theta}{2\tau}. \tag{B5}$$

In all other cases, we can look for a quadratic relation linking $g$ and $s$:

$$A_1 g^2 + 2A_2 gs + A_3 s^2 + 2A_4 g + 2A_5 s + A_6 = 0 \tag{B6}$$

where the coefficients $\{A_i\},\ 1 \le i \le 6$ are constants. We obtain the following equation:

$$\begin{aligned}
&A_1 + 2A_4 + A_6 + \\
&\left(-2(1+\cos\alpha)A_1 + 2\sin\alpha A_2 - 2(1+3\cos\alpha)A_4 + 2\sin\alpha A_5 - 4\cos\alpha A_6\right)x + \\
&\begin{pmatrix} \left(1+4\cos\alpha+\cos^2\alpha\right)A_1 - 2\sin\alpha\left(2+\cos\alpha\right)A_2 + \sin^2\alpha A_3 \\ +2\left(1+3\cos\alpha+2\cos^2\alpha\right)A_4 - 2\sin\alpha\left(1+2\cos\alpha\right)A_5 + 2\left(1+2\cos^2\alpha\right)A_6 \end{pmatrix} x^2 + \\
&\begin{pmatrix} -2\cos\alpha\left(1+\cos\alpha\right)A_1 + 2\sin\alpha\left(1+2\cos\alpha\right)A_2 - 2\sin^2\alpha A_3 \\ -2\left(1+\cos\alpha+2\cos^2\alpha\right)A_4 + 2\sin\alpha\left(1+2\cos\alpha\right)A_5 - 4\cos\alpha A_6 \end{pmatrix} x^3 + \\
&\left(\cos^2\alpha A_1 - 2\sin\alpha\cos\alpha A_2 + \sin^2\alpha A_3 + 2\cos\alpha A_4 - 2\sin\alpha A_5 + A_6\right)x^4 = 0
\end{aligned} \tag{B7}$$

where we have used the notation $x = e^{-\theta/\tau}$ introduced above. For this equation to be verified for all values of $x$ (i.e. for all values of $\tau$), all coefficients of powers of $x$ in Eq. (B7) need to be equal to zero. This yields a system of 5 equations for the 6 unknowns $\{A_i\}_{1\le i\le 6}$ (it is clear from Eq. (B6) that only 5 of these parameters are independent, as a mere rescaling of all parameters by a constant preserve the identity). A solution of this equation can be found using a symbolic programming language such as Mathematica:

$$\begin{cases} A_1 = -2A_4 = \dfrac{2\sin\alpha}{1-\cos\alpha} \\ A_2 = A_6 = 0 \\ A_3 = \dfrac{2(1+\cos\alpha)}{\sin\alpha} \\ A_5 = 1 \end{cases} \tag{B8}$$

Eq. (B6) can then be simplified into:

$$\begin{cases} (g-g_c)^2 + (s-s_c)^2 = r^2 \\ g_c = \dfrac{1}{2} \\ s_c = -\dfrac{1}{2}\tan(\alpha/2) \\ r = \dfrac{1}{2\left|\cos(\alpha/2)\right|} \end{cases} \tag{B9}$$

showing that, unless $\cos\frac{\alpha}{2} = 0$ (*i.e.* $\frac{n}{N} = q + \frac{1}{2},\ q \in \mathbb{N}$), the locus of the discrete phasor of PSEDs is an arc of circle centered on $(g_c, s_c)$, with radius *r* (Fig. 4).

The special case $\cos(\alpha/2) = 0$ is encountered for instance for *n* = 1, *N* = 2, in which case the locus of phasor is a line connecting 0 and 1, as discussed above.

In all other cases, as the number of gates *N* increases, $\alpha = 2\pi f \theta$ decreases and *a* tends to ½, while $s_c$ tends to 0, showing that the locus of the phasors of PSEDs approaches the standard $\mathcal{L}_\infty$. In all cases, $z_N(0) = 1$ and $z_N(\infty) = 0$ as for the standard $\mathcal{L}_\infty$.

The two possible values of $s_*$ such that $g_* = \frac{1}{2}$ are given by:

$$\begin{cases} s_*^{(1)} = \dfrac{1}{2}\dfrac{\cos\frac{\alpha}{2}}{1+\sin\frac{\alpha}{2}} \\ s_*^{(2)} = -\dfrac{1}{2}\dfrac{\cos\frac{\alpha}{2}}{1-\sin\frac{\alpha}{2}} \end{cases} \tag{B10}$$

which tend to $\pm\frac{1}{2}$ when $\alpha \to 0$ (*i.e.* $N \to \infty$). The corresponding solutions for *x* are given by:

$$\begin{cases} x_*^{(1)} = \left(1 + 2\sin\frac{\alpha}{2}\right)^{-1} \\ x_*^{(2)} = \left(1 - 2\sin\frac{\alpha}{2}\right)^{-1} \end{cases} \tag{B11}$$

only one of which corresponds to a lifetime $\tau_*$ given by $x_* = e^{-\theta/\tau_*}$ (the root such that $x_* \le 1$):

$$\tau_* = \frac{\theta}{\ln\left(1 + 2\left|\sin\frac{\alpha}{2}\right|\right)} \tag{B12}$$

$\tau_*$ is an increasing function of *θ* (or equivalently, a decreasing function of *N*), its minimum being obtained in the continuum limit ($\theta \to 0,\ N \to \infty$) where it takes the value $\tau_* = 1/(2\pi f)$ (Fig. S1). Its maximum is obtained for the largest meaningful interval *θ* = *T/N*, *i.e.* the smallest meaningful number of samples, *N* = 2. In this latter case, however, we have seen that the locus of phasors of single exponential decays is either a single point or a straight segment (depending on the harmonic *n*).

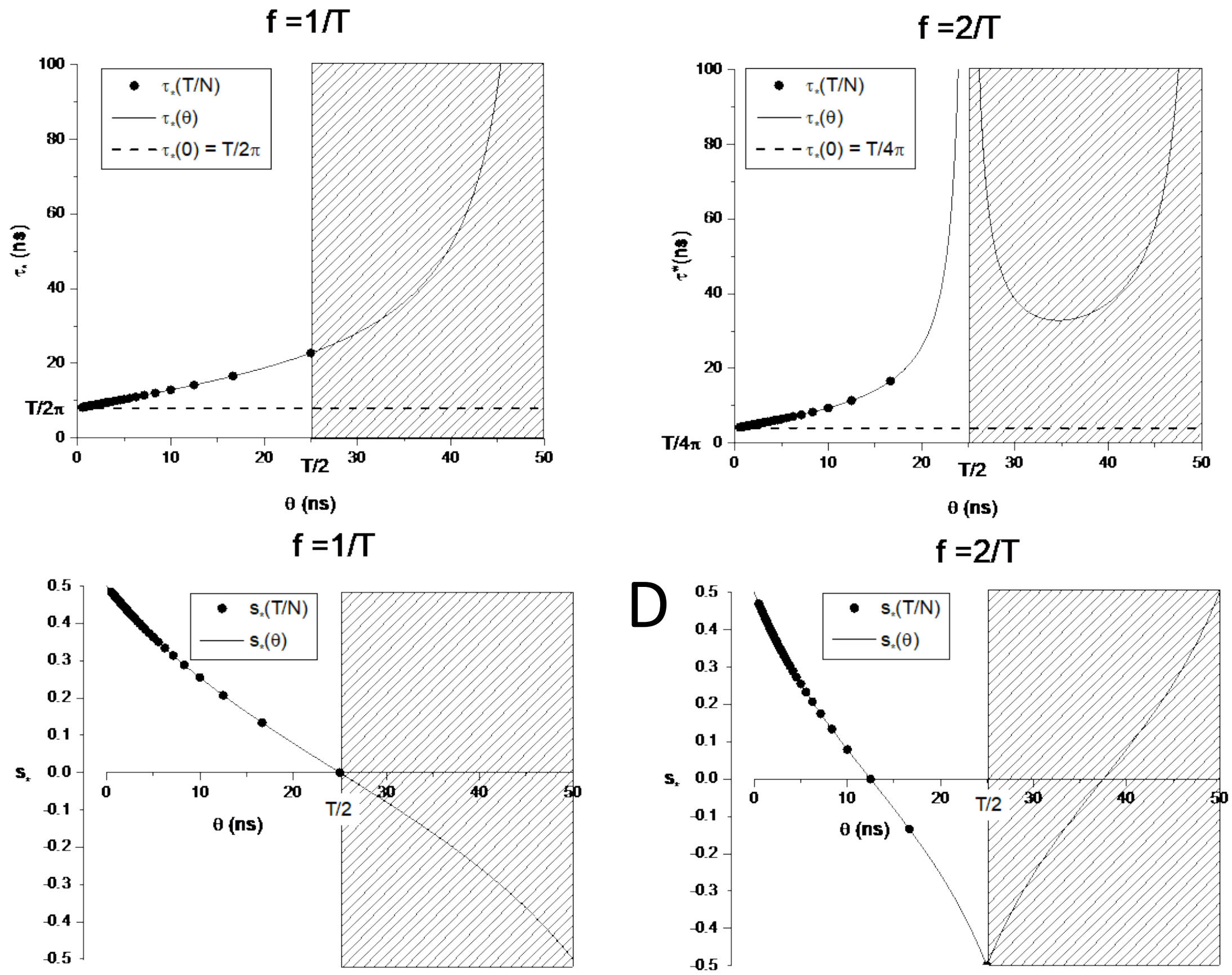

**Fig. S1**: Dependence of the lifetime $\tau_*$ and coordinate $s_*$ of the discrete phasor characterized by $g$ = ½ on the gate step $\theta$. A laser period $T$ = 50 ns is used for illustration. A, C: For a phasor harmonic $f = 1/T$, $s_*$ increases monotonically up to its maximum value obtained at $T/2$ ($N$ = 2 gates covering the laser period). The hatched area represents the region $\theta > T/2$ which does not have practical meaning. Plain curve in A, C corresponds to Eq. (B12), and symbols represent the data for actual gate step values $T/N$. The dashed line in A corresponds to the standard UC value. B, D: For a phasor harmonic $f = 2/T$, $s_*$ still increases monotonically but diverges at its maximum value $T/2$ where the locus of phasors is reduced to a single point $z$ = 1. The hatched areas represent the region $\theta > T/2$ which does not have practical meaning.

Note that for some choices of ($n$, $N$), $\mathcal{L}_N$ can be on the opposite side of the axis compared to $\mathcal{L}_\infty$ (*e.g.* $n = 2, N = 3 \Rightarrow x_*^{(1)}$ is the valid root, and the corresponding $s_*^{(1)}$ is negative, see Fig. 4B).

### *B.1.2. With offset*

We use Eq. (120) for the definition of a PSED with offset $t_0$, $\Lambda_{\tau,T|t_0}(t)$ in Eq. (92). The denominator is:

$$\left\|\Lambda_{\tau,T|t_0}\left(t_p\right)\right\|_N = \frac{\theta}{\tau\left(1-e^{-T/\tau}\right)}\sum_{p=1}^{N} e^{-\left(t_p-t_0-\lfloor\left(t_p-t_0\right)/T\rfloor T\right)/\tau} \tag{B13}$$

where $t_p = (p-1)\theta$. To obtain a simpler expression for the sum, we need to split it into two parts, each with a single value for the integer part $n = \lfloor (t-t_0)/T \rfloor$. We can assume that $t_0 \in [0,T[$ for simplicity[1] and rewrite:

$$\begin{aligned} & t_0 = \theta_0 + r\theta \\ & \theta_0 \in [0,\theta[, \ r = \lfloor t_0/\theta \rfloor \in \mathbb{Z} \end{aligned} \tag{B14}$$

With this definition,

$$t - t_0 = (p-1-r)\theta - \theta_0 \tag{B15}$$

and the integer part takes the values:

$$\begin{aligned} & \theta_0 = 0 \Rightarrow \begin{cases} p \le r \Rightarrow n = -1 \\ p > r \Rightarrow n = 0 \end{cases} \\ & \theta_0 \ne 0 \Rightarrow \begin{cases} p \le r+1 \Rightarrow n = -1 \\ p > r+1 \Rightarrow n = 0 \end{cases} \end{aligned} \tag{B16}$$

It appears advantageous from Eq. (B16) to introduce the following integer:

$$q = \left\lceil \frac{t_0}{\theta} \right\rceil = \begin{cases} r & \text{if} \quad \theta_0 = 0 \\ r+1 & \text{if} \quad \theta_0 > 0 \end{cases} \tag{B17}$$

with which Eq. (B16) can be rewritten:

$$\begin{cases} p \le q \Rightarrow n = -1 \\ p > q \Rightarrow n = 0 \end{cases} \tag{B18}$$

We obtain the following expressions for the sum in Eq. (B13):

$$\sum_{p=1}^{N} e^{-\left(t_p - t_0 - \lfloor (t_p - t_0)/T \rfloor \mathrm{T}\right)/\tau} = \frac{e^{t_0/\tau}}{1-x}\left( y\left(1-x^q\right) + x^q - x^N \right) \tag{B19}$$

where we have used the previous notations:

$$\begin{cases} x(t) = e^{-\theta/\tau} \\ y(t) = e^{-T/\tau} \end{cases} \tag{B20}$$

In the special case where the $N$ gates cover the whole laser period exactly ($T = N\theta$), these expressions simplify into:

$$T = N\theta \Rightarrow \sum_{p=1}^{N} e^{-(t_p - t_0)[T]/\tau} = \frac{1-y}{1-x} x^q e^{t_0/\tau} \tag{B21}$$

Returning to the general case, the numerator in Eq. (92):

---

[1] If $t_0 \in [0,T[$ , $t_0 - \lfloor t_0/T \rfloor T \in [0,T[$ can be used instead by periodicity.

$$\left\| \Lambda_{\tau,T|t_0}\left(t_p\right) e^{i2\pi f t_p} \right\|_N = \frac{\theta}{\tau\left(1-e^{-T/\tau}\right)} \sum_{p=1}^{N} e^{-\left(t_p - t_0\right)[T]/\tau} e^{i2\pi f t_p} . \tag{B22}$$

can be computed in the same manner. The sum in this expression reads:

$$\sum_{p=1}^{N} e^{-\left(t_p - t_0\right)[T]/\tau} e^{i2\pi f t_p} = \frac{e^{t_0/\tau}}{1-xe^{i\alpha}} \left( y\left(1 - x^q e^{iq\alpha}\right) + x^q e^{iq\alpha} - x^N e^{iN\alpha} \right) \tag{B23}$$

where we have used the previous notation:

$$\alpha = 2\pi f \theta . \tag{B24}$$

If we assume that the $N$ gates cover the whole laser period exactly ($T = N\theta$), these expressions simplify into:

$$T = N\theta \Rightarrow \sum_{p=1}^{N} e^{-\left(t_p - t_0\right)[T]/\tau} e^{i2\pi f t_p} = \frac{1-y}{1-xe^{i\alpha}} x^q e^{iq\alpha} e^{t_0/\tau} \tag{B25}$$

The phasor is given by the ratio of Eq. (B23) and Eq. (B19) in the general case (discussed further below), or, if we assume that the $N$ gates cover the whole laser period exactly, by the ratio of Eq. (B25) and Eq. (B21).

In that latter case ($T = N\theta$), this reads:

$$T = N\theta \Rightarrow \quad z_N\left[\Lambda_{\tau,T|t_0}\right] = \frac{1-x}{1-xe^{i\alpha}} e^{i\left\lceil \frac{t_0}{\theta} \right\rceil \alpha} = z_N\left[\Lambda_{\tau,T}\right] e^{i\left\lceil \frac{t_0}{\theta} \right\rceil \alpha} \tag{B26}$$

which shows that the discrete phasor of an ungated PSED with offset, is obtained by rotation of the phasor in the absence of offset (Eq. (B3)). It is easy to verify that this formula leads to Eq. (131).

For $T \neq N\theta$, we obtain:

$$z_N\left[\Lambda_{\tau,T|t_0}\right] = \frac{1-x}{1-xe^{i\alpha}} \frac{(1-y)x^q e^{iq\alpha} + y - x^N e^{iN\alpha}}{(1-y)x^q + y - x^N} \tag{B27}$$

Interestingly, if $f$ is chosen such that $N\alpha = 2\pi n,\ n \in \mathbb{N}$, i.e. $f = n/\Theta$, where $\Theta = N\theta$, then some simplification of this expression is possible:

$$f = \frac{n}{N\theta} \Rightarrow z_N\left[\Lambda_{\tau,T|t_0}\right] = \frac{1-x}{1-xe^{i\alpha}} \frac{(1-y)x^q e^{iq\alpha} + y - x^N}{(1-y)x^q + y - x^N} \tag{B28}$$

The term $y - x^N = e^{-T/\tau} - e^{-N\theta/\tau}$ cannot be simplified unless $\Theta = T$, and its presence is due to the $T$-periodicity of the PSED. However, if we assume that there is no decay offset ($\theta_0 = 0,\ r = 0$), then we recover the expression for $z_N\left[\Lambda_{\tau,T}\right]$:

$$f = \frac{n}{N\theta}, t_0 = 0 \Rightarrow z_N\left[\Lambda_{\tau,T}\right] = \frac{1-x}{1-xe^{i\alpha}} \tag{B29}$$

In other words, in the special cases where there is no decay offset, and the $N$ gates do not cover the whole period, it may be advantageous to use a non-standard phasor harmonic frequency $f$, defined as in Eq. (B29), in order to obtain a simple analytical form for the phasor.

### *B.2. Discrete phasor of square-gated PSEDs*

*B.2.1. General case (no offset)*

For a square-gated PSED $\Lambda_{\tau,T,W}(t)$ with lifetime $\tau$ and gate width $W = \omega + kT$ (Eq. (47) **Error! Reference source not found.**):

$$\begin{cases} \left\| \Lambda_{\tau,T,W}(t_p) \right\|_N = \theta \sum_{p=1}^{N} \Lambda_{\tau,T,W}(t_p) \\ t_p = (p-1)\theta \\ \theta = \dfrac{T}{N} \end{cases} \tag{B30}$$

Because the expression of $\Lambda_{\tau,T,W}(t)$ depends on the location of $t$ in [0, $T$], we need to find index $r$ in [1, $N$] such that $t_r = (r-1)\theta < T - \omega$ and $t_{r+1} \geq T - \omega$. The result is:

$$r = \left\lceil \frac{T-\omega}{\theta} \right\rceil \tag{B31}$$

where $\lceil x \rceil$ designates the smallest integer larger than or equal to $x$ (`ceil` function).

The sum in Eq. (B30) can be expanded as:

$$\left\| \Lambda_{\tau,T,W}(t_p) \right\|_N = \theta \left\{ \sum_{p=1}^{r} \left( \frac{1-e^{-\omega/\tau}}{1-e^{-T/\tau}} e^{-t_p/\tau} + k \right) + \sum_{p=r+1}^{N} \left( \frac{1-e^{-(\omega-T)/\tau}}{1-e^{-T/\tau}} e^{-t_p/\tau} + k + 1 \right) \right\} \tag{B32}$$

The final result, assuming $T = N\theta$, is:

$$\left\| \Lambda_{\tau,T,W}(t) \right\|_N = \theta \left\{ (k+1)N - r + \frac{1-ux^r y^{-1}}{1-x} \right\}. \tag{B33}$$

where we have used the notations:

$$\begin{cases} x(\tau) = e^{-\theta/\tau} \\ y(\tau) = e^{-T/\tau} \\ u(\tau) = e^{-\omega/\tau} \end{cases} \tag{B34}$$

The discrete phasor of a square-gated PSED is given by Eq. (92), where the denominator is given by Eq. (B33) and the numerator by:

$$\begin{aligned} \left\| \Lambda_{\tau,T,W}(t_p) e^{i2\pi f t_p} \right\|_N &= \theta \sum_{p=1}^{N} \Lambda_{\tau,T,W}(t_p) e^{2\pi i f t_p} \\ &= \theta \left\{ \sum_{p=1}^{r} \left( \frac{1-e^{-\omega/\tau}}{1-e^{-T/\tau}} e^{-t_p/\tau} + k \right) e^{2\pi i f t_p} + \sum_{p=r+1}^{N} \left( \frac{1-e^{-(\omega-T)/\tau}}{1-e^{-T/\tau}} e^{-t_p/\tau} + k + 1 \right) e^{2\pi i f t_p} \right\} \end{aligned} \tag{B35}$$

Assuming $T = N\theta$ and noting $\alpha = 2\pi f \theta$, we obtain:

$$\left\| \Lambda_{\tau,T,W}\left(t_p\right) e^{i2\pi f t_p} \right\|_N = \theta \left\{ -\frac{1-e^{ir\alpha}}{1-e^{i\alpha}} + \frac{1-u}{1-y}\frac{1-x^r e^{ir\alpha}}{1-xe^{i\alpha}} + \frac{1-uy^{-1}}{1-y}\frac{x^r e^{ir\alpha}-y}{1-xe^{i\alpha}} \right\}$$
$$= \theta \left\{ -\frac{1-e^{ir\alpha}}{1-e^{i\alpha}} + \frac{1-ux^r y^{-1} e^{ir\alpha}}{1-xe^{i\alpha}} \right\} \quad \text{(B36)}$$

and the expression for the discrete phasor of a square-gated PSED reads, assuming $T = N\theta$ :

$$\begin{cases} z_{N[W]}\left[\Lambda_{\tau,T}\right] = \dfrac{-\dfrac{1-e^{ir\alpha}}{1-e^{i\alpha}} + \dfrac{1-\beta e^{ir\alpha}}{1-xe^{i\alpha}}}{(k+1)N - r + \dfrac{1-\beta}{1-x}} \\ \beta(\tau) = u(\tau)x(\tau)^r y(\tau)^{-1} \end{cases} \quad \text{(B37)}$$

*B.2.2. Special cases (no offset)*

<u>a. Gate width *W* proportional to gate step *θ*</u>

For cases where $T - \omega$ is proportional to $\theta$, Eq. (B31) simplifies into:

$$r = \frac{T-\omega}{\theta}, \quad \text{(B38)}$$

and $uy^{-1}x^r = 1$, resulting in a simpler version of Eq. (B37):

$$z_{N[W]}\left[\Lambda_{\tau,T}\right] = \frac{1-e^{ir\alpha}}{(k+1)N-r}\left(\frac{1}{1-xe^{i\alpha}} - \frac{1}{1-e^{i\alpha}}\right) = -\frac{e^{i(r+1)\alpha/2}}{(k+1)N-r}\frac{\sin\frac{r\alpha}{2}}{\sin\frac{\alpha}{2}} z_N\left[\Lambda_{\tau,T}\right] \quad \text{(B39)}$$

which is a rotated and dilated version of the discrete phasor of ungated PSEDs ( $z_N\left[\Lambda_{\tau,T}\right]$, Eq. (B3)), and thus is an arc of circle. Rewriting $W = q\theta$, Eq. (B39) takes the form:

$$z_{N[W]}\left[\Lambda_{\tau,T}\right] = \frac{\sin q\frac{\alpha}{2}}{q\sin\frac{\alpha}{2}} e^{-i(q-1)\frac{\alpha}{2}} \, z_N\left[\Lambda_{\tau,T}\right] \quad \text{(B40)}$$

<u>b. Value for *τ* = 0 and *τ* = ∞</u>

In the limit $\tau = 0$, one obtains:

If $r \neq \dfrac{T-\omega}{\theta}$:

$$z_{N[W]}\left[\Lambda_{0,T}\right] = \frac{-\dfrac{1-e^{ir\alpha}}{1-e^{i\alpha}}+1}{(k+1)N-r+1} = -\frac{1}{(k+1)N-r+1}\frac{\sin(r-1)\frac{\alpha}{2}}{\sin\frac{\alpha}{2}} e^{ir\frac{\alpha}{2}} \quad \text{(B41)}$$

and if $r = \frac{T-\omega}{\theta}$:

$$z_{N[W]}\left[\Lambda_{0,T}\right] = \frac{-\frac{1-e^{ir\alpha}}{1-e^{i\alpha}} + 1 - e^{ir\alpha}}{(k+1)N - r} = -\frac{1}{(k+1)N-r}\frac{\sin r\frac{\alpha}{2}}{\sin\frac{\alpha}{2}} e^{i(r+1)\frac{\alpha}{2}} \tag{B42}$$

In other words, the value of the phasor for $\tau = 0$ only depends on $r = \left\lceil \frac{T-\omega}{\theta} \right\rceil$.

In all cases, in the limit $\tau = \infty$ ($x = y = u = 1$), one obtains:

$$z_{N[W]}\left[\Lambda_{\infty,T}\right] = 0 . \tag{B43}$$

Note that in the special cases were $\alpha$ is a multiple of $2\pi$, *i.e.* $f = q\ N/T$, $q \in \mathbb{N}$, the above expressions can be further simplified using:

$$\frac{\alpha}{2} = q\pi,\ q \in \mathbb{N} \Rightarrow (\forall s \in \mathbb{N}), \frac{\sin s\frac{\alpha}{2}}{\sin\frac{\alpha}{2}} = s \text{ and } e^{is\frac{\alpha}{2}} = (-1)^{sq} \tag{B44}$$

This situation is however unlikely to be encountered, since it means that the harmonic $m = qN$, where the number of gates $N$ is in general not very small, would be quite large.

c. Gate width *W* smaller than the gate step *θ*

In this case, Eq. (104) reads:

$$r = \left\lceil \frac{T-\omega}{\theta} \right\rceil = \left\lceil N - \frac{W}{\theta} \right\rceil = N \tag{B45}$$

from which it results that Eq. (103) reads:

$$z_{N[W]}\left[\Lambda_{\tau,T}\right] = \frac{-\frac{1-e^{ir\alpha}}{1-e^{i\alpha}} + \frac{1-e^{ir\alpha}}{1-xe^{i\alpha}}}{\frac{1}{1-x}} = z_N\left[\Lambda_{\tau,T}\right] \tag{B46}$$

In other words, in these cases, the discrete phasor of square-gated PSEDs does not depend on the gate width and is identical to the discrete phasor of ungated PSEDs.

d. Gate with reduced width *ω* smaller than the gate step *θ*

This case encompasses the previous one, but is a bit more general (if not common). Returning to Eq. (B31) defining *r*, the point at which the expression for $\Lambda_{\tau,T,W}(t)$ changes form, it has no solution in [1, *N*[ if the gate step $\theta > \omega$, or in other words, if the gates are not overlapping or contiguous but instead are separated by a finite gap. In this case, only the first sum in Eqs. (B32) and (B35) is involved and:

$$\left\|\Lambda_{\tau,T,W}\left(t_p\right)\right\|_N = \theta\sum_{p=1}^{N}\left(\frac{1-e^{-\omega/\tau}}{1-e^{-T/\tau}}e^{-t_p/\tau}+k\right)$$
$$= \theta\left\{kN + \frac{1-u}{1-y}\frac{1-x^N}{1-x}\right\} \tag{B47}$$
$$= \theta\left(kN + \frac{1-u}{1-x}\right)$$

where we have used $x^N = y$ , since we are assuming $T = N\theta$ , to obtain the last identity. Similarly, the numerator of *z* is given by:

$$\left\|\Lambda_{\tau,T,W}\left(t_p\right)e^{i2\pi f t_p}\right\|_N = \theta\sum_{p=1}^{N}\Lambda_{\tau,nT,W}\left(t_p\right)e^{2\pi i f t_p} = \theta\sum_{p=1}^{N}\left(\frac{1-u}{1-y}e^{-t_p/\tau}+k\right)e^{i(p-1)\alpha}$$
$$= \theta\left\{k\frac{1-e^{iN\alpha}}{1-e^{i\alpha}} + \frac{1-u}{1-y}\frac{1-x^N e^{iN\alpha}}{1-xe^{i\alpha}}\right\} = \theta\frac{1-u}{1-xe^{i\alpha}} \tag{B48}$$

where we have used $x^N = y$ and $e^{iN\alpha} = 1$ to obtain the last identity. Finally, we obtain the discrete phasor of a square-gated PSED when the gate step *θ* is larger than the reduced gate width *ω* as:

$$\theta > \omega \Rightarrow z_{N[W]}\left[\Lambda_{\tau,T}\right] = \frac{1}{kN + \frac{1-u}{1-x}}\frac{1-u}{1-xe^{i\alpha}} = \frac{\frac{1-u}{1-x}}{kN + \frac{1-u}{1-x}}z_N\left[\Lambda_{\tau,T}\right] \tag{B49}$$

where $z_N\left[\Lambda_{\tau,T}\right]$ is the discrete phasor of ungated PSEDs given by Eq. (B3). Note that because the prefactor of $z_N\left[\Lambda_{\tau,T}\right]$ depends on *τ*, Eq. (B49) does *not* represent an arc of circle unless *k* = 0, in which case the phasor of square-gated decays is identical to that of the ungated decays as seen above. *k* = 0 is in fact the most likely case, as there is hardly any justification for choosing a gate width *W* > *T* (unless the hardware prevents obtaining gates with *W* < *T*).

It is easy to verify that in the special case *θ* = *ω*, for which *x* = *u*, the previous formula also applies and:

$$\theta = \omega \;\Rightarrow\; z_{N[W]}\left[\Lambda_{\tau,T}\right] = \frac{1}{kN+1}\frac{1-x}{1-xe^{i\alpha}} \tag{B50}$$

This formula is identical to Eq. (B3) obtained for the discrete phasor of an ungated PSED, except for the prefactor $\left(1+kN\right)^{-1}$, showing that Eq. (B50) describes an arc of circle centered on $\left(g_c, s_c\right)$ given by:

$$\begin{cases} g_c = \frac{1}{2}\frac{1}{kN+1} \\ s_c = -\frac{1}{2}\frac{1}{kN+1}\tan\left(\alpha/2\right) \end{cases} \tag{B51}$$

and with radius *r* equal to:

$$r = \frac{1}{kN+1}\frac{1}{2\left|\cos\left(\alpha/2\right)\right|}, \tag{B52}$$

e. $N = 2$ gates

A last special case of interest is when the number of gates $N = 2$ and the gates cover the whole laser period, $\theta = T/2$. Here again, the final formula depends on the gate width $W$ (or more exactly its reduced width, $\omega$):

$$z_{N[W]}\left[\Lambda_{\tau,T}\right] = \begin{cases} \dfrac{\dfrac{1-u}{1-y}\left(1+(-1)^n x\right)+k\left(1+(-1)^n\right)}{2k+\dfrac{1-u}{1-y}(1+x)}, \ \omega < \dfrac{T}{2} \\ \dfrac{\dfrac{1-u}{1-y}+k+(-1)^n\left(\dfrac{1-uy^{-1}}{1-y}x+k+1\right)}{2k+1+\dfrac{1-u+\left(1-uy^{-1}\right)x}{1-y}}, \omega \geq \dfrac{T}{2} \end{cases} \tag{B53}$$

where we have used the definition $f = nT^{-1}$ for the harmonic frequency.

The main cases of interest are obtained for $k = 0$ (gate of width $W = \omega < T$). Depending on whether the harmonic is an odd (e.g. $n = 1$) or even (e.g. $n = 2$) multiple of the fundamental frequency $T^{-1}$, one obtains:

$$z_{N[W]}\left[\Lambda_{\tau,T}\right] = \begin{cases} \tanh\dfrac{\theta}{2\tau}, & W < \dfrac{T}{2},\ k=0,\ n \text{ odd} \\ 1, & W < \dfrac{T}{2},\ k=0,\ n \text{ even} \\ \dfrac{u\left(xy^{-1}-1\right)+y-x}{2-u\left(xy^{-1}+1\right)-y+x}, & W \geq \dfrac{T}{2},\ k=0,\ n \text{ odd} \\ 1, & W \geq \dfrac{T}{2},\ k=0,\ n \text{ even} \end{cases} \tag{B54}$$

In all cases, the phasor is a real number, meaning the locus of the phasor of square-gated PSEDs is a segment of the $s = 0$ axis (or a single point $z = 1$, in case $n$ is even). However, in case $W < \frac{T}{2}$, the whole [0,1] segment is covered ($z_{N[W]}\left[\Lambda_{0,T}\right] = 1$ and $z_{N[W]}\left[\Lambda_{\infty,T}\right] = 0$), whereas in case $W \geq \frac{T}{2}$, both $z_{N[W]}\left[\Lambda_{0,T}\right] = 0$ and $z_{N[W]}\left[\Lambda_{\infty,T}\right] = 0$, which shows that the locus of the phasor of square-gated PSEDs is a segment occupying only *a fraction* of [0,1] segment, as illustrated in Fig. 5. The phasor first increases with $\tau$, to reach a value $z_{max}$ for a particular value $\tau_{max}$, before decreasing back to 0 as $\tau \to \infty$.

*B.2.3. General case (with offset)*

As in the ungated PSED case, the presence of an offset $t_0$ introduces an additional subtlety in the calculation, with the expression $(t-t_0)[T]$ intervening in definition (121). As indicated in Fig. S2, depending on the gates location $t_p$ in the period ($p = 1,\ldots, N$), this expression takes either one of the two forms:

$$q = \left\lceil \frac{t_0}{\theta} \right\rceil \rightarrow \begin{cases} p \le q \Rightarrow (t_p - t_0)[T] = t_p - t_0 - T \\ p > q \Rightarrow (t_p - t_0)[T] = t_p - t_0 \end{cases} \tag{B55}$$

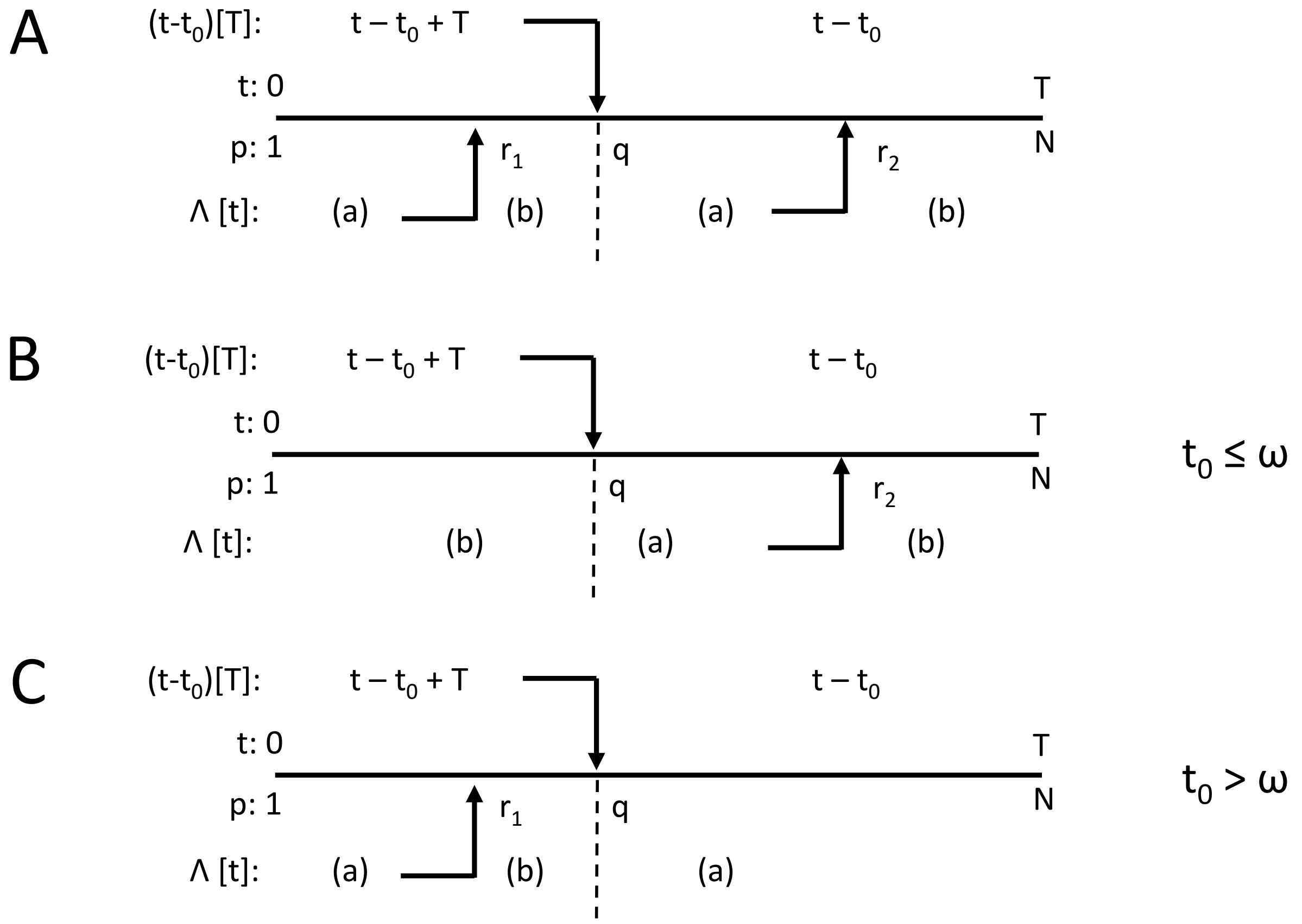

**Fig. S2**: A: Depending on the gate index $p$, the expression $(t_p - t_0)[T]$ takes the form $t_p - t_0 + T \ (p \le q)$ or $t_p - t_0 \ (p > q)$, where $q$ is defined in the text. Likewise, depending on the gate index $p$ position with respect to index $r_1$ and $r_2$ defined in the text, the square-gated PSED expression to use is form (a) or (b) of Eq. (121). Specifically, (B) for $t_0 \le \omega$, a single index, $r_2 \ge q$ is necessary, while (C) when $t_0 > \omega$, a single index $r_1 \le q$ is needed.

To determine which form (a) or (b) of Eq. (121) to use for $\Lambda_{\tau,T,W,t_0}(t)$, we need to look into each of the two domains defined in Eq. (B55) and determine index $r_1$ and $r_2$ such that:

$$
\begin{aligned}
p \le q &\Rightarrow \begin{cases} p \le r_1 \Rightarrow \text{form } (a) \\ p > r_1 \Rightarrow \text{form } (b) \end{cases} \\
p > q &\Rightarrow \begin{cases} p \le r_2 \Rightarrow \text{form } (a) \\ p > r_2 \Rightarrow \text{form } (b) \end{cases}
\end{aligned}
\tag{B56}
$$

It is straightforward to establish that there are two main situations, $t_0 \le \omega,\;\; r_2 = N + \lceil \frac{t_0 - \omega}{\theta} \rceil$ and $t_0 > \omega,\; r_1 = \lceil \frac{t_0 - \omega}{\theta} \rceil$ (see Fig. S2B & C), with $t_0 = \omega$ being a special case of $t_0 \le \omega$ for which $r_2 = N$. Calculation of the denominator of Eq. (92) thus involves:

$$
\begin{aligned}
& t_0 \le \omega \Rightarrow \\
& \quad \left\| \Lambda_{\tau,T,W|t_0}\left(t_p\right) \right\|_N = \theta \left\{ \sum_{p=1}^{q} (b)\left(t_p - t_0 + T\right) + \sum_{p=q+1}^{r_2} (a)\left(t_p - t_0\right) + \sum_{p=r_2+1}^{N} (b)\left(t_p - t_0\right) \right\} \\
& t_0 > \omega \Rightarrow \\
& \quad \left\| \Lambda_{\tau,T,W|t_0}\left(t_p\right) \right\|_N = \theta \left\{ \sum_{p=1}^{r_1} (a)\left(t_p - t_0 + T\right) + \sum_{p=r_1+1}^{q} (b)\left(t_p - t_0 + T\right) + \sum_{p=q+1}^{N} (a)\left(t_p - t_0\right) \right\}
\end{aligned}
\tag{B57}
$$

where the notation $(b)\left(t_p - t_0 + T\right)$ means that form (*b*) of Eq. (121) needs to be used, with $\left(t - t_0\right)\left[T\right] = t - t_0 + T$ and so on. Similar sums are involved for the numerator of Eq. (92).

After some lengthy but straightforward calculations, one obtains:

$$
z_N\left[\Lambda_{\tau,T,W|t_0}\right] = \frac{\dfrac{e^{ir_2\alpha} - e^{iq\alpha}}{1 - e^{i\alpha}} + \dfrac{x^q e^{iq\alpha} - u y^{-1} x^{r_2} e^{ir_2\alpha}}{1 - x e^{i\alpha}} e^{t_0/\tau}}{\left(k+1\right)N + q - r_2 + \dfrac{x^q - u y^{-1} x^{r_2}}{1 - x} e^{t_0/\tau}},\; t_0 \le \omega \tag{B58}
$$

and:

$$
z_N\left[\Lambda_{\tau,T,W|t_0}\right] = \frac{\dfrac{e^{ir_1\alpha} - e^{iq\alpha}}{1 - e^{i\alpha}} + \dfrac{x^q e^{iq\alpha} - u x^{r_1} e^{ir_1\alpha}}{1 - x e^{i\alpha}} e^{t_0/\tau}}{kN + q - r_1 + \dfrac{x^q - u x^{r_1}}{1 - x} e^{t_0/\tau}},\; t_0 > \omega \tag{B59}
$$

which can be combined into a single form:

$$
\begin{cases}
z_{N[W]}\left[\Lambda_{\tau,T|t_0}\right] = z_N\left[\Lambda_{\tau,T,W|t_0}\right] = \dfrac{\dfrac{e^{ir\alpha} - e^{iq\alpha}}{1 - e^{i\alpha}} + \dfrac{x^q e^{iq\alpha} - u x^{r} e^{ir\alpha}}{1 - x e^{i\alpha}} e^{t_0/\tau}}{kN + q - r + \dfrac{x^q - u x^{r}}{1 - x} e^{t_0/\tau}} \\[2ex]
q = \lceil \dfrac{t_0}{\theta} \rceil, \quad r = \lceil \dfrac{t_0 - \omega}{\theta} \rceil
\end{cases}
\tag{B60}
$$

As discussed in the main text (Section 4), this equation does not in general represent an arc of circle.

It is easy to verify that for $\tau = \infty$, $z\left[\Lambda_{\infty,T,W|t_0}\right] = 0$

For $\tau = 0$, we need to distinguish between 4 cases depending on the respective values of:

$$\begin{cases} \theta_q = q\theta - t_0 \\ \theta_r = r\theta - (t_0 - \omega) \end{cases} \tag{B61}$$

The final result for $z\left[\Lambda_{0,T,W|t_0}\right]$ is easy to verify and can be tabulated as:

| | $\theta_r = 0$ | $\theta_r > 0$ |
|---|---|---|
| $\theta_q = 0$ | $\frac{1}{\left(kN + (q+1) - (r+1)\right)} \frac{e^{i(r+1)\alpha} - e^{i(q+1)\alpha}}{1 - e^{i\alpha}}$ | $\frac{1}{\left(kN + q + 1 - r\right)} \frac{e^{ir\alpha} - e^{i(q+1)\alpha}}{1 - e^{i\alpha}}$ |
| $\theta_q > 0$ | $\frac{1}{\left(kN + q - (r+1)\right)} \frac{e^{i(r+1)\alpha} - e^{iq\alpha}}{1 - e^{i\alpha}}$ | $\frac{1}{\left(kN + q - r\right)} \frac{e^{ir\alpha} - e^{iq\alpha}}{1 - e^{i\alpha}}$ |

(B62)

These formulas are undefined if $k = 0$ and $q = r$ ($\theta_q = \theta_r = 0$ or $\theta_q > 0$ and $\theta_r > 0$), or $k = 0$ and $q = r + 1$ ($\theta_q > 0$ and $\theta_r = 0$), or $k = 0$ and $q = r - 1$ ($\theta_q = 0$ and $\theta_r > 0$). In those cases, the result is instead:

| | $\theta_r = 0$ | $\theta_r > 0$ |
|---|---|---|
| $\theta_q = 0$ | $k = 0,\ q = r \to e^{ir\alpha}$ | $k = 0,\ q = r - 1 \to e^{ir\alpha}$ |
| $\theta_q > 0$ | $k = 0,\ q = r + 1 \to e^{ir\alpha}$ | $k = 0,\ q = r \to e^{ir\alpha}$ |

(B63)

In other words, in these situations (which also includes $\theta_q = \theta_r = 0$, which strictly speaking is not a square-gated case), the phasor of the square-gated decay with offset at $\tau = 0$ is equal to $e^{ir\alpha}$.

*B.2.4. Special cases (with offset)*

a. Gate width equal to gate step: $W = \theta$

In the particular case where the gate width $W$ equals the gate step $\theta$ (a case relevant for instance to binned TCSPC data where the time bin can be identified to a gate), the analytical expression (B62) takes the simpler form:

$$z_N\left[\Lambda_{\tau,T,\theta|t_0}\right] = \frac{\left(1 - xe^{i\alpha} - \left(1 - e^{i\alpha}\right)x^q e^{t_0/\tau}\right)}{1 - xe^{i\alpha}} e^{i(q-1)\alpha} \tag{B64}$$

We can distinguish two general subcases: (i) $t_0 = q\theta$ and (ii) $t_0 = q\theta - \theta_0,\ \theta_0 \in \left]0, \theta\right[$.

(i) In the first case, Eq. (B64) simplifies into:

$$t_0 = q\theta \quad \Rightarrow \quad z_{N[W]}\left[\Lambda_{\tau,T|t_0}\right] = \frac{1-x}{1-xe^{i\alpha}} e^{i2\pi f t_0} = z_N\left[\Lambda_{\tau,T|t_0}\right] \tag{B65}$$

In other words, for a square-gated ($W = \theta$) PSED with offset $t_0$ proportional to the gate step $\theta$, the discrete phasor is equal to the discrete phasor of the ungated PSED with the same offset.

(ii) In the second case, Eq. (B64) can be rewritten:

$$t_0 = q\theta - \theta_0,\ \theta_0 \in \left]0,\theta\right[ \ \Rightarrow\ z_N\left[\Lambda_{\tau,T,\theta|t_0}\right] = \frac{\left(1-x+\left(1-e^{i\alpha}\right)\left(x-e^{-\theta_0/\tau}\right)\right)}{1-xe^{i\alpha}} e^{i(q-1)\alpha} \tag{B66}$$

We recognize the term $z_N\left[\Lambda_{\tau,T}\right] = (1-x)/(1-xe^{i\alpha})$ in this expression, but this is where the similarity with the previous case ends, since there are additional terms as well, all depending on $\tau$. The term $e^{i(q-1)\alpha}$ introduces a constant rotation depending on the offset $t_0$, which is constant for sets of offsets $t_0$ within $\theta$ from one another. However, because of the additional term involving $\theta_0$ in Eq. (B66), this does not mean that the corresponding SEPLs are identical, contrary to the case of the ungated decays with offset.

b. Truncated decay with offset: $t_1 = t_0$ and $t_N < T - \omega$

Because of the assumption that all $t_p$s are such that $t_0 \le t_p = t_0 + (p-1)\theta < T - \omega$, only the first expression in Eq. (121) for the square-gated PSED needs to be used, and the time argument reads:

$$t_p' = t_p - t_0 - \left\lfloor (t_p - t_0)/T \right\rfloor T = (p-1)\theta \tag{B67}$$

This simplifies the calculation of the numerator and denominator in the phasor formula (Eq.(92)). We obtain:

$$\left\| \Lambda_{\tau,T,W|t_0}(t_p) \right\|_N = \theta\left\{ \frac{1-u}{1-y} \sum_{p=1}^{N} e^{-t_p'/\tau} + kN \right\} = \theta\left\{ kN + \frac{1-u}{1-y}\frac{1-x^N}{1-x} \right\} \tag{B68}$$

and:

$$\begin{aligned} \left\| \Lambda_{\tau,T,W|t_0}(t_p) e^{i2\pi f t_p} \right\|_N &= \theta \sum_{p=1}^{N} \left( \frac{1-u}{1-y} e^{-t_p'/\tau} + k \right) e^{i2\pi f t_p} \\ &= \theta\left( k\frac{1-e^{iN\alpha}}{1-e^{i\alpha}} + \frac{1-u}{1-y}\frac{1-x^N e^{iN\alpha}}{1-xe^{i\alpha}} \right) e^{i2\pi f t_0} \end{aligned} \tag{B69}$$

This expression simplifies greatly if we assume that the phasor frequency $f$ is chosen such that:

$$f = \frac{n}{N\theta} = \frac{n}{D} \tag{B70}$$

In that case:

$$\left\| \Lambda_{\tau,T,W|t_0}(t_p) e^{i2\pi f t_p} \right\|_N = \theta \frac{1-u}{1-y} \frac{1-x^N}{1-xe^{i\alpha}} e^{i2\pi f t_0} \tag{B71}$$

and the phasor reads:

$$z_{N[W]}\left[\Lambda_{\tau,T|t_0}\right]=\frac{\left\|\Lambda_{\tau,T,W}\left(t_p\right)e^{i2\pi ft_p}\right\|_N}{\left\|\Lambda_{\tau,T,W,t_0}\left(t_p\right)\right\|_N}=\frac{\dfrac{1-x^N}{1-xe^{i\alpha}}}{kN\dfrac{1-y}{1-u}+\dfrac{1-x^N}{1-x}}e^{i2\pi ft_0} \tag{B72}$$

Unless $k = 0$ (i.e. $W < T$), this expression does not describe a simple curve. However, $k = 0$ (*i.e.* $W < T$) is the most common case, for which Eq. (B72) simplifies into:

$$z_{N[W]}\left[\Lambda_{\tau,T|t_0}\right]=\frac{1-x}{1-xe^{i\alpha}}e^{i2\pi ft_0}=z_N\left[\Lambda_{\tau,T|t_0}\right] \tag{B73}$$

which is the same formula as for a discrete phasor of an ungated PSED with offset $t_0$, and corresponds to an arc of circle rotated about the origin.

### *B.3. Discrete phasor of arbitrary T-periodic decays*

#### *B.3.1. General case*

By analogy with the continuous phasor case discussed in Section 3.2.4, the discrete phasor of an arbitrary *T*-periodic decay $S_T\left(t\right)$ can be written as a weighted sum of the discrete phasors of PSEDS, $\zeta_{f,N}\left(\tau\right)$. As argued in that discussion, it is advantageous to look at the $\left\|\;\right\|_N$*-normalized* decay $\sigma_T\left(t\right)$ now defined (because we are dealing with discrete phasors) as:

$$\sigma_T\left(t\right)=\frac{S_T\left(t\right)}{\left\|S_T\left(t_p\right)\right\|_N} \tag{B74}$$

By definition, we shall call $\phi_0\left(\tau\right)$ the weight function such that:

$$\sigma_T\left(t\right)=\int_0^\infty d\tau\phi_0\left(\tau\right)\Lambda_{\tau,T}\left(t\right) \tag{B75}$$

The discrete phasor of $\sigma_T\left(t\right)$ is given by:

$$z_N\left[\sigma_T\right]=\frac{\left\|\sigma_T\left(t_p\right)e^{i2\pi ft_p}\right\|_N}{\left\|\sigma_T\left(t_p\right)\right\|_N} \tag{B76}$$

The two terms in this ratio are:

$$\begin{cases} \left\|\sigma_T\left(t_p\right)\right\|_N = \int_0^\infty d\tau\phi_0\left(\tau\right)\left\|\Lambda_{\tau,T}\left(t_p\right)\right\|_N = 1 \\ \left\|\sigma_T\left(t_p\right)e^{i2\pi f t_p}\right\|_N = \int_0^\infty d\tau\phi_0\left(\tau\right)\left\|\Lambda_{\tau,T}\left(t_p\right)e^{i2\pi f t_p}\right\|_N = \int_0^\infty d\tau\phi_0\left(\tau\right)\frac{\theta}{\tau\left(1-xe^{i\alpha}\right)} \\ \qquad = \int_0^\infty d\tau\phi_0\left(\tau\right)\frac{\theta}{\tau\left(1-x\right)}\zeta_{f,N}\left(\tau\right) \\ x = 1-e^{-\theta/\tau}; \alpha = 2\pi f\theta \end{cases} \tag{B77}$$

The first result comes from the definition of $\sigma_T\left(t\right)$(Eq. (B74)) and the second from Eq. (B1) and the definition of the discrete PSED phasor (Eq. (B3)).

Introducing the weight function $\mu_0\left(\tau\right)$ and the $\| \ \|_N$-normalized basis $\left\{\Lambda_{\tau,T,N}\left(t\right)\right\}_{\tau>0}$ :

$$\begin{cases} \mu_0\left(\tau\right) = \frac{\theta}{\tau\left(1-x\right)}\phi_0\left(\tau\right) \\ \Lambda_{\tau,T,N}\left(t\right) = \frac{\tau\left(1-x\right)}{\theta}\Lambda_{\tau,T}\left(t\right) \\ \left\|\Lambda_{\tau,T,N}\left(t_p\right)\right\| = 1 \end{cases} \tag{B78}$$

we can rewrite:

$$\begin{cases} \sigma_T\left(t\right) = \int_0^\infty d\tau\mu_0\left(\tau\right)\Lambda_{\tau,T,N}\left(t\right) \\ z_N\left[\sigma_T\right] = \int_0^\infty d\tau\mu_0\left(\tau\right)\zeta_{f,N}\left(\tau\right) \\ z_N\left[\Lambda_{\tau,T,N}\right] = \zeta_{f,N}\left(\tau\right) \end{cases} \tag{B79}$$

The last identity in Eq. (B79) follows from Eq. (B78) and the discrete phasor invariance by dilation (Eq. (89)).

Eq. (B79) shows that when decomposing a normalized *T*-periodic function in the basis of $\left\{\Lambda_{\tau,T,N}\left(t\right)\right\}_{\tau>0}$, its discrete phasor takes the same functional form in terms of the discrete phasors $\zeta_{f,N}\left(\tau\right)$.

*B.3.2. Special case: linear combination of exponentials*

Starting from the same definition for decay $S\left(t\right)$and its weight function $\phi_0\left(\tau\right)$as in the continuous case (Eq. (85)), we obtain for its *T*-periodic version:

$$S_T(t)=\sum_{i=1}^{n}a_i\tau_i\Lambda_{\tau_i,T}(t) \tag{B80}$$

Eq. (B78) for $\mu_0(\tau)$ yields:

$$\begin{cases}\mu_0(t)=\sum_{i=1}^{n}\mu_i\delta(\tau-\tau_i)\\ \mu_i=\dfrac{a_i}{\left(1-e^{-\theta/\tau_i}\right)}\Big/\sum_{j=1}^{n}\dfrac{a_j}{\left(1-e^{-\theta/\tau_j}\right)}\end{cases} \tag{B81}$$

Note that for $N\to\infty$ (*i.e.* $\theta\to 0$ ), the $\mu_i$s defined by Eq. (B81) are identical to those obtained in the continuous case (Eq. (86)), as expected.

With these definitions, the $\| \|_N$-normalized decay and its discrete phasor are given by:

$$\begin{cases}\sigma_T(t)=\sum_{i=1}^{n}\mu_i\Lambda_{\tau_i,T,N}(t)\\ z_N[\sigma_T]=\sum_{i=1}^{n}\mu_i\zeta_{f,N}(\tau_i)\end{cases} \tag{B82}$$

In other words, the discrete phasor of a linear combination of single-exponential decays can be expressed as a linear combination of phasors, but, in order for the same functional form to be used, the discrete decay needs to be expressed in the basis of $\left\{\Lambda_{\tau,T,N}(t)\right\}_{\tau>0}$ (which are proportional to the pure exponentials). The discrete phasor of the total decay is then expressed in the same manner as a function of their individual phasors.

## Appendix C: Phasor of the convolution product of periodic decays

### *C.1. Relation between convolution $*$ and cyclic convolution $\underset{T}{*}$*

Let $f(t)$ be a non-periodic function. The *T-periodic summation* of $f$ is defined as the infinite sum of shifted versions of the original function:

$$f_T(t)=\sum_{i=-\infty}^{+\infty}f(t-iT) \tag{C1}$$

Let $g_T(t)$ be another $T$-periodic function obtained as the *T-periodic summation* of $g$, a non-periodic function, similarly to the process described for $f$ in Eq. (C1). The convolution of $f$ and $g_T$ verifies:

$$
\begin{aligned}
f * g_T(t) \quad &= \int_{-\infty}^{+\infty} du\, f(u)\, g_T(t-u) \\
&= \sum_{i=-\infty}^{+\infty} \int_{iT}^{iT+T} du\, f(u)\, g_T(t-u) = \sum_{i=-\infty}^{+\infty} \int_{iT}^{iT+T} du\, f(u)\, g_T(t-u+iT) \\
(u = v + iT) \quad &= \sum_{i=-\infty}^{+\infty} \int_{0}^{T} dv\, f(v+iT)\, g_T(t-v) = \int_{0}^{T} dv \sum_{i=-\infty}^{+\infty} f(v+iT) g_T(t-v) \\
&= \int_{0}^{T} dv\, f_T(v)\, g_T(t-v) = f_T \underset{T}{*} g_T(t)
\end{aligned}
\tag{C2}
$$

This establishes the identity between the convolution product involving an integral with infinite bounds of a non-periodic function with a *T*-periodic one and the *cyclic* (or *circular*) *convolution product*, denoted by a $\underset{T}{*}$ symbol, involving an integral of two *T*-periodic functions over a single period *T*.

Note that the convolution product as defined in the last line of Eq. (C2) is commutative:

$$
\begin{aligned}
f_T \underset{T}{*} g_T(t) &= g_T \underset{T}{*} f_T(t) \\
&= f * g_T(t) = f_T * g(t) = g_T * f(t) = g * f_T(t)
\end{aligned}
\tag{C3}
$$

as can be easily verified with the example of the cyclic convolution of two PSEDs defined by Eq. (17):

$$
\Lambda_{\tau_0,T} \underset{T}{*} \Lambda_{\tau,T}(t) = \Psi_{\tau,\tau_0,T}(t) \tag{C4}
$$

Finally, it is useful to notice that the convolution product of two non-periodic functions is unrelated to those discussed above (as can be easily verified using two exponential function such as $I_0$ and $F_0$ defined in Eq. (19)):

$$
\begin{aligned}
f_T * g(t) &= (f * g)_T(t) = (g * f)_T(t) \\
&\neq f * g(t)
\end{aligned}
\tag{C5}
$$

### *C.2. Cyclic convolution of the mirrored gate function and detected periodic decay*

In the case of a *T*-periodic decay $S_T(t)$ (given by Eq. (14)) convolved with a *nT*-periodic function ($n \geq 1$) $\overline{\Gamma}_{W,nT}(t)$ (Eq. (38)):

$$
\begin{aligned}
S_{W,T}(s) = \overline{\Gamma}_{W,nT} \underset{nT}{*} S_T(t) &= \int_0^{nT} du\, \overline{\Gamma}_{W,nT}(t-u) \int_0^T dv\, I_T(v)\, f_{0,T}(u-v) \\
(w = u - v) \qquad &= \int_0^{nT} du\, \overline{\Gamma}_{W,nT}(t-u) \int_{u-T}^{u} dw\, I_T(u-w)\, f_{0,T}(w) \\
(x = u - w) \qquad &= \int_0^{T} dw\, f_{0,T}(w) \int_{-w}^{nT-w} dx\, \overline{\Gamma}_{W,nT}(t-w-x)\, I_T(x) \\
&= \int_0^{T} dw\, f_{0,T}(w)\, \overline{\Gamma}_{W,nT} \underset{T}{*} I_T(t-w) \\
&= \int_0^{T} dw\, f_{0,T}(w)\, I_{T,W}(t-w) = f_{0,T} \underset{T}{*} I_{T,W}(t) \\
&= I_{T,W} \underset{T}{*} f_{0,T}(t)
\end{aligned}
\tag{C6}
$$

where we have used the $T$-periodicity of $I_T(t)$ and $f_{0,T}(t)$ and the $nT$-periodicity of $\overline{\Gamma}_{W,nT}(t)$ and $I_T(t)$ to change the bounds of integration. The cyclic convolution product in the next-to-last line involves an integral over [0, $nT$], but the resulting gated instrument response function $I_{T,W}(t)$ is $T$-periodic:

$$
I_{T,W}(t) = \int_0^{nT} du\, \overline{\Gamma}_{W,nT}(u)\, I_T(t-u) \tag{C7}
$$

The last cyclic convolution product in Eq. (C6) involves the standard integration over [0,$T$].

### *C.3. Continuous phasor convolution rule*

With this definition, the numerator of the continuous phasor expression for a $T$-periodic function (Eq. **Error! Reference source not found.**(63)) reads:

$$
\begin{aligned}
\left\| f_T \underset{T}{*} g_T(t) e^{i2\pi ft} \right\|_T &= \int_0^T dt\, f_T \underset{C}{*} g_T(t) e^{i2\pi ft} \\
&= \int_0^T dt \int_0^T du\, f_T(u)\, g_T(t-u)\, e^{i2\pi ft} \\
&= \int_0^T du\, f_T(u)\, e^{i2\pi fu} \int_0^T dt\, g_T(t-u)\, e^{i2\pi f(t-u)}
\end{aligned}
\tag{C8}
$$

Due to the $T$-periodicity of $g_T$ and the exponential term, we can rewrite:

$$\int_0^T dt\, g_T(t-u)e^{i2\pi f(t-u)} = \int_{-u}^{T-u} dv\, g_T(v)e^{i2\pi fv}$$
$$= \int_{-u}^{0} dv\, g_T(v)e^{i2\pi fv} + \int_{0}^{T-u} dv\, g_T(v)e^{i2\pi fv}$$
$$(w = v+T) \qquad = \int_{T-u}^{T} dw\, g_T(T+w)e^{i2\pi f(T+w)} + \int_{0}^{T-u} dv\, g_T(v)e^{i2\pi fv} \tag{C9}$$
$$= \int_{T-u}^{T} dw\, g_T(w)e^{i2\pi fw} + \int_{0}^{T-u} dv\, g_T(v)e^{i2\pi fv} = \int_0^T dv\, g_T(v)e^{i2\pi fv}$$

from which it follows that:

$$\begin{aligned}\left\| f_T \underset{T}{*} g_T(t)e^{i2\pi ft} \right\|_T &= \int_0^T dt\, {}_T g_T(t)e^{i2\pi ft} \int_0^T du\, f_T(u)e^{i2\pi fu} \\ &= \left\| f_T(t)e^{i2\pi ft} \right\|_T \left\| g_T(t)e^{i2\pi ft} \right\|_T\end{aligned} \tag{C10}$$

Similarly, we obtain for the numerator:

$$\left\| f_T \underset{T}{*} g_T \right\|_T = \| f_T \|_T \, \| g_T \|_T \tag{C11}$$

from which it results that the continuous phasor of the convolution of two $T$-periodic decays is the product of their individual continuous phasors (‘*continuous phasor convolution rule*’):

$$z\left[ f_T \underset{T}{*} g_T \right] = z[f_T]\, z[g_T] \tag{C12}$$

### *C.4. Discrete phasor convolution rule*

As mentioned in Section 3, the discrete phasor is related to the DFT of the $N$-periodic sequence $\left\{S_T\left(t_p\right)\right\}, 1 \le p \le N$:

$$z_N[S_T](f) = \frac{\mathcal{F}^*[S_T](f)}{\mathcal{F}^*[S_T](0)} \tag{C13}$$

and as such inherits the properties of DFTs, in particular that related to the convolution of two *discrete* periodic functions. However, the convolution products involved in time-resolved spectroscopy are for the most part not involving discrete functions (discretization applies only at the data recording level), and thus, this property is of little use. Instead, it is replaced by a different phasor convolution rule. Because this distinction is important, we will first recall the convolution rule as it applies to discrete periodic functions and their DFT before examining the phasor itself.

#### *C.4.1. DFT convolution rule*

Lets start with the definition of the *discrete cyclic convolution* of two $T$-periodic functions $f_T$ and $g_T$ sampled at $N$ equidistant time points $\left\{t_p = (p-1)\theta\right\}, 1 \le p \le N$:

$$f_T *_N g_T(t_p) = \sum_{m=1}^{N} f_T(t_m) g_T(t_{p-m}) = \sum_{m=1}^{N} f_m g_{p-m} \quad \text{(C14)}$$

with obvious definitions for $f_m$ and $g_m$ and the index $N$ placed next to the convolution product symbol ($*_N$) indicates a discrete cyclic convolution. In the above definition, $p - m \leq 0 \Rightarrow g_{p-m} = g_{N+p-m}$ accounts for the $T$-periodicity (or equivalently, in terms of indices, $N$-periodicity) of $g_T$.

The convolution theorem for DFTs is easily verified by plugging Eq. (C14) in the definition of the DFT (Eq. (94)**Error! Reference source not found.**) and reads:

$$\mathcal{DF}\left[f_T *_N g_T\right] = D\mathcal{F}\left[f_T\right]\mathcal{DF}\left[g_T\right] \quad \text{(C15)}$$

where the frequency $f_n = n/T$, $0 \leq n \leq N-1$ (or equivalently the harmonic $n$) is omitted in the DFT notation, but is implicit.

*C.4.2. Discrete phasor convolution rule*

Eq. (C15) results in the following identity:

$$z_N\left[f_T *_N g_T\right] = z_N\left[f_T\right] z_N\left[g_T\right] \quad \text{(C16)}$$

which is not particularly useful, because the discrete cyclic convolution product does not intervene in any physical process. Instead, one is usually interested in the discrete phasor of a *continuous* cyclic convolution product, for which we have in general ('negative *discrete phasor convolution rule*'):

$$z_N\left[f_T \underset{T}{*} g_T\right] \neq z_N\left[f_T\right] z_N\left[g_T\right]. \quad \text{(C17)}$$

because the cyclic convolution products in Eqs. (C16) & (C17) are in general different:

$$f_T \underset{T}{*} g_T(t_p) \neq f_T *_N g_T(t_p) \quad \text{(C18)}$$

This 'negative' discrete phasor convolution rule (Eq. (C17)) has a direct implication for the phasor of samples measured with a setup that results in a measured signal equal to a *continuous* convolution product (e.g. Eq. (14) or Eq. **Error! Reference source not found.**(42)):

$$S_T(t) = I_T \underset{T}{*} F_{0,T}(t) \quad \text{or} \quad S_{T,W}(t) = I_{T,W} \underset{T}{*} F_{0,T}(t) \quad \text{(C19)}$$

Indeed, the discrete phasor convolution rule states that, in general, the discrete phasor of the convolution products in Eq. (C19) *cannot* be written as a product of the respective phasors:

$$z_N\left[I_T \underset{T}{*} F_{0,T}\right] \neq z_N\left[I_T\right] z_N\left[F_{0,T}\right] \quad \text{or} \quad z_N\left[I_{T,W} \underset{T}{*} F_{0,T}\right] \neq z_N\left[I_{T,W}\right] z_N\left[F_{0,T}\right] \quad \text{(C20)}$$

This has the very important implication that standard phasor calibration, as implemented for continuous phasors, in general *does not work*.

However, in some particular cases, the discrete phasor convolution rule can be rewritten:

$$z_N\left[I_T \underset{T}{*} F_{0,T}\right] = \kappa\, z_N\left[I_T\right] z_N\left[F_{0,T}\right] \quad \text{(C21)}$$

where $\kappa$ is constant for a family of decays $\left\{F_{0,T,\lambda}(t)\right\}$, $\lambda \in \Omega$, where $\Omega$ is a subset of $\mathbb{R}$ (generally the family of PSEDs) and $I_T(t)$ represents the instrument response function. This '*weak discrete phasor convolution rule*' allows using a modified phasor calibration approach, as is discussed in Section 8.

We shall review a few examples of the weak as well as the negative discrete phasor convolution rule in the following sub-sections.

*C.4.3. Examples of the discrete phasor convolution rule*

C.4.3.1. Discrete phasor of ungated PSEDs with single-exponential IRF

To make this clear, lets consider the case of a *T*-periodic single-exponential IRF (time constant $\tau_0$) convolved with a single-exponential decay with lifetime $\tau$ (sample emitted signal $\Psi_{\tau,\tau_0,T}(t)$ defined in Section 2.1.6). The convolution product defining this sample emitted signal has to be a continuous convolution product, as this is how the physics of the process is defined:

$$\Psi_{\tau,\tau_0,T}(t) = \Lambda_{\tau_0,T} \underset{T}{*} \Lambda_{\tau,T}(t) \tag{C22}$$

As shown in Appendix D.8, the discrete phasor of $\Psi_{\tau,\tau_0,T}(t)$, is given by (Eq. (D32)):

$$\begin{aligned} z_N\left[\Psi_{\tau,\tau_0,T}\right] &= z_N\left[\Lambda_{\tau,T}\right] z_N\left[\Lambda_{\tau_0,T}\right] e^{i2\pi f\theta} \\ &\neq z_N\left[\Lambda_{\tau,T}\right] z_N\left[\Lambda_{\tau_0,T}\right] \end{aligned} \tag{C23}$$

While the discrete phasor of the continuous convolution product is not equal to the product of the respective discrete phasors in this case, the result shows that it is *proportional* to the product, with a constant proportionality factor $\kappa = e^{i2\pi f\theta}$ where $\theta = T/N$ is the gate step. This result is at the basis of the phasor calibration strategy for ungated decays discussed in Section 8.3. Note however that Eq. (C23) does not apply to decays other than PSEDs

By contrast, it is easy to verify directly that Eq. (C16) holds for $\Lambda_{\tau_0,T}$ and $\Lambda_{\tau,T}$, but once again, the discrete convolution product:

$$\begin{aligned} \Lambda_{\tau_0,T} *_N \Lambda_{\tau,T}(t_p) &= \frac{1}{\tau_0\tau\left(e^{-\theta/\tau} - e^{-\theta/\tau_0}\right)}\left(\frac{e^{-p\theta/\tau}}{1-e^{-T/\tau}} - \frac{e^{-p\theta/\tau_0}}{1-e^{-T/\tau_0}}\right) \\ &= \frac{1}{x - x_0}\left(\frac{x}{\tau_0}\Lambda_{\tau,T}(t_p) - \frac{x_0}{\tau}\Lambda_{\tau_0,T}(t_p)\right) \end{aligned} \tag{C24}$$

(where $\theta = T/N;\ x(\tau) = e^{-\theta/\tau};\ x_0 = x(\tau_0) = e^{-\theta/\tau_0}$) defined at discrete points $\left\{t_p = (p-1)\theta\right\}, 1 \le p \le N$ only, is different from the physical signal incident on the detector, $\Psi_{\tau,\tau_0,T}(t)$ (Eq. (20)), and has in fact no physical interpretation.

C.4.3.2. Discrete phasor of square-gated PSEDs

The proportionality observed in Eq. (C23) is an exception rather than the rule. For instance, as soon as a square-gate of width $W$ is added to the acquisition scheme (with a Dirac IRF), the discrete phasor of a square-gated PSED (convolution product of a mirrored square-gate function and a PSED) defined by:

$$\Lambda_{\tau,nT,W}(t) = \bar{\Pi}_{W,nT} \underset{T}{*} \Lambda_{\tau,T}(t) \tag{C25}$$

is given by Eq. (103), which, in general cannot be expressed as a product of terms involving the discrete phasor of a PSED ($z_N\left[\Lambda_{\tau,T}\right]$) and that of a square gate ($z_N\left[\bar{\Pi}_{W,nT}\right]$). However, in the particular case where the gate width $W$ is proportional to the gate step $\theta$, Eq. (106) or (B39) for the discrete phasor of a square-gated PSED can be rewritten:

$$\begin{aligned} r = \frac{T-\omega}{\theta} \Rightarrow z_N\left[\bar{\Pi}_{W,nT} \underset{T}{*} \Lambda_{\tau,T}\right] &= z_{N[W]}\left[\Lambda_{\tau,T}\right] = -\frac{e^{i\alpha}}{(k+1)N-r}\frac{1-e^{ir\alpha}}{1-e^{i\alpha}}\frac{1-x}{1-xe^{i\alpha}} \\ &= e^{i\alpha} z_N\left[\bar{\Pi}_{W,nT}\right] z_N\left[\Lambda_{\tau,T}\right] \end{aligned} \tag{C26}$$

where we have used Eq. (C36) for the expression of $z_N\left[\bar{\Pi}_{W,nT}\right]$ and $\alpha = 2\pi f\theta$.
Eq. (C26) is clearly of the form defined in Eq. (C21) with $\kappa = e^{i2\pi f\theta}$.

C.4.3.3. Discrete phasor of square-gated PSED with single-exponential IRF

The phasor of a decay given by Eq. (D19) was calculated in Section D.9, and even in the special case where the width $W$ is proportional to the gate step $\theta$ (eq. (D49), the result shows that:

$$z_N\left[\Psi_{\tau,\tau_0,T,W}\right] \neq \kappa\, z_N\left[\bar{\Pi}_{W,nT}\right] z_N\left[\Psi_{\tau,\tau_0,T}\right] \tag{C27}$$

### *C.5. Phasors of mirrored square gates*

Here we derive a couple of expressions used in the previous sections.

*C.5.1. Continuous phasor of a mirrored square-gate*

The continuous phasor of the mirrored square-gate defined by Eq. (36) is given by:

$$z\left[\bar{\Pi}_{W,nT}\right] = \frac{\left\|\bar{\Pi}_{W,nT}(t)e^{i2\pi ft}\right\|_{nT}}{\left\|\bar{\Pi}_{W,nT}(t)\right\|_{nT}} \tag{C28}$$

where $\bar{\Pi}_{W,nT}(t)$ is defined in Eq. (37). It follows:

$$\begin{cases} \left\|\bar{\Pi}_{W,nT}(t)\right\|_{nT} = \int_0^{nT} dt\bar{\Pi}_{W,nT}(t) = \int_{nT-W}^{nT} dt = W \\ \left\|\bar{\Pi}_{W,nT}(t)e^{i2\pi ft}\right\|_{nT} = \int_0^{nT} dt\bar{\Pi}_{W,nT}(t)e^{i2\pi ft} = \int_{nT-W}^{nT} dt\, e^{i2\pi ft} = \frac{1-e^{-i2\pi fW}}{i2\pi f} \end{cases} \tag{C29}$$

and the continuous phasor:

$$z\left[\overline{\Pi}_{W,nT}\right]=\frac{\sin\pi fW}{\pi fW}e^{-i\pi fW}=M_W e^{-i\varphi_W} \tag{C30}$$

using the definitions of Eq. (75).

*C.5.2. Discrete phasor of a mirrored square-gate*

The discrete phasor of the mirrored square-gate defined by Eq. (36) is given by:

$$z_N\left[\overline{\Pi}_{W,nT}\right]=\frac{\left\|\overline{\Pi}_{W,nT}\left(t_p\right)e^{i2\pi ft_p}\right\|_N}{\left\|\overline{\Pi}_{W,nT}\left(t_p\right)\right\|_N} \tag{C31}$$

To simplify, we will first assume that $n$ = 1, $W < T$. In that case,

$$\left\|\overline{\Pi}_{W,T}\left(t_p\right)\right\|_N=\theta\sum_{p=1}^{N}\overline{\Pi}_{W,T}\left(\left(p-1\right)\theta\right) \tag{C32}$$

where:

$$\overline{\Pi}_{W,T}\left(\left(p-1\right)\theta\right)=\Pi_{0,W,T}\left(T-\left(p-1\right)\theta\right)=\begin{cases}0 \text{ if } T-\left(p-1\right)\theta<0\\ 1 \text{ if } 0\le T-\left(p-1\right)\theta\le W\\ 0 \text{ if } T-\left(p-1\right)\theta>W\end{cases} \tag{C33}$$

Calling $r=\left\lceil\left(T-W\right)/\theta\right\rceil$, $\overline{\Pi}_{W,T}\left(\left(p-1\right)\theta\right)=1$ for $p\in\left[r+1,N+1\right]$ and we obtain:

$$\begin{cases}\left\|\overline{\Pi}_{W,T}\left(t\right)\right\|_N=\left(N-r\right)\theta\\ \left\|\overline{\Pi}_{W,T}\left(t\right)e^{i2\pi ft}\right\|_N=-\theta\dfrac{1-e^{ir\alpha}}{1-e^{i\alpha}}\end{cases} \tag{C34}$$

where $\alpha=2\pi f\theta$. If follows that the discrete phasor of the mirrored square-gate is given by:

$$z_N\left[\overline{\Pi}_{W,T}\right]=-\frac{1}{N-r}\frac{1-e^{ir\alpha}}{1-e^{i\alpha}}=-\frac{1}{N-r}\frac{\sin\left(r\alpha/2\right)}{\sin\left(\alpha/2\right)}e^{i(r-1)\alpha/2} \tag{C35}$$

If we dont assume $W < T$, then the sum in Eq. (C32) needs to be extended up to $nN$ and it is easy to verify that $\overline{\Pi}_{W,nT}\left(\left(p-1\right)\theta\right)=1$ only for $p\in\left[\left(n-k-1\right)N+r+1,nN+1\right]$ where we have used the definition $k=\left\lfloor W/T\right\rfloor, W=\omega+kT$ introduced in Eq. (47) and the modified definition $r=\left\lceil\left(T-\omega\right)/\theta\right\rceil$. After some simple algebra, we obtain:

$$z_N\left[\overline{\Pi}_{W,nT}\right]=-\frac{1}{\left(k+1\right)N-r}\frac{1-e^{ir\alpha}}{1-e^{i\alpha}}=-\frac{1}{\left(k+1\right)N-r}\frac{\sin\left(r\alpha/2\right)}{\sin\left(\alpha/2\right)}e^{i(r-1)\alpha/2} \tag{C36}$$

When $W$ is proportional to $\theta$ ($W = q\theta$), $r=\left(k+1\right)N-q$, Eq. (C36) can be rewritten:

$$z_N\left[\overline{\Pi}_{W,nT}\right]=\frac{\sin\left(q\alpha/2\right)}{q\sin\left(\alpha/2\right)}e^{-i(q+1)\alpha/2} \tag{C37}$$

## Appendix D: Some properties of the convolution of two *T*-periodic single-exponential functions

Decays of the type discussed in Section 2.1.6 (convolution of a periodic single-exponential IRF with characteristic time $\tau_*$ and a single-exponential emission with lifetime $\tau$) are interesting as they showcase a number of properties of *T*-periodic decays that are lost when the periodicity is not considered. Here, we review some useful properties of this family of decays and derive a few results used in the main text. For simplicity, we will refer to this type of decays as *convolution of PSEDs*.

### *D.1. Calculation of the convolution product*

Note: In the following discussion, we will assume that one of the time constants is not equal to zero (namely the time constant representing the excitation function, $\tau_\times \neq 0$ ).

#### *D.1.1. Case $\tau \neq \tau_x$*

a. No offset

The *T*-periodic single-exponential IRF $I_T(t)$ is given by:

$$\begin{aligned} I_T(t) &= \frac{1}{\tau_\times}\sum_{i=-\infty}^{+\infty} e^{-(t-iT)/\tau_\times} H(t-iT) = \frac{1}{\tau_\times}\sum_{i=-\infty}^{k=\lfloor t/T \rfloor} e^{-(t-iT)/\tau_\times} \\ &= \frac{1}{\tau_\times\left(1-e^{-T/\tau_\times}\right)} e^{-\left(t-\lfloor t/T \rfloor T\right)/\tau_\times} = \frac{1}{\tau_\times\left(1-e^{-T/\tau_\times}\right)} e^{-t[T]/\tau_\times} = \Lambda_{\tau_\times,T}(t) \end{aligned} \tag{D1}$$

and the convolution with the single-exponential decay $F_0(t)$ with lifetime $\tau$ is given by:

$$\begin{aligned} I_T * F_0(t) &= \Lambda_{\tau_\times,T} * \Lambda_\tau(t) \\ &= \int_{-\infty}^{+\infty} du\, I_T(u) F_0(t-u) = \frac{1}{\tau\tau_\times}\int_{-\infty}^{t} du\, e^{-(t-u)/\tau}\sum_{i=-\infty}^{k} e^{-(u-iT)/\tau_\times} H(u-iT) \\ &= \frac{e^{-t/\tau}}{\tau\tau_\times}\sum_{i=-\infty}^{k} e^{iT/\tau_\times}\int_{-\infty}^{t} du\, e^{-u\left(\frac{1}{\tau_\times}-\frac{1}{\tau}\right)} H(u-iT) \\ &= \frac{e^{-t/\tau}}{\tau\tau_\times}\sum_{i=-\infty}^{k} e^{iT/\tau_\times}\int_{iT}^{t} du\, e^{-u\left(\frac{1}{\tau_\times}-\frac{1}{\tau}\right)} = \frac{1}{\tau-\tau_\times}\sum_{i=-\infty}^{k}\left(e^{-(t-iT)/\tau} - e^{-(t-iT)/\tau_\times}\right) \end{aligned} \tag{D2}$$

This immediately gives:

$$\begin{aligned} \Psi_{\tau,\tau_\times,T}(t) &= \Lambda_{\tau_\times,T} * \Lambda_\tau(t) = \Lambda_{\tau_\times,T} \underset{T}{*} \Lambda_{\tau,T}(t) \\ &= \frac{1}{\tau-\tau_\times}\left(\frac{e^{-t[T]/\tau}}{1-e^{-T/\tau}} - \frac{e^{-t[T]/\tau_\times}}{1-e^{-T/\tau_\times}}\right) = \frac{\tau\Lambda_{\tau,T}(t) - \tau_\times\Lambda_{\tau_\times,T}(t)}{\tau-\tau_\times} \end{aligned} \tag{D3}$$

Its integral over [0, *T*] is equal to 1.

b. With offset

In the presence of an offset $t_0$, the *T*-periodic single-exponential IRF is given by:

$$\Lambda_{\tau_\times,T|t_0}(t)=\Lambda_{\tau_\times,T}(t-t_0) \tag{D4}$$

where $\Lambda_{\tau,T|t_0}(t)$ is defined in Eq. (120). It is straightforward to verify that the convolution of this IRF and the single-exponential decay $F_0(t)$ with lifetime $\tau$ is given by:

$$\begin{aligned}\Psi_{\tau,\tau_\times,T|t_0}(t)=\Lambda_{\tau_\times,T|t_0}*F_0(t)=\frac{1}{\tau-\tau_\times}\left(\frac{e^{-t'/\tau}}{1-e^{-T/\tau}}-\frac{e^{-t'/\tau_\times}}{1-e^{-T/\tau_\times}}\right)=\frac{\tau\Lambda_{\tau,T}(t')-\tau_\times\Lambda_{\tau_\times,T}(t')}{\tau-\tau_\times}\\ =\frac{\tau\Lambda_{\tau,T|t_0}(t)-\tau_\times\Lambda_{\tau_\times,T|t_0}(t)}{\tau-\tau_\times}\end{aligned} \tag{D5}$$

which is the formula obtained in the absence of offset with the replacement $t\mapsto t'=t-t_0$.

*D.1.2. Case $\tau = \tau_x$*

a. No offset

When $\tau=\tau_\times$, Eq. (D3) is replaced by:

$$\Lambda_{\tau_\times,T}*\Lambda_{\tau_\times}(t)=\Psi_{\tau_\times,\tau_\times,T}(t)=\frac{1}{\tau_\times}\frac{e^{-t[T]/\tau_\times}}{1-e^{-T/\tau_\times}}\left(\frac{t[T]}{\tau_\times}+\frac{e^{-T/\tau_\times}}{1-e^{-T/\tau_\times}}\frac{T}{\tau_\times}\right) \tag{D6}$$

which reaches its maximum at $t=t_M$ given by:

$$t_M=\tau_\times-\frac{e^{-T/\tau_\times}}{1-e^{-T/\tau_\times}}T \tag{D7}$$

and its minimum at $t = 0$.

b. With offset

Similarly, Eq. (D5) is replaced by:

$$\Psi_{\tau_\times,\tau_\times,T|t_0}(t)=\frac{1}{\tau_\times}\frac{e^{-t'/\tau_\times}}{1-e^{-T/\tau_\times}}\left(\frac{t'}{\tau_\times}+\frac{e^{-T/\tau_\times}}{1-e^{-T/\tau_\times}}\frac{T}{\tau_\times}\right) \tag{D8}$$

We will not consider this case any further in the remainder, but it should be reminded that it needs to be treated separately, as formulas for $\tau\neq\tau_\times$ will not apply when $\tau=\tau_\times$.

### *D.2. Position of the maximum and minimum of the decay*

One verify easily that the maximum of the convolution of PSEDs defined by Eq. (D3) and represented in Fig. S3A for a few values of $\tau$ is located at $t_M$ given by ($\tau \neq \tau_\times$):

$$t_M = \frac{1}{1/\tau_\times - 1/\tau} \ln \frac{\tau}{\tau_\times} \frac{\left(1-e^{-T/\tau}\right)}{\left(1-e^{-T/\tau_\times}\right)} \tag{D9}$$

$t_M$ is an increasing function of $\tau$ and tends to $t_{M,\infty}$ when $\tau \to \infty$ given by:

$$t_{M,\infty} = -\tau_\times \ln\left[\frac{\tau_\times}{T}\left(1-\mathrm{e}^{-T/\tau_\times}\right)\right] \tag{D10}$$

In other words, as the lifetime of the emitting species increases, the location of the peak of the recorded signal is shifted towards larger values, but remains bounded by $t_{M,\infty}$.

The decays minimum is attained at $t = 0$ and is equal to:

$$F_T(0) = \frac{1}{\tau-\tau_\times}\left(\frac{1}{1-e^{-T/\tau}} - \frac{1}{1-e^{-T/\tau_\times}}\right) \tag{D11}$$

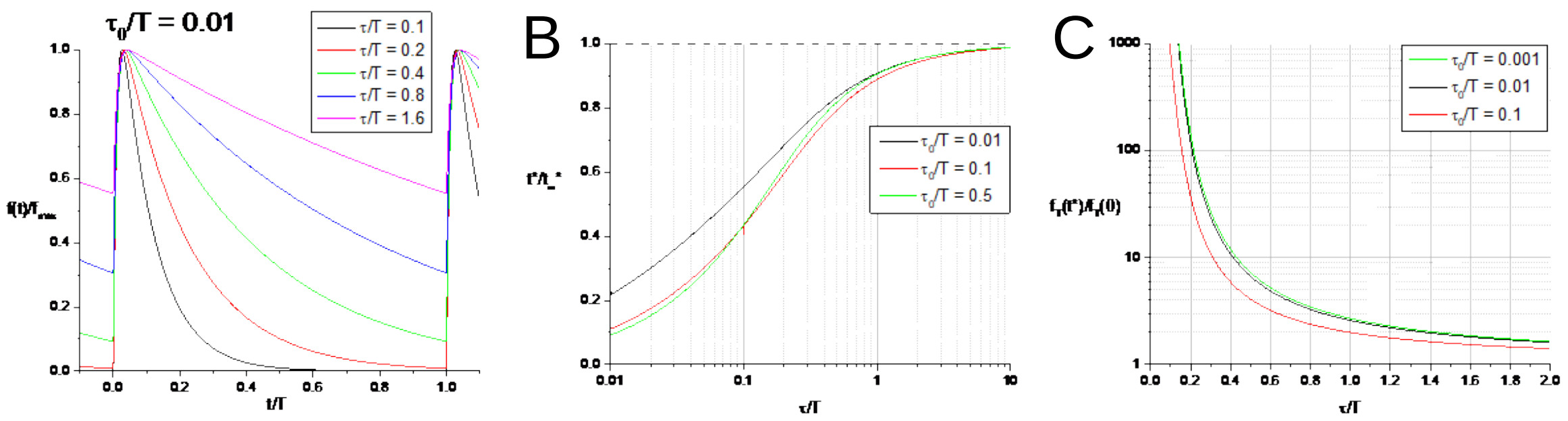

**Fig. S3**: Convolution of an exponential excitation function ($\tau_0$) with an exponential emission function ($\tau$). A: Representative curves for an excitation function characterized with a decay time $\tau_0$ equal to 1% of the period duration $T$. The decays are normalized to their maximum value, set to 1. As the lifetime $\tau$ increases, the maximum location, $t^*$, increases (see panel B), while the contrast ratio (max/min) decreases (see panel C). B: Peak location relative to the asymptotic peak location ($t^*_\infty$) as a function of lifetime $\tau$ for different values of $\tau_0/T$. C: Contrast ratio as a function of lifetime $\tau$ for different values of $\tau_0/T$. As the lifetime increases, the apparent baseline (minimum decay value) increases, making the decay look like it comprises a constant background component (see panel A).

### *D.3. Contrast ratio*

An interesting feature of convolution of PSEDs is the fact that, as $\tau$ increases, the ratio between the maximum and minimum of these functions, which we can call a *contrast ratio*, decreases:

$$\frac{\Psi_{\tau,\tau_\times,T}(t_M)}{\Psi_{\tau,\tau_\times,T}(0)} = \frac{\left(1-e^{-T/\tau_\times}\right)e^{-t_M/\tau} - \left(1-e^{-T/\tau}\right)e^{-t_M/\tau_\times}}{e^{-T/\tau} - e^{-T/\tau_\times}} \tag{D12}$$

This property is clearly visible in Fig. S3 in which different versions of Eq. (D12) **Error! Reference source not found.** for a fixed ratio $\tau_\times / T$ and variable lifetime $\tau$, normalized to their maximum, are represented. The minimum of the decays with larger lifetimes are getting closer to the maximum and away from 0, as $\tau$ increases. This is in the absence of additional background, and therefore shows that there is information in the minimum of a periodic decay function.

### *D.4. Continuous phasor of the decay*

*D.4.1. No offset*

Plugging in Eq. (D3)**Error! Reference source not found.** in Eq. (63) **Error! Reference source not found.**, we obtain:

$$\begin{cases} \left\| \Psi_{\tau,\tau_\times,T}(t) e^{i2\pi ft} \right\|_T = \dfrac{1}{1-i2\pi f\tau} \dfrac{1}{1-i2\pi f\tau_\times} \\ \left\| \Psi_{\tau,\tau_\times,T}(t) \right\|_T = 1 \end{cases} \tag{D13}$$

from which is results that the continuous phasor of $F_T(t)$ is given by:

$$z\left[\Psi_{\tau,\tau_\times,T}\right] = \frac{1}{1-i2\pi f\tau} \frac{1}{1-i2\pi f\tau_\times} = \zeta_f(\tau_\times)\zeta_f(\tau) \tag{D14}$$

which also follows from the continuous phasor convolution rule (Eq. (69)) and the definition of $\Psi_{\tau,\tau_\times,T}(t)$ as a convolution product (Eq. (D3)).

As a result, the locus of continuous phasors of convolution of PSEDs is a scaled and rotated semicircle, the scaling factor and rotation angle being given by (Eq. (71)):

$$\begin{cases} m(\tau_\times) = \dfrac{1}{\sqrt{1+(2\pi f\tau_\times)^2}} \\ \varphi(\tau_\times) = \tan^{-1}(2\pi f\tau_\times) \end{cases} \tag{D15}$$

*D.4.2. With offset*

It is straightforward to verify that, in the presence of an offset $t_0$, Eq. (D14) is replaced by:

$$z\left[\Psi_{\tau,\tau_\times,T|t_0}\right] = \frac{1}{1-i2\pi f\tau} \frac{1}{1-i2\pi f\tau_\times} e^{i2\pi ft_0} = \zeta_f(\tau_\times)\zeta_f(\tau) e^{i2\pi ft_0} \tag{D16}$$

which allows writing:

$$z\left[\Psi_{\tau,\tau_\times,T|t_0}\right] = z\left[\Lambda_{\tau_\times,T|t_0}\right]\zeta_f(\tau) \tag{D17}$$

where $z\left[\Lambda_{\tau_\times,T|t_0}\right]$, the calibration phasor, is given by Eq. (126).

As a result, the locus of continuous phasors of convolution of PSEDs with offset is also a scaled and rotated semicircle, the scaling factor and rotation angle being given by:

$$\begin{cases} m(\tau_\times, t_0) = \dfrac{1}{\sqrt{1+(2\pi f \tau_\times)^2}} \\ \varphi(\tau_\times, t_0) = \tan^{-1}(2\pi f \tau_\times) + 2\pi f t_0 \end{cases} \tag{D18}$$

## *D.5. Square-gated periodic single-exponential decay*

### *D.5.1. No offset*

Using Eq. (D3) for $\Psi_{\tau,\tau_\times,T}(t)$, the signal accumulated during a square gate of width $W$ is given by Eq. (33) with $\Gamma_{s,W,nT}(t)$ replaced by the boxcar function $\Pi_{s,W}(t)$ (Eq. (29)):

$$\begin{aligned} \Psi_{\tau,\tau_\times,T,W}(s) &= \int_s^{s+W} dt \Psi_{\tau,\tau_\times,T}(t) \Pi_{s,W}(t) = \int_s^{s+W} dt \Pi_{s,W}(t) \frac{\tau \Lambda_{\tau,T}(t) - \tau_\times \Lambda_{\tau_\times,T}(t)}{\tau - \tau_\times} \\ &= \frac{1}{\tau - \tau_\times}\left(\tau \Lambda_{\tau,T,W}(s) - \tau_\times \Lambda_{\tau_\times,T,W}(s)\right) \end{aligned} \tag{D19}$$

where we have used the notation $\Lambda_{\tau,T,W}(t)$ of Section 2.2.3 for the square-gated integral of $\Lambda_{\tau,T}(t)$. Using the results of Eq. (47), we obtain the explicit formula:

$$\Psi_{\tau,\tau_\times,T,W}(t) = \begin{cases} \dfrac{1}{\tau-\tau_\times}\left(\tau \dfrac{1-u}{1-y} e^{-t'/\tau} - \tau_\times \dfrac{1-u_\times}{1-y_\times} e^{-t'/\tau_\times}\right) + k, & \text{if } t' \in [0, T-\omega[ \\ \dfrac{1}{\tau-\tau_\times}\left(\tau \dfrac{1-uy^{-1}}{1-y} e^{-t'/\tau} - \tau_\times \dfrac{1-u_\times y_\times^{-1}}{1-y_\times} e^{-t'/\tau_\times}\right) + k+1, & \text{if } t' \in [T-\omega, T[ \end{cases}$$

$$\text{where} \quad \begin{cases} k = \lfloor W/T \rfloor \\ \omega = W[T] = W - kT \\ t' = t[T] = t - \lfloor t/T \rfloor T \\ y = e^{-T/\tau};\ y_\times = e^{-T/\tau_\times};\ u = e^{-\omega/\tau};\ u_\times = e^{-\omega/\tau_\times} \end{cases} \tag{D20}$$

### *D.5.2. With offset*

As in the case of the ungated decay, the expression for the square-gated decay in the presence of offset is obtained by a simple replacement $t \mapsto t - t_0$ in the previous expression (Eq. (D20)):

$$\Psi_{\tau,\tau_\times,T,W|t_0}(t) = \Psi_{\tau,\tau_\times,T,W}(t - t_0) \tag{D21}$$

### *D.6. Continuous phasor of the square-gated decay*

*D.6.1. No offset*

The continuous phasor of $\Psi_{\tau,\tau_\times,T,W}(t)$ given by Eq. (D19) is given by definition by (Eq. (63)):

$$z\left[\Psi_{\tau,\tau_\times,T,W}\right]=\frac{\left\|\Psi_{\tau,\tau_\times,T,W}(t)e^{i2\pi ft}\right\|_T}{\left\|\Psi_{\tau,\tau_\times,T,W}(t)\right\|_T} \tag{D22}$$

Using Eq. (D19) and the fact (used to derive Eq. (75)) that:

$$\begin{cases}\left\|\Lambda_{\tau,T,W}(t)\right\|_T=W\\ \left\|\Lambda_{\tau,T,W}(t)e^{i2\pi ft}\right\|_T=WM_We^{-i\varphi_W}\zeta_f(\tau)\end{cases} \tag{D23}$$

where $M_W$ and $\varphi_W$ where defined in Eq. (75), we obtain:

$$z\left[\Psi_{\tau,\tau_\times,T,W}\right]=z\left[\overline{\Pi}_{W,nT}\right]\zeta_f(\tau)\zeta_f(\tau_\times)=z_{[W]}\left[\Lambda_{\tau_\times,T}\right]\zeta_f(\tau) \tag{D24}$$

$z_{[W]}\left[\Lambda_{\tau_\times,T}\right]=z\left[I_{T,W}\right]$ is the continuous phasor of the square-gated instrument response function (calibration phasor), while $\zeta_f(\tau)=z\left[\Lambda_{\tau,T}\right]$ is the continuous phasor of the samples response to a periodic Dirac IRF. We thus recover by a direct calculation, that the locus of phasors of square-gated PSEDs convolved with an exponential IRF is a semicircle rotated by an angle $\varphi_\times-\varphi_W=\tan^{-1}(2\pi f\tau_\times)-\pi fW$ and dilated by a factor $m_*M_W$ (given in Eqs. (74) & (75)).

*D.6.2. With offset*

To compute the continuous phasor of $\Psi_{\tau,\tau_\times,T,W|t_0}(t)$, it is easiest to start from $\Psi_{\tau,\tau_\times,T|t_0}(t)$ (Eq. (D5)) and write:

$$\begin{aligned}\Psi_{\tau,\tau_\times,T,W|t_0}(s)&=\int_s^{s+W}dt\Psi_{\tau,\tau_\times,T|t_0}(t)\Pi_{s,W}(t)=\int_s^{s+W}dt\Pi_{s,W}(t)\frac{\tau\Lambda_{\tau,T|t_0}(t)-\tau_\times\Lambda_{\tau_\times,T|t_0}(t)}{\tau-\tau_\times}\\&=\frac{1}{\tau-\tau_\times}\left(\tau\Lambda_{\tau,T,W|t_0}(s)-\tau_\times\Lambda_{\tau_\times,T,W|t_0}(s)\right)\end{aligned} \tag{D25}$$

The phasor:

$$z\left[\Psi_{\tau,\tau_\times,T,W|t_0}\right]=\frac{\left\|\Psi_{\tau,\tau_\times,T,W|t_0}(t)e^{i2\pi ft}\right\|_T}{\left\|\Psi_{\tau,\tau_\times,T,W|t_0}(t)\right\|_T} \tag{D26}$$

involves two quantities that are easily obtained from previous results (Eqs. (A5) & (A10)):

$$\left\|\Psi_{\tau,\tau_\times,T,W|t_0}(t)\right\|_T = \frac{1}{\tau-\tau_\times}\left(\tau\left\|\Lambda_{\tau,T,W|t_0}(t)\right\|_T - \tau_\times\left\|\Lambda_{\tau_\times,T,W|t_0}(t)\right\|_T\right) = W$$

$$\left\|\Psi_{\tau,\tau_\times,T,W|t_0}(t)e^{i2\pi ft}\right\|_T = \frac{1}{\tau-\tau_\times}\left(\tau\left\|\Lambda_{\tau,T,W|t_0}(t)e^{i2\pi ft}\right\|_T - \tau_\times\left\|\Lambda_{\tau_\times,T,W|t_0}(t)e^{i2\pi ft}\right\|_T\right) \tag{D27}$$

$$= \frac{\sin\pi f\omega}{\pi f}e^{-i\pi f\omega}\left(\frac{\tau}{1-i2\pi f\tau} - \frac{\tau_\times}{1-i2\pi f\tau_\times}\right)e^{i2\pi ft_0}$$

The final result can be written in any of the following equivalent forms:

$$\begin{aligned} z\left[\Psi_{\tau,\tau_\times,T,W,t_0}\right] &= z\left[\overline{\Pi}_{W,nT}\right]\zeta_f(\tau)\zeta_f(\tau_\times)e^{i2\pi ft_0} \\ &= z_{[W]}\left[\Lambda_{\tau_\times,T}\right]\zeta_f(\tau)e^{i2\pi ft_0} \\ &= z_{[W]}\left[\Lambda_{\tau_\times,T|t_0}\right]\zeta_f(\tau) \\ &= z\left[\Psi_{\tau,\tau_\times,T,W}\right]e^{i2\pi ft_0} \end{aligned} \tag{D28}$$

Which shows that the locus of phasors of square-gated PSEDs convolved with a single-exponential IRF with offset is a semicircle rotated by an angle $\varphi_* - \varphi_W + 2\pi ft_0$ and dilated by a factor $m_* M_W$ (given in Eqs. (74) & (75)).

## *D.7. Discrete phasor of the ungated decay*

### *D.7.1. No offset*

The discrete phasor of $\Psi_{\tau,\tau_\times,T}(t)$ given by Eq. (D3) is obtained from definition (92):

$$z_N\left[\Psi_{\tau,\tau_\times,T}\right] = \frac{\left\|\Psi_{\tau,\tau_\times,T}(t_p)e^{i2\pi ft_p}\right\|_N}{\left\|\Psi_{\tau,\tau_\times,T}(t_p)\right\|_N} \tag{D29}$$

where $t_p = (p-1)\theta, 1 \le p \le N$ and we assume $T = N\theta$.

Using the results of Eq. (B1), we obtain:

$$\begin{cases} \left\|\Psi_{\tau,\tau_\times,T}(t_p)\right\|_N = \dfrac{1}{(\tau-\tau_\times)}\left(\tau\left\|\Lambda_{\tau,T}(t_p)\right\|_N - \tau_\times\left\|\Lambda_{\tau_\times,T}(t_p)\right\|_N\right) \\ \qquad = \dfrac{\theta}{(\tau-\tau_\times)}\left(\dfrac{1}{1-e^{-\theta/\tau}} - \dfrac{1}{1-e^{-\theta/\tau_\times}}\right) = \theta\dfrac{x-x_\times}{\tau-\tau_\times}\dfrac{1}{1-x}\dfrac{1}{1-x_\times} \\ \left\|\Psi_{\tau,\tau_\times,T}(t_p)e^{i2\pi ft_p}\right\|_N = \dfrac{1}{(\tau-\tau_\times)}\left(\tau\left\|\Lambda_{\tau,T}(t_p)e^{i2\pi ft_p}\right\|_N - \tau_\times\left\|\Lambda_{\tau_\times,T}(t_p)e^{i2\pi ft_p}\right\|_N\right) \\ \qquad = \dfrac{\theta}{(\tau-\tau_\times)}\left(\dfrac{1}{1-e^{(-1/\tau+i2\pi f)\theta}} - \dfrac{1}{1-e^{(-1/\tau_\times+i2\pi f)\theta}}\right) \\ \qquad = \theta e^{i\alpha}\dfrac{x-x_\times}{\tau-\tau_\times}\dfrac{1}{1-xe^{i\alpha}}\dfrac{1}{1-x_\times e^{i\alpha}} \end{cases} \tag{D30}$$

extending the notations of Eq. (B3) to encompass the terms involving $\tau_\times$. We get the final result:

$$\begin{cases} z_N\left[\Psi_{\tau,\tau_\times,T}\right] = \dfrac{1-x}{1-xe^{i\alpha}}\dfrac{1-x_\times}{1-x_\times e^{i\alpha}}e^{i\alpha} \\ x = x(\tau) = e^{-\theta/\tau};\ x_\times = x(\tau_\times) = e^{-\theta/\tau_\times};\ \alpha = 2\pi f\theta \end{cases} \tag{D31}$$

We can rewrite Eq. (D31) as:

$$z_N\left[\Psi_{\tau,\tau_\times,T}\right] = e^{i\alpha} z_N\left[\Lambda_{\tau,T}\right] z_N\left[\Lambda_{\tau_\times,T}\right] \tag{D32}$$

This identity states that the *discrete* phasor of the *continuous* cyclic convolution product of two single-exponential functions is *distinct* from the product of their individual *discrete* phasors but can be rewritten in the special form of Eq. (C21) (weak discrete phasor convolution rule).

*Note*: It will not escape the reader's attention that the limit when $\tau_\times \to 0$ of Eq. (D32) is:

$$\lim_{\tau_\times\to 0} z_N\left[\Psi_{\tau,\tau_\times,T}\right] = e^{i2\pi f\theta} z_N\left[\Lambda_{\tau,T}\right] \lim_{\tau_\times\to 0} z_N\left[\Lambda_{\tau_\times,T}\right] = e^{i2\pi f\theta} z_N\left[\Lambda_{\tau,T}\right] = e^{i2\pi f\theta}\zeta_{f,N}(\tau) \tag{D33}$$

However,

$$\lim_{\tau_\times\to 0} \Psi_{\tau,\tau_\times,T}(t) = \Lambda_{\tau,T}(t) \tag{D34}$$

which would suggest that:

$$\lim_{\tau_\times\to 0} z_N\left[\Psi_{\tau,\tau_\times,T}\right] = z_N\left[\lim_{\tau_\times\to 0}\Psi_{\tau,\tau_\times,T}\right] = z_N\left[\Lambda_{\tau,T}\right] = \zeta_{f,N}(\tau) \tag{D35}$$

Comparison of Eqs. (D33) and (D35) shows an additional phase term in the former. Its origin comes from the way $z_N\left[\Psi_{\tau,\tau_\times,T}\right]$ is calculated (see Eq. (D30) assuming $\tau_\times \neq 0$. Indeed this assumption results in the first term of the sum in $\left\|\Lambda_{\tau_*,T}\left(t_p\right)\right\|_N$ (resp. $\left\|\Lambda_{\tau_\times,T}\left(t_p\right)e^{i2\pi f t_p}\right\|_N$) to be equal to 1, because $t_1 = 0 \Rightarrow e^{-t_1/\tau_\times} = 1$. The remaining terms do tend to 0 when $\tau_\times \to 0$, but that single 1 value remains in the limit. In other words:

$$\lim_{\tau_\times\to 0} z_N\left[\Psi_{\tau,\tau_\times,T}\right] \neq z_N\left[\lim_{\tau_\times\to 0}\Psi_{\tau,\tau_\times,T}\right] \tag{D36}$$

which is due to the fact that $\Psi_{\tau,\tau_\times,T}(t)$ is a function with a value of 0 for $t=0$, except for $\tau_\times = 0$, in which case $\Psi_{\tau,0,T}(0) = \Lambda_{\tau,T}(0) = \tau^{-1}\left(1-e^{-T/\tau}\right)^{-1}$.

*D.7.2. With offset*

In the presence of offset, we start from Eq. (D5) for $\Psi_{\tau,\tau_\times,T|t_0}(t)$ and need to compute:

$$z_N\left[\Psi_{\tau,\tau_\times,T|t_0}\right] = \frac{\left\|\Psi_{\tau,\tau_\times,T|t_0}\left(t_p\right)e^{i2\pi f t_p}\right\|_N}{\left\|\Psi_{\tau,\tau_\times,T|t_0}\left(t_p\right)\right\|_N} \tag{D37}$$

The two quantities in Eq. (D37) are easily obtained from previous results for the discrete phasor of ungated PSEDs with offset (Appendix B.1.2):

$$\begin{aligned}\left\|\Psi_{\tau,\tau_\times,T|t_0}\left(t_p\right)\right\|_N &= \frac{1}{\tau-\tau_\times}\left(\tau\left\|\Lambda_{\tau,T|t_0}\left(t_p\right)\right\|_N - \tau_\times\left\|\Lambda_{\tau_\times,T|t_0}\left(t_p\right)\right\|_N\right)\\ &= \frac{\theta}{\tau-\tau_\times}\left(\frac{x^q e^{t_0/\tau}}{1-x} - \frac{x_\times^q e^{t_0/\tau_\times}}{1-x_\times}\right)\end{aligned} \quad \text{(D38)}$$

and similarly:

$$\begin{aligned}\left\|\Psi_{\tau,\tau_\times,T|t_0}\left(t_p\right)e^{i2\pi f t_p}\right\|_N &= \frac{1}{\tau-\tau_\times}\left(\tau\left\|\Lambda_{\tau,T|t_0}\left(t_p\right)e^{i2\pi f t_p}\right\|_N - \tau_\times\left\|\Lambda_{\tau_\times,T|t_0}\left(t_p\right)e^{i2\pi f t_p}\right\|_N\right)\\ &= \frac{\theta}{\tau-\tau_\times}\left(\frac{x^q e^{t_0/\tau}}{1-xe^{i\alpha}} - \frac{x_\times^q e^{t_0/\tau_\times}}{1-x_\times e^{i\alpha}}\right)e^{iq\alpha}\end{aligned} \quad \text{(D39)}$$

where $q = \left\lceil \frac{t_0}{\theta} \right\rceil$ (Eq. (B17)). We obtain:

$$\begin{cases} z_N\left[\Psi_{\tau,\tau_\times,T|t_0}\right] = z_N\left[\Lambda_{\tau,T}\right] z_N\left[\Lambda_{\tau_\times,T}\right]\Omega\left(\tau,\tau_\times,t_0\right)e^{iq\alpha} \\ \Omega\left(\tau,\tau_\times,t_0\right) = \dfrac{x^q e^{t_0/\tau}\left(1-x_\times e^{i\alpha}\right) - x_\times^q e^{t_0/\tau_\times}\left(1-xe^{i\alpha}\right)}{x^q e^{t_0/\tau}\left(1-x_\times\right) - x_\times^q e^{t_0/\tau_\times}\left(1-x\right)} \end{cases} \quad \text{(D40)}$$

which once again shows that the discrete phasor of the *continuous* cyclic convolution of two *T*-periodic functions is in general different from the product of their discrete phasors.

Note however that if $t_0 = q\theta$ (*i.e.* the offset corresponds to the start of a gate), the expression of factor *Ω* in Eq. (D40) simplifies into:

$$t_0 = q\theta \Rightarrow \Omega\left(\tau,\tau_\times,t_0\right) = e^{i\alpha} \quad \text{(D41)}$$

which yield the following simple result:

$$\begin{aligned} t_0 = q\theta \Rightarrow z_N\left[\Psi_{\tau,\tau_\times,T|t_0}\right] &= z_N\left[\Lambda_{\tau,T}\right] z_N\left[\Lambda_{\tau_\times,T}\right] e^{i(q+1)\alpha}\\ &= z_N\left[\Lambda_{\tau,T|t_0}\right] z_N\left[\Lambda_{\tau_\times,T}\right] e^{i\alpha}\\ &= z_N\left[\Lambda_{\tau,T}\right] z_N\left[\Lambda_{\tau_\times,T|t_0}\right] e^{i\alpha}\end{aligned} \quad \text{(D42)}$$

This result shows that when the decay offset corresponds to the start of one of the gates, the resulting discrete phasor is obtained as the product of the discrete phasor of the PSED, $z_N\left[\Lambda_{\tau,T}\right]$, and the discrete phasor of the 'instrument response function', $z_N\left[\Lambda_{\tau_\times,T|t_0}\right]$ multiplied by a constant, $\kappa = e^{i\alpha}$ (weak discrete phasor convolution rule). As a consequence, 'standard' calibration of the phasor of this type of decays will work as intended.

In the general case, the values of factor *Ω* in Eq. (D40) when $\tau \to 0$ and $\tau \longrightarrow \infty$ are given by:

$$\begin{cases} \Omega\left(0,\tau_\times,t_0\right) = 1 \\ \Omega\left(\infty,\tau_\times,t_0\right) = \dfrac{\left(1-x_\times e^{i\alpha}\right) - x_\times^q e^{t_0/\tau_\times}\left(1-e^{i\alpha}\right)}{1-x_\times} \end{cases} \quad \text{(D43)}$$

### *D.8. Discrete phasor of the square-gated decay*

*D.8.1. No offset*

The discrete phasor of $\Psi_{\tau,\tau_\times,T,W}(t)$ is given by:

$$z_{N[W]}\left[\Psi_{\tau,\tau_\times,T}\right] \triangleq z_N\left[\Psi_{\tau,\tau_\times,T,W}\right] = \frac{\left\|\Psi_{\tau,\tau_\times,T,W}\left(t_p\right)e^{i2\pi f t_p}\right\|_N}{\left\|\Psi_{\tau,\tau_\times,T,W}\left(t_p\right)\right\|_N} \tag{D44}$$

The two terms involved in this expression can be calculated based on the definition of $\Psi_{\tau,\tau_\times,T,W}(t)$ (Eq. (D19)):

$$\begin{cases} \left\|\Psi_{\tau,\tau_\times,T,W}\left(t_p\right)\right\|_N = \dfrac{1}{\tau-\tau_\times}\left(\tau\left\|\Lambda_{\tau,T,W}\left(t_p\right)\right\|_N - \tau_\times\left\|\Lambda_{\tau_\times,T,W}\left(t_p\right)\right\|_N\right) \\ \left\|\Psi_{\tau,\tau_\times,T,W}\left(t_p\right)e^{i2\pi f t_p}\right\|_N = \dfrac{1}{\tau-\tau_\times}\left(\tau\left\|\Lambda_{\tau,T,W}\left(t_p\right)e^{i2\pi f t_p}\right\|_N - \tau_\times\left\|\Lambda_{\tau_\times,T,W}\left(t_p\right)e^{i2\pi f t_p}\right\|_N\right) \end{cases} \tag{D45}$$

The necessary expressions have been calculated in Section B.2 (Eqs. (B32) & (B36)). We obtain:

$$\begin{cases} \left\|\Psi_{\tau,\tau_\times,T,W}\left(t_p\right)\right\|_N = \theta\left\{(k+1)N - r + \dfrac{1}{\tau-\tau_\times}\left(\tau\dfrac{1-\beta}{1-x} - \tau_\times\dfrac{1-\beta_\times}{1-x_\times}\right)\right\} \\ \left\|\Psi_{\tau,\tau_\times,T,W}\left(t_p\right)e^{i2\pi f t_p}\right\|_N = \theta\left\{-\dfrac{1-e^{ir\alpha}}{1-e^{i\alpha}} + \dfrac{1}{\tau-\tau_\times}\left(\tau\dfrac{1-\beta e^{ir\alpha}}{1-xe^{i\alpha}} - \tau_\times\dfrac{1-\beta_\times e^{ir\alpha}}{1-x_\times e^{i\alpha}}\right)\right\} \\ r = \left\lceil\dfrac{T-\omega}{\theta}\right\rceil \\ y = e^{-T/\tau};\ y_\times = e^{-T/\tau_\times};\ u = e^{-\omega/\tau};\ u_\times = e^{-\omega/\tau_\times};\ \beta = u\ x^r y^{-1};\quad \beta_\times = u_\times x_\times^r y_\times^{-1} \end{cases} \tag{D46}$$

which leads to the following expression for $z_{N[W]}\left[\Psi_{\tau,\tau_\times,T}\right]$:

$$\begin{cases} z_{N[W]}\left[\Psi_{\tau,\tau_\times,T}\right] = \dfrac{-\dfrac{1-e^{ir\alpha}}{1-e^{i\alpha}} + \dfrac{1}{\tau-\tau_\times}\left(\tau\dfrac{1-\beta e^{ir\alpha}}{1-xe^{i\alpha}} - \tau_\times\dfrac{1-\beta_\times e^{ir\alpha}}{1-x_\times e^{i\alpha}}\right)}{(k+1)N - r + \dfrac{1}{\tau-\tau_\times}\left(\tau\dfrac{1-\beta}{1-x} - \tau_\times\dfrac{1-\beta_\times}{1-x_\times}\right)} \\ k = \left\lfloor\dfrac{W}{T}\right\rfloor;\ r = \left\lceil\dfrac{T-\omega}{\theta}\right\rceil \\ y = e^{-T/\tau};\ y_\times = e^{-T/\tau_\times};\ u = e^{-\omega/\tau};\ u_\times = e^{-\omega/\tau_\times};\ \beta = u\ x^r y^{-1};\quad \beta_\times = u_\times x_\times^r y_\times^{-1} \end{cases} \tag{D47}$$

The phasor values for $\tau = 0$ and $\tau \to \infty$ are given by:

$$
\begin{cases}
z_{N[W]}\left[\Psi_{0,\tau_\times,T}\right]=\dfrac{-\dfrac{1-e^{ir\alpha}}{1-e^{i\alpha}}+\dfrac{1-\beta_\times e^{ir\alpha}}{1-x_\times e^{i\alpha}}}{\left(k+1\right)N-r+\dfrac{1-\beta_\times}{1-x_\times}} \\
z_{N[W]}\left[\Psi_{\infty,\tau_\times,T}\right]=0
\end{cases}
\tag{D48}
$$

In the special case where $W$ is a multiple of $\theta$, we have seen that $\beta=\beta_\times=1$, which leads to some simplifications for Eq. (D47):

$$
z_{N[W]}\left[\Psi_{\tau,\tau_\times,T}\right]=-\frac{1}{\left(k+1\right)N-r}\frac{1-e^{ir\alpha}}{1-e^{i\alpha}}\left[1-\frac{1-e^{i\alpha}}{\tau-\tau_\times}\left(\frac{\tau}{1-xe^{i\alpha}}-\frac{\tau_\times}{1-x_\times e^{i\alpha}}\right)\right]
\tag{D49}
$$

where we recognize the phasor $z_N\left[\overline{\Pi}_{W,nT}\right]$ (Eq. (C36)) as a prefactor, but the remainder of the expression is not factorizable with $z_N\left[\Psi_{\tau,\tau_\times,T}\right]$ (Eq. (D32)):

$$
z_{N[W]}\left[\Psi_{\tau,\tau_\times,T}\right]=z_N\left[\overline{\Pi}_{W,nT}\right]\left[1-\frac{1-e^{i\alpha}}{\tau-\tau_\times}\left(\frac{\tau}{1-xe^{i\alpha}}-\frac{\tau_\times}{1-x_\times e^{i\alpha}}\right)\right]
\tag{D50}
$$

*D.8.2. With offset*

The discrete phasor of $\Psi_{\tau,\tau_\times,T,W|t_0}\left(t\right)$ is given by:

$$
z_{N[W]}\left[\Psi_{\tau,\tau_\times,T|t_0}\right]\triangleq z_N\left[\Psi_{\tau,\tau_\times,T,W|t_0}\right]=\frac{\left\langle\Psi_{\tau,\tau_\times,T,W|t_0}\left(t_p\right)e^{i2\pi ft_p}\right\rangle_N}{\left\langle\Psi_{\tau,\tau_\times,T,W|t_0}\left(t_p\right)\right\rangle_N}
\tag{D51}
$$

The two terms involved in this expression can be computed based on the definition of $\Psi_{\tau,\tau_\times,T,W|t_0}\left(t\right)$ (Eq. (D21)):

$$
\begin{cases}
\left\|\Psi_{\tau,\tau_\times,T,W|t_0}\left(t_p\right)\right\|_N=\dfrac{1}{\tau-\tau_\times}\left(\tau\left\|\Lambda_{\tau,T,W|t_0}\left(t_p\right)\right\|_N-\tau_\times\left\|\Lambda_{\tau_\times,T,W|t_0}\left(t_p\right)\right\|_N\right) \\
\left\|\Psi_{\tau,\tau_\times,T,W|t_0}\left(t_p\right)e^{i2\pi ft_p}\right\|_N=\dfrac{1}{\tau-\tau_\times}\left(\tau\left\|\Lambda_{\tau,T,W|t_0}\left(t_p\right)e^{i2\pi ft_p}\right\|_N-\tau_\times\left\|\Lambda_{\tau_\times,T,W|t_0}\left(t_p\right)e^{i2\pi ft_p}\right\|_N\right)
\end{cases}
\tag{D52}
$$

The necessary expressions have been calculated in Section B.2 (Eq. (B60)). We obtain:

$$
\begin{cases}
z_{N[W]}\left[\Psi_{\tau,\tau_\times,T|t_0}\right]=\dfrac{\dfrac{e^{ir\alpha}-e^{iq\alpha}}{1-e^{i\alpha}}+\dfrac{1}{\tau-\tau_{*\times}}\left(\tau e^{t_0/\tau}\dfrac{x^qe^{iq\alpha}-ux^re^{ir\alpha}}{1-xe^{i\alpha}}-\tau_\times e^{t_0/\tau_\times}\dfrac{x_\times^qe^{iq\alpha}-u_\times x_\times^re^{ir\alpha}}{1-x_\times e^{i\alpha}}\right)}{kN+q-r+\dfrac{1}{\tau-\tau_\times}\left(\tau e^{t_0/\tau}\dfrac{x^q-ux^r}{1-x}-\tau_\times e^{t_0/\tau_\times}\dfrac{x_\times^q-u_\times x_\times^r}{1-x_\times}\right)} \\
q=\left\lceil\dfrac{t_0}{\theta}\right\rceil;\ r=\left\lceil\dfrac{t_0-\omega}{\theta}\right\rceil \\
y=y\left(\tau\right)=e^{-T/\tau};\ y_\times=y\left(\tau_\times\right)=e^{-T/\tau_\times};\ u=u\left(\tau\right)=e^{-\omega/\tau};\ u_\times=u\left(\tau_\times\right)=e^{-\omega/\tau_\times}
\end{cases}
\tag{D53}
$$

Notice that there are several terms involving the offset $t_0$ in this expression, preventing any kind of simplification.

The phasor values for $\tau = 0$ and $\tau \to \infty$ are given by:

$$\begin{cases} z_{N[W]}\left[\Psi_{0,\tau_\times,T|t_0}\right] = \dfrac{\dfrac{e^{ir\alpha} - e^{iq\alpha}}{1 - e^{i\alpha}} + e^{t_0/\tau_\times}\dfrac{x_\times^q e^{iq\alpha} - u_\times x_\times^r e^{ir\alpha}}{1 - x_\times e^{i\alpha}}}{kN + q - r + e^{t_0/\tau_\times}\dfrac{x_*^q - u_\times x_\times^r}{1 - x_\times}} \\ z_{N[W]}\left[\Psi_{\infty,\tau_\times,T|t_0}\right] = 0 \end{cases} \tag{D54}$$

**Appendix E: SEPL computation from an experimental IRF curve**

As discussed in Section 8.3.5, in the general experimental case, the exact functional form of the IRF might not be known, or if known, not match any of the examples detailed in this article. In these cases, it may be useful to compute pseudo-calibrated phasors as discussed in that section, and relate them to the loci of PSEDs. By definition, the pseudo-calibration factor is given by the computed phasor (which will assume to be a discrete phasor, since it is an experimental one), Eq. (88):

$$z_N\left[I_T \underset{T}{*} \Lambda_{0,T}\right] = z_N\left[I_T\right] = \frac{\left\| I_T\left(t_p\right) e^{i2\pi f t_p} \right\|_N}{\left\| I_T\left(t_p\right) \right\|_N} \tag{E1}$$

The phasors of PSEDs with nonzero lifetimes are then given by:

$$z_N\left[I_T \underset{T}{*} \Lambda_{\tau,T}\right] = \frac{\left\| I_T \underset{T}{*} \Lambda_{\tau,T}\left(t_p\right) e^{i2\pi f t_p} \right\|_N}{\left\| I_T \underset{T}{*} \Lambda_{\tau,T}\left(t_p\right) \right\|_N} \tag{E2}$$

where the continuous cyclic convolution product involved in Eq. (E2) is formally given by:

$$I_T \underset{T}{*} \Lambda_{\tau,T}\left(t_p\right) = \int_0^T du\, I_T(u)\, \Lambda_{\tau,T}\left(t_p - u\right) \tag{E3}$$

This integral requires knowledge of $I_T(t)$ over the whole interval [0, *T*], but in practice, only a finite number of values $\left\{I_T\left(t_p\right)\right\}$, $p = 1, \ldots, N$ are known. $\Lambda_{\tau,T}(t)$ given by Eq. (17) can be computed for any value of *t*.

Although this lack of information is preventing an exact calculation of Eq. (E3), any numerical method to approximate this integral can be used instead. For instance, the Phasor Explorer software uses a Lobatto quadrature algorithm when choosing an adaptive integration option or a simple trapezoidal rule when using a fixed number of integration steps to compute the integral.

Computation of Eq. (E2) for a sufficient number of PSEDs (for instance the default in Phasor Explorer is to compute 1,000 PSED phasors with lifetimes logarithmically spaced between 1 ps and 1 µs) will provide a good representation of the SEPL.

## Appendix F: Phasor Explorer and other software

### *F.1. Overview*

This Appendix provides a brief overview of the Phasor Explorer software capabilities. Most calculations reported here were performed with it. A software installer can be found at https://sites.google.com/a/g.ucla.edu/phasor-explorer/. A detailed online manual is available on that website, as well as download and installation instructions. The LabVIEW source code is available on Github at https://doi.org/10.5281/zenodo.3884101 (reference number [38]).

Some of the curves presented in the figures were computed with *AlliGator*, a free Microsoft Windows 64 bit executable available at (https://sites.google.com/a/g.ucla.edu/alligator/) [15] together with an extensive online manual.

Publication quality graphs were generated with OriginPro 9.1 (OriginLab, Northampton, MA). The corresponding .opj file can be found in the Figshare repository mentioned in the Data Availability section (https://doi.org/10.6084/m9.figshare.11653182, reference number [37]) and opened with the free OriginViewer software (https://www.originlab.com/viewer/), which allows exporting data and figures.

The user interface consists in a main window with a top menu bar (File, Edit, Analysis, Windows). The user interface elements (graphs, buttons, numeric controls and indicators, etc.) displayed on the window can be modified with the help of a pull-down menu below the menu bar, offering the following options: ‘Phasor Plot’, ‘Pseudo Phasor Lifetime’, and ‘Decay, Gate & Gated Decay’. They will be briefly discussed next.

### *F.2. Windows*

In addition to the main window, the Windows menu gives access to 4 other windows: Notebook, Settings, Context Help and About.

#### *F.2.1. Phasor Explorer main window*

As mentioned earlier, the main windows appearance depends on the selected pull-down menu option located below the menu bar. Generally, it shows one graph and a few controls (numeric controls or indicators, check boxes). Each graph is a rich LabVIEW object allowing many interactions with each individual or groups of plots, including saving or loading them as ASCII files. Right-clicking a graph and choosing ‘Copy Data’ will copy a bitmap image of the Graph, which can then be pasted into the Notebook (or in another rich text format document) for documentation. Because of the finite resolution of such an image, it is generally recommended to save the raw data contained in each Graph (or individual plots) as ASCII files, and re-open them in a dedicated data analysis and visualization software for publication quality results (OriginPro was used for this purpose in this article).

### *F.2.2. Notebook*

The Notebook window logs all user actions and allows user-typed notes to be recorded, even if the window is closed. Upon quitting the software, a reminder will be displayed, asking users whether or not they want to save the Notebook before quitting. This is the recommended choice, as it contains a lot of useful information which would be otherwise difficult if not impossible to recover from saved plot files, for instance. Graphs can be copied from the main window to the Notebook as bitmap. The Notebooks content can be saved at all times as a rich text format (RTF) file, which can then be opened by most text editors (as well as Phasor Explorer).

### *F.2.3. Settings*

The Settings window gives access to all parameters/options controlling the different types of calculations: Display, Phasor, Decay, Gates, Detector. Once options and parameters have been set in the Settings window, calculations performed in the main window will use these parameters/options. The whole set of options/parameter is also printed in the Notebook.

### *F.2.4. Context Help and About*

While each control and indicator generally shows a ‘tip strip’ next to it when the mouse hovers over them, a more extended description can be found in the floating Context Help window, which can be shown or hidden as needed.

The About dialog window provides legal disclaimer and copyright information.

## ***F.3. Analysis***

We provide here a brief summary of the type of analysis currently possible with *Phasor Explorer*. More details can be found in the online manual.

### *F.3.1. Phasor Plot*

The different ‘Single-Exponential Phasor Loci’ (SEPL) discussed in this article can be reproduced by setting the gate parameters, laser period, phasor frequency and calibration lifetime in the respective panels of the Settings window, and then selecting Analysis>> Phasor Plot. This adds the corresponding plot to the graph.

Using a cursor linked to that plot, and moving it along the plotted SEPL, the three indicators labelled ‘tau_i’, ‘tau_ph’ and ‘tau_m’ will be updated according to the cursor location. ‘tau_i’ represents the actual lifetime used to compute the phasor at that location, while ‘tau_ph’ and ‘tau_m’ represent the calculated pseudo phase and modulus lifetimes corresponding to that phasor. Note that these values are computed only for the finite number of lifetime values used to build the SEPL (defined in the Display panel of the Settings window).

*F.3.2. Pseudo Phase Lifetime*

The dependence of the pseudo phase lifetime $\tau_\varphi$ of PSED as defined in this article can be studied using Analysis>>Pseudo Phase Lifetime Plot. The gate parameters, laser period, phasor frequency and calibration lifetime used in this calculation are those defined in the respective panels of the Settings window. A checkbox on the front panel (Compute Delta Tau) allows calculating the difference $\tau_\varphi - \tau$ instead.

*F.3.3. Decay & Gates*

The effect of gating, decay offset, truncation, etc. on the shape of the T-periodic decay (which is used to perform all other computations) can be studied by representing (i) the original PSED, (ii) the selected gate and (iii) the gated-decay using Analysis>>Decay, Gate & Gated Decay. Checkboxes on the front panel allow specifying which of the 3 curves to represent.